\documentclass{aa}

\usepackage{graphicx}
\usepackage{txfonts}
\usepackage{xcolor}
\usepackage{caption}
\usepackage{makecell}
\usepackage{placeins}
\usepackage{float}
\usepackage{geometry}  
\usepackage{hyperref} 
\usepackage{dblfloatfix}

\usepackage{xcolor}
\definecolor{movaki}{rgb}{1, 0.0, 0.8}
\hypersetup
{
    colorlinks,
    linkcolor=[rgb]{0.06, 0.1, 0.7},
    citecolor=[rgb]{0.06, 0.1, 0.7},
    urlcolor=[rgb]{0.06, 0.2, 0.9},
}

\newcommand{\teff}{$T_{\rm eff}$}
\newcommand{\kms}{\ensuremath{\mathrm{km\,s^{-1}}}}
\newcommand{\logg}{\ensuremath{\mathrm{\log g}}}
\newcommand{\vt}{\ensuremath{\mathrm{v_{turb}}}}

\newcommand*\mean[1]{\bar{#1}}

\usepackage[normalem]{ulem}

\begin{document}

   \title{Large databases of metal-poor stars corrected for \\three-dimensional and/or non-local thermodynamic \\equilibrium effects}

\titlerunning{Large databases of metal-poor stars corrected for 3D and/or NLTE effects}

   \author{I. Koutsouridou\thanks{\email{ioanna.koutsouridou@unifi.it }}
          \inst{1}
          \and
          Á. Skúladóttir\inst{1}
          \and
          S. Salvadori\inst{1}
          }

   \institute{Dipartimento di Fisica e Astronomia, Universit\'a degli Studi di Firenze, Via G. Sansone 1, I-50019 Sesto Fiorentino, Italy.
             }

   \date{}

  \abstract{Early chemical enrichment processes can be revealed by the careful study of metal-poor stars. In our Local Group, we can obtain spectra of individual stars to measure their precise, but not always accurate, chemical abundances. Unfortunately, stellar abundances are typically estimated under the simplistic assumption of local thermodynamic equilibrium (LTE). This can systematically alter both the abundance patterns of individual stars and the global trends of chemical enrichment. The SAGA database compiles the largest catalogue of metal-poor stars in the Milky Way. For the first time, we provide the community with the SAGA catalogue fully corrected for non-LTE (NLTE) effects, using state-of-the-art publicly available grids. In addition, we present an easy-to-use online tool \texttt{NLiTE} that quickly provides NLTE corrections for large stellar samples.
  For further scientific exploration, \texttt{NLiTE} facilitates the comparison of different NLTE grids to investigate their intrinsic uncertainties. Finally, we compare the NLTE-SAGA catalogue with our cosmological galaxy formation and chemical evolution model,  \texttt{NEFERTITI}. By accounting for NLTE effects, we can solve the long-standing discrepancy between models and observations in the abundance ratio of [C/Fe], which is the best tracer of the first stellar populations. At low $\rm[Fe/H]<-3.5$, models are unable to reproduce the high measured [C/Fe] in LTE, which are lowered in NLTE, aligning with simulations. Other elements are a mixed bag, where some show improved agreement with the models (e.g.~Na) and others appear even worse (e.g.~Co). Few elemental ratios do not change significantly (e.g. [Mg/Fe], [Ca/Fe]). Properly accounting for NLTE effects is fundamental for correctly interpreting the chemical abundances of metal-poor stars. Our new \texttt{NLiTE} tool, thus, enables a meaningful comparison of stellar samples with chemical and stellar evolution models as well as with low-metallicity gaseous environments at higher redshift.}

\keywords{Galaxy: abundances - stars: abundances - 
          stars: atmospheres - stars: Population II - line: formation }

\maketitle

\section{Introduction}

Metal-poor (MP) stars (${\rm [Fe/H]<-1}$; \citealt{Beers2005}) are ancient relics of the early chemical enrichment in the Universe, providing a unique window into the conditions and processes that shaped the formation of the first stellar populations \citep[e.g.][]{Tumlinson2007,Salvadori2007,Hartwig2015,Sarmento2018, Koutsouridou2024}. These stars, observed today in our Local Group, retain in their photospheres the chemical signatures of the gas clouds from which they formed. Unravelling their detailed and accurate elemental abundances is therefore key to answering numerous scientific questions. In particular, MP stars provide valuable insights into the nature of the first  metal-free Population III (Pop~III) stars (their masses, supernova explosion energies, rotation rates, and mixing processes), and into the transition to normal Population II (Pop~II) star formation \citep[e.g.][]{deBen2017, Ishigaki2018,Hartwig2019,Vanni2023,Koutsouridou2023,Sestito2024}. They can be used to test theoretical predictions of Big Bang nucleosynthesis (e.g. the lithium problem; see \citealt{Fields2011}), stellar nucleosynthesis and galactic chemical evolution models \citep[e.g.][]{Matteucci2021,Rossi2024b,Brauer2025}.

In addition, when paired with kinematic data, MP stars can offer a unique perspective on the accretion and early formation history of the Milky Way and its satellite galaxies \citep[e.g.][]{Gaia2018_HRD, Gaia2018_dwarfs}.

 However, confirming stellar candidates as very metal poor (VMP, ${\rm [Fe/H] < -2}$) or extremely metal poor (EMP, ${\rm [Fe/H] < -3}$) requires medium- to high-resolution spectroscopic follow-up observations, which are resource intensive. Consequently, the number of confirmed VMP stars remains significantly lower than the available candidate pool \citep[e.g.][]{Xylakis-Dornbusch2022,Martin2024}. Databases such as the Stellar Abundances for Galactic Archaeology (SAGA)\footnote{\url{http://sagadatabase.jp/}}  compile such follow-up data, currently including thousands of MP stars observed at high or medium resolution \citep{Suda2008, Suda2011, Yamada2013, Suda2017}.
 
Almost all of the stars in the SAGA database have chemical abundances  determined using one-dimensional (1D) model atmospheres and the assumption of local thermodynamic equilibrium (LTE). In many cases, the basic atmospheric stellar parameters, such as the effective temperature, $T_{\rm eff}$; surface gravity, $\logg$; and micro-turbulence velocity, $v_{\rm turb}$, are also determined spectroscopically within the LTE framework. The LTE assumption is generally valid when frequent particle collisions maintain a Maxwellian velocity distribution in the system, such that the energy level populations are determined solely by the local temperature and electron density, as dictated by the Saha and Boltzmann equations. However, in realistic stellar atmospheres the gas density is low, collisions between particles are rare, and radiative processes — absorption, emission, and scattering of photons — play a dominant role in determining the energy level populations, causing departures from equilibrium. These non-LTE (NLTE) effects can significantly affect the derived chemical abundances, with discrepancies ranging from negligible to over an order of magnitude, depending on the stellar atmospheric conditions and the spectral line analysed (e.g. \citealt{Asplund2005, Mashonkina2014, Amarsi2020, Lind2022,Lind2024}). The problem is compounded at low metallicities, where NLTE effects become progressively stronger due to the decreased collisional rates and increased radiative rates caused by low ultraviolet (UV) opacity \citep{Mashonkina2023}. These metallicity-dependent NLTE effects, can create artificial abundance trends with metallicity in the LTE assumption, potentially distorting our understanding of early chemical evolution.

Additionally, 1D atmosphere models, which assume static and homogeneous layers, overlook dynamic 3D phenomena, for example stellar granulation caused by convection. These processes create temperature and density inhomogeneities that impact spectral line formation, often in a direction opposite to NLTE effects \citep[e.g.][]{Asplund2005}. Moreover, 3D and NLTE effects are non-linearly coupled, and therefore only models that account simultaneously for the two effects can provide highly accurate and reliable abundance determinations \citep{Lind2024}. Accurate chemical abundances, taking into account 3D and/or NLTE effects, are therefore fundamental so that the stellar observations can be contrasted against models in a meaningful way \citep[e.g.][]{Cayrel2004,Kobayashi2020,Skuladottir2024,Storm2025} and can be compared to higher-redshift observations of gaseous absorbers, which do not undergo the same effects \citep[e.g.][]{Cooke2011,Skuladottir2018,Welsh2022,Saccardi2023,Vanni2024}.

To address these challenges, much effort has been put in the development of sophisticated 1D and 3D\,NLTE models, including high-quality model atoms, atmospheres and spectral synthesis techniques (for more details see the comprehensive reviews by \citealt{Asplund2005} and \citealt{Lind2024}, and references therein). Currently, 1D\,NLTE abundance corrections for specific spectral lines as a function of atmospheric parameters ($T_{\rm eff}$, $\logg$, [Fe/H] and in cases $v_{\rm turb}$, [X/Fe] or line equivalent width) are readily available on the following websites:

 \begin{enumerate}
     \item MPIA\footnote{\url{https://nlte.mpia.de/gui-siuAC_secE.php}} for lines of \ion{O}{I}, \ion{Mg}{I}, \ion{Si}{I}, \ion{Ca}{I-II}, \ion{Ti}{I-II}, \ion{Cr}{I}, \ion{Mn}{I}, \ion{Fe}{I-II}, and \ion{Co}{I}
     \item INASAN\footnote{\url{https://spectrum.inasan.ru/nLTE/}} for lines of  \ion{Na}{I}, \ion{Mg}{I}, \ion{Ca}{I} and \ion{Ca}{II}, \ion{Ti}{II}, \ion{Fe}{I}, \ion{Zn}{I-II}, \ion{Sr}{II}, \ion{Ba}{II} and \ion{Eu}{II}  
     \item INSPECT\footnote{\url{http://www.inspect-stars.com/}} for lines of \ion{Li}{I}, \ion{O}{I}, \ion{Na}{I}, \ion{Mg}{I}, \ion{Ti}{I}, \ion{Fe}{I-II},  \ion{Sr}{II}
 \end{enumerate}

\noindent In addition, various grids of 1D\,NLTE corrections for individual elements are available in the literature \citep[e.g.][]{Takeda2005,Korotin2015_Ba2,Nordlander2017}. Due to the large computational cost involved, grids of 3D\,NLTE corrections are currently available only for a few chemical species and in most cases a limited range of stellar parameters \citep[e.g.][]{Sbordone2010,Amarsi19,Amarsi2022,Gallagher2020}.

Although a plethora of studies have corrected the abundances of various stellar samples for one or more chemical elements \citep[e.g.][]{Andrievsky2007, Andrievsky2008, Andrievsky2009, Andrievsky2010, Zhao2016, Mashonkina2017,Mashonkina2019, Kovalev2019, Mashonkina2022,Shen2023}, a unified catalogue encompassing a large sample of NLTE-corrected abundances for MP stars, extending down to the lowest metallicities, is still missing. As a result, theoretical predictions are commonly compared with uncorrected datasets, such as the SAGA database \citep{Hartwig2018, Kobayashi2020, Koutsouridou2023, Vanni2023, Rossi2024}. This practice can lead to flawed conclusions regarding  stellar nucleosynthesis, galactic chemical evolution, and the properties of the first stars.

In this work, our aim is to apply NLTE corrections to the entire SAGA catalogue of Galactic MP stars by utilizing all available NLTE grids. The main challenge with this endeavour is that SAGA does not include chemical abundance measurements for specific spectral lines, which are necessary for calculating precise and accurate NLTE corrections. To tackle this, we identified the most commonly observed spectral lines for each element, as a function of stellar atmospheric parameters ($T_{\rm eff}$, $\logg$, [Fe/H] and in cases [X/Fe]), and computed average NLTE corrections for these lines from the available grids. While these corrections are approximate — due to differences in the lines used across observational studies and the fact that NLTE grids are sometimes available only for a subset of the lines commonly used by observers — they are statistically robust and provide a reliable representation of general NLTE effects. Thus, they are adequate for analysing large datasets to be compared with theoretical chemical evolution models.

Currently, SAGA is the largest available catalogue for chemical abundances of MP stars, but this is likely to change in the near future with the large ongoing and upcoming spectroscopic surveys, such as GALAH, LAMOST, WEAVE and 4MOST \citep{Martell2017,Zhao2012,Jin2024,deJong2019}. Some of these surveys (GALAH, 4MOST) are aiming to release state-of-the-art NLTE abundances for all chemical elements, while other surveys might mainly rely on the LTE approach. Thus, the need is evident for an efficient tool to easily correct large stellar databases for NLTE effects. As part of this effort, we therefore developed the online tool \texttt{NLiTE}, which is optimized for MP stars analysed through optical spectra. The tool interpolates within precomputed average NLTE grids to provide corrections given the atmospheric parameters of each individual star. It is particularly useful for quickly providing NLTE-corrections for large stellar samples and/or when the information about individual spectral lines is unavailable. Furthermore, \texttt{NLiTE} facilitates direct comparisons between different NLTE studies, as it includes multiple NLTE grids for the same element where available. 

With this work, we therefore provide the astronomic community with a full NLTE-corrected SAGA catalogue of MP stars (online Table~\ref{table:Fiducial}), as well as the \texttt{NLiTE} tool\footnote{\url{https://nlite.pythonanywhere.com/}} to easily correct large databases and to compare different studies of NLTE effects. Finally, we compare this new NLTE SAGA database with the predictions of our cosmological galaxy formation model of the Milky Way halo, {\sc NEFERTITI} \citep{Koutsouridou2023}. Thus, we show the importance of using accurate chemical abundances when trying to understand early chemical enrichment and the properties of the first stars in the Universe.

\section{The \texttt{NLiTE} online tool}
\label{Nlite}

We present the online tool \texttt{NLiTE}, designed to provide NLTE abundance corrections for MP stars, which have been analysed using optical spectra ($3\,500\,\AA\lesssim\lambda\lesssim10\,000\,\AA$). The main goals of this tool are: a)~Construct a fully corrected NLTE-SAGA database (Sec.~\ref{sec:SAGA}) to contrast against models (Sec.~\ref{sec:nefertiti}); b)~Provide an easy way to compare different grids of NLTE corrections, and establish which elemental ratios [X/Fe] are the least or most affected; c)~Have a readily available tool for the community to use and correct large databases of MP stars.
For the public use, two modes of \texttt{NLiTE} are available: 
1)~Correcting a single element with a grid of choice; or  2)~Correcting all elements using our fiducial grids (Table~\ref{table:Grids}). Both modes include the option to receive a ready-to-use bib file, to facilitate and encourage citations to the original NLTE grids.

By interpolating within precomputed grids, \texttt{NLiTE} computes corrections based on given stellar atmospheric parameters: $T_{\rm eff}$, $\log g$, [Fe/H], and either [Element/Fe] or [Element/H]. 
Unlike other tools, \texttt{NLiTE} does not take spectral line data as input. The precomputed grids are built from publicly available NLTE datasets (Sec.~\ref{grids}) that are originally tied to specific spectral lines. These grids are averaged using the corrections of the lines most representative of MP stars at given atmospheric conditions. As a result, while \texttt{NLiTE} provides approximate corrections, it can be particularly useful for analysing large stellar samples or when equivalent width measurements are unavailable. Furthermore, by incorporating multiple grids per element when available, \texttt{NLiTE} facilitates direct comparisons between NLTE corrections from different studies.

\subsection{Line lists}
\label{Linelists}

For each element, we assume a line list that is the intersection between the lines most frequently used in the chemical abundance analysis of MP stars and those for which NLTE corrections exist. The selection of the commonly used lines is largely based on the studies of \citet{Fulbright2000, Cayrel2004, Barklem2005,  Ishigaki2012, Cohen2013, Roederer14, Jacobson2015, Sakari2018}; and \citet{Li2022}. 
Among these, we select the lines with available NLTE corrections. In cases where multiple NLTE grids are available for the same element, we choose when possible to adopt lines that are common across all grids. This allows a direct comparison between the published NLTE grids, which is presented in Sec.~\ref{sec:grids}, and in Appendix~\ref{app: NLTEcorr}. Our analysis shows that in most cases, the NLTE corrections of different lines (at given stellar parameters) are in good agreement, with standard deviation $\sigma \lesssim0.1$\,dex (see Sec.~\ref{Scatter}). This indicates that our adopted line list is a reliable way to calculate NLTE corrections for large samples, since in general these are not strongly dependent on the exact line list. The adopted line list is published in an online Table~\ref{table: linelist}.

\renewcommand{\arraystretch}{1.5}
\begin{table}[t]
\centering
\footnotesize
\caption{Line list}
\tabcolsep=0.48cm
\begin{tabular}{c c}
\hline
\hline
Species & Wavelength \\
&  \multicolumn{1}{c}{(\AA)} \\
\hline
\ion{Li}{I} & 6707.8 \\
\ion{C}{I} & 9094.8 \\
\ion{C}{I} & 9111.8 \\
\ion{C}{I} & 9405.7 \\
\ion{O}{I} & 7772.0 \\
\ion{O}{I} & 7774.2 \\
... & ... \\

\hline
\end{tabular}

\noindent\parbox{0.23\textwidth}{
    \vspace{2.5mm} 
    \small (Full table available at the CDS.)}

\label{table: linelist}
\end{table}

Depending on the set of stellar parameters ($T_{\rm eff}$, \logg, ${\rm [Fe/H]}$, and ${\rm [X/H]}$), certain lines may be very weak (making them challenging to measure) or severely saturated (making them less sensitive to abundance) and are, thus, commonly discarded in observational studies. To account for this, for each stellar parameter set we compute the average NLTE correction using only lines with predicted equivalent widths (EW) in the range $5\:{\rm m\AA}< {\rm EW}<200\:{\rm m\AA}$. Exceptions to this are elements that have only a few lines available, which are therefore rarely discarded. Thus, we put no upper limit on the EW for: Li, Al, Na, K, Sr, and Ba.

\subsection{NLTE grids}
\label{grids}

The NLTE grids are adopted from the literature, in particular from the databases of MPIA,\footnote{For MPIA the option of the plane-parallel 1D atmosphere model has been chosen.} INASAN, INSPECT, that of Anish M. Amarsi,\footnote{\url{https://www.astro.uu.se/~amarsi/}} as well as individual studies for specific elements (Table~\ref{table:Grids}).
For each element X, the grids are given as a function of $T_{\rm eff}$, $\logg$, ${\rm [Fe/H]}$, and in some cases elemental abundance ${\rm [X/H]}$. The range of the parameter space for each grid is given in Table~\ref{table:Grids}. 

Through \texttt{NLiTE} we facilitate the use of all available grids for individual elements. However, for our NLTE-SAGA database we choose a fiducial grid for each element, as listed in Table~\ref{table:Grids}. Since our final goal is to be able to correct large databases with stars in various evolutionary phases, we typically adopt as our fiducial grids those that cover the largest parameter space.

\label{sec:grids}

\subsection{Impact of microturbulence}
\label{Microturbulence}

\begin{figure}
\begin{center}
\includegraphics[width=0.99\hsize]{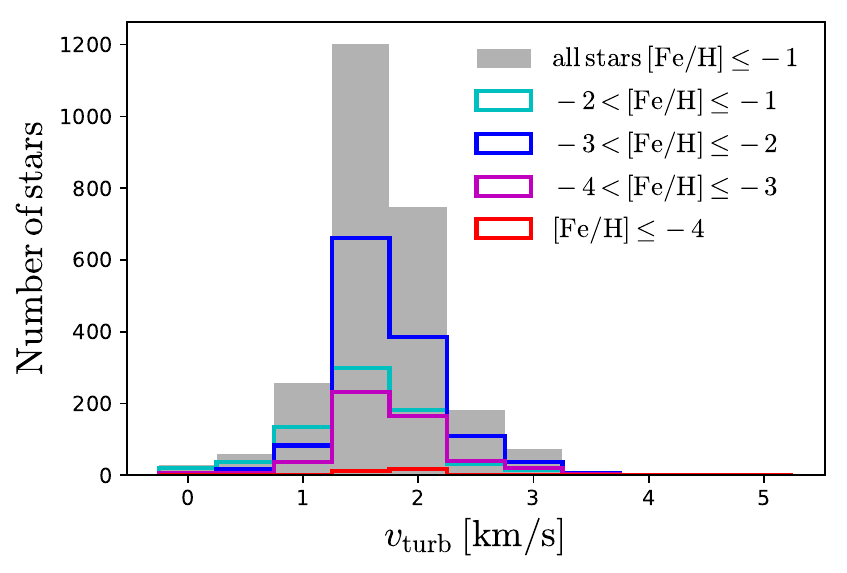} 
\caption{All MP SAGA stars with microturbulence velocity, $v_{\rm turb}$, estimates in different metallicity bins. In the case of multiple entries from different surveys and/or authors for the same star, the microturbulence and [Fe/H] here are equal to the mean values.}
\label{fig: SAGA_microturbulence}
\end{center}
\end{figure}

Several NLTE grids include microturbulent velocity, $v_{\rm turb}$, as an input parameter. By inspecting the distribution of $v_{\rm turb}$ in all the observed metal-poor MW stars in the SAGA database (Fig.~\ref{fig: SAGA_microturbulence}), we find that it peaks at $v_{\rm turb}=1.5\,$km/s in all metallicity bins, except from the lowest one, $\rm [Fe/H]\leq-4$ where it peaks at $v_{\rm turb}=2\,$km/s. But these stars represent only $\sim1\%$ of the total MP population.

\begin{figure}
\begin{center}
\includegraphics[width=0.99\hsize]{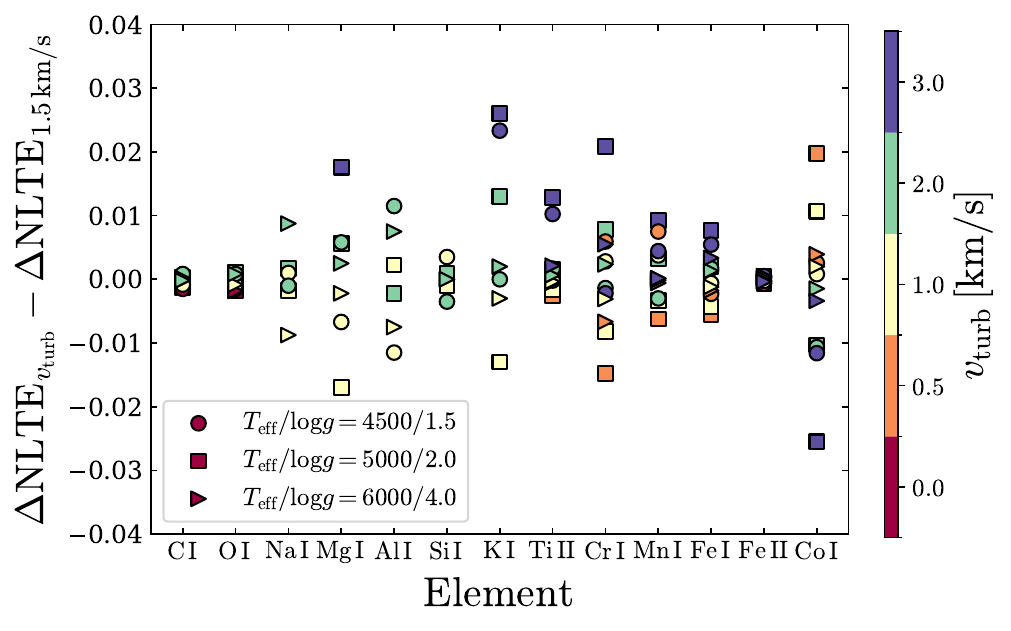} 
\caption{Difference in NLTE corrections, when assuming $ v_{\rm turb}=0,0.5,1,2$ and 3\,km/s, compared to the most common value for MP SAGA stars, $v_{\rm turb}=1.5$\,km/s. Three representative stellar model atmospheres at $\rm [Fe/H]=-2$ are shown (circles, squares, triangles).
}
\label{fig: micro}
\end{center}
\end{figure}

To assess the impact of $v_{\rm turb}$ on the NLTE corrections, we investigate three representative stars at $\rm [Fe/H]=-2$: (i)~$T_{\rm eff} = 4500$ K, $\log g = 1.5$; (ii)~$T_{\rm eff} = 5000$ K, $\log g = 2.0$; and (iii)~$T_{\rm eff} = 6000$ K, $\log g = 4.0$. 
Fig.~\ref{fig: micro} compares the mean NLTE corrections (averaged across all spectral lines considered; see Table~\ref{table: linelist}) as a function of $v_{\rm turb}$, ranging from 0-3 km/s, to the case where the most common $v_{\rm turb}=1.5\,$km/s is assumed. The results are based on our fiducial grids for each element (see Table~\ref{table:Grids}) and the assumed abundance ratios correspond to typical MW values at this metallicity: [C/Fe] = 0, [O/Fe] +0.6, [Na/Fe] = 0, [Mg/Fe] = +0.4, [Al/Fe] = -0.5, and [Si/Fe] = +0.5.
It should be noted that not all grids offer the full range of 0-3\,km/s. Furthermore, the elements \ion{Li}{I}, \ion{Ca}{I}, \ion{Zn}{I}, \ion{Sr}{II}, \ion{Ba}{II}, and \ion{Eu}{II} are missing because \citet{Wang2021} and the INASAN database do not include $v_{\rm turb}$ as an input parameter. Similarly, \citet{Norris19} do not account for $v_{\rm turb}$ in their corrections of CH.

Fig.~\ref{fig: micro} shows that our fiducial NLTE corrections do not depend strongly on $v_{\rm turb}$. In all cases, the deviations from the standard $v_{\rm turb} = 1.5$\,km/s case remain within 0.03\,dex. Therefore we adopt this fixed value as a standard in our NLTE grids whenever $v_{\rm turb}$ is an input parameter. Further discussion on other NLTE grids and broader parameter sets can be found in the respective sections for each element (Sec.~\ref{sec:grids}).

\subsection{Interpolation-extrapolation} 
\label{sec:code}

For each element and corresponding set of NLTE corrections (see Table~\ref{table:Grids}) we construct an interpolation function in a three- or four-dimensional space ($T_{\rm eff}$, $\logg$, [Fe/H] and when available [X/Fe] or A(X)), using the linear \texttt{scipy.interpolate.Rbf} function in Python. 

We note that, for stellar parameters within each $\Delta$NLTE grid, the scipy multiquadric fitting function gives corrections that differ less than $0.1\,$dex from those obtained with the linear method. For stellar parameters outside the grids, the differences tend to be larger. Therefore, rather than extrapolating beyond the available grids, \texttt{NLiTE} applies the NLTE corrections corresponding to the nearest grid boundary.

\section{NLTE grids for individual elements}
\label{sec:grids}

\subsection{Lithium I}
\label{Lithium}

\begin{figure*}
\begin{center}
\includegraphics[width=0.9\hsize]{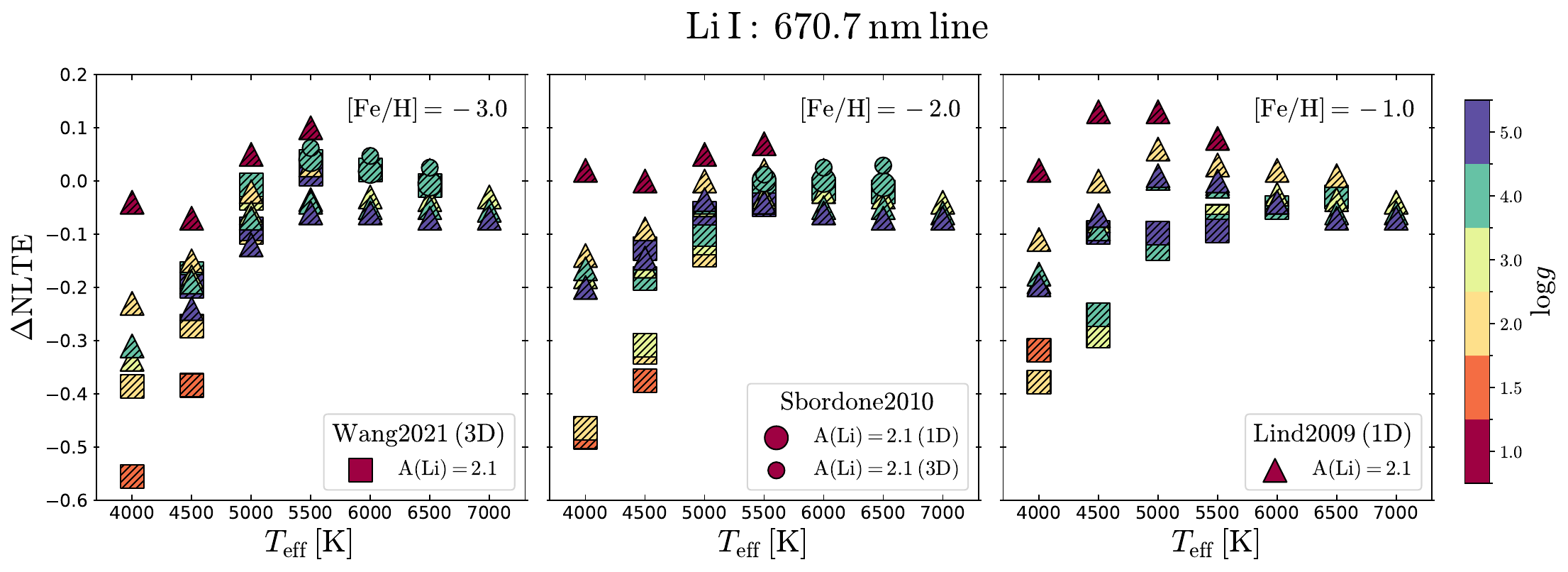} 
\caption{NLTE corrections for \ion{Li}{I}, colour-coded by $\logg$: squares are from \citet{Wang2021}, triangles from \citet{Lind2009_Li1} and circles from \cite{Sbordone2010}. Three metallicities are shown: $\rm [Fe/H]=-3$ (left), $\rm [Fe/H]=-2$ (middle), and $\rm [Fe/H]=-1$ (right).}
\label{fig: Li1_lines}
\end{center}
\end{figure*}

We computed NLTE corrections for the \ion{Li}{I} resonance line at 670.7 nm adopting the grids from \citet{Lind2009_Li1}, \citet{Sbordone2010} and \citet{Wang2021}, see Fig.~\ref{fig: Li1_lines}.

The \citet{Lind2009_Li1} corrections were derived using 1D-model atmospheres, and are available for $v_{\rm turb}=1, 2\,$and 5$\,$km/s. We adopted the average between the $v_{\rm turb}=1$ and $v_{\rm turb}=2\,$km/s corrections, noting that their differences do not exceed $0.1\,$dex, at fixed $T_{\rm eff}$, $\logg$ and [Fe/H].

\citet{Sbordone2010} computed corrections for both 1D- and 3D-hydrodynamical model atmospheres and provided analytical fits of A(Li)$_{\rm 3D,NLTE}$, A(Li)$_{\rm 1D,NLTE}$ and A(Li)$_{\rm 3D,NLTE}$ as functions of equivalent width, $T_{\rm eff}$, $\logg$, and [Fe/H] but restricted only to dwarf stars. In the overlapping parameter space, we find that the 1D NLTE corrections from \citet{Sbordone2010} are weaker than those of \citet{Lind2009_Li1}, which are more negative by approximately 0.05 dex. The 3D NLTE corrections of \citet{Sbordone2010} are slighlty positive, about 0.03\,dex higher than their 1D values, in accordance with previous studies who reported $<0.1$~dex differences in lithium abundance between 3D NLTE and 1D NLTE \citep{Asplund2003,Barklem2003}.

\citet{Wang2021} provided 3D NLTE corrections covering a broader parameter range along with an interpolation routine\footnote{\url{https://github.com/ellawang44/Breidablik}} based on multilayer perceptrons (a class of fully connected
feedforward neural networks), which we used to construct our NLTE grid. We note that using our linear interpolation method, we find corrections for individual stars that differ by no more than 0.018 dex from those obtained with \citet{Wang2021}'s interpolation routine.

The \citet{Wang2021} corrections are generally more negative than those of \citet{Lind2009_Li1} by approximately 0.1\,dex at $T_{\rm eff} \leq 4500$ K. At higher $T_{\rm eff}$, this trend reverses. Compared to the 3D corrections from \citet{Sbordone2010}, the \citet{Wang2021} values are, on average, about 0.05\,dex more negative. These differences may arise from a NLTE effect identified by \citet{Wang2021} that was previously overlooked, involving the blocking of UV lithium lines by background opacities.

Overall, the NLTE corrections span the range 
[$-1.1$,+0.45]\,dex (not all values are shown in Fig.~\ref{fig: Li1_lines})  and tend to be more positive at lower A(Li) values in all grids. Given the broader parameter coverage of \citet{Wang2021}, we adopt this as our fiducial grid.

\citet{Mott2020} computed also 1D and 3D NLTE A(Li) abundance corrections. We did not incorporate those here, as the authors provide their own Python script that evaluates them as a function of $T_{\rm eff}$, $\logg$, [Fe/H], lithium isotopic ratio $^{6}\mathrm{Li}/^{7}\mathrm{Li}$ and A(Li).\footnote{\url{https://gitlab.aip.de/mst/Li_FF/}} However, we note that their 1D corrections are typically 0.02\,dex lower than their 3D counterparts, with the latter being, on average, 0.06\,dex more positive than those of \citet{Wang2021}.

Finally, \citet{Shi2007} reported 1D NLTE abundances for Li in 19 stars with $ \rm -2.5<[Fe/H]<-1$, and $3<\logg<5$. Their corrections are small, $ \rm 0<|\Delta_{\rm1D\,NLTE}|<0.03$, for the majority of the sample, and generally in good agreement with those of \citet{Sbordone2010}, while \citet{Lind2009_Li1} and \citet{Wang2021} present slightly stronger negative corrections. However, the work of \citet{Shi2007} does not include a grid of corrections, and is thus not implemented here.

\subsection{Molecular carbon (CH)}

Due to complexities in modelling molecular spectra, which involve numerous energy levels and lines, a full grid of NLTE corrections for the molecular CH G-band is not currently available. We, therefore, adopted the empirical CH corrections from \citet{Norris19}, who considered the analysis of near-infrared high-excitation \ion{C}{I} lines in metal-poor stars, to assess the role of NLTE effects in determining A(C)$_{\rm 3D,NLTE}$ values from G-band data.

The authors, initially, compiled 9 ($-5.7\leq {\rm [Fe/H]}\leq-1$) stars with existing 3D-1D LTE CH corrections \citep{Collet2006, Collet2007, Collet2018, Frebel2008, Spite2013,Gallagher2016} and computed the linear least-squares best fit to the data:

\begin{equation}
    {\rm A(CH)_{\rm 3D,LTE}} = {\rm A(CH)_{\rm 1D,LTE}} + 0.087 + 0.170\,{\rm [Fe/H]_{1D,LTE}}.
\label{e:CH3D}
\end{equation}
They then determined A(CH)$_{\rm 1D, LTE}$ abundances for 23 stars (dwarfs and subgiants) using high-resolution, high signal-to-noise spectra from the literature, with A(CI)$_{\rm 1D, NLTE}$ abundances previously provided by \citet{Fabbian2009}. 
Using Equation~\ref{e:CH3D}, they converted A(CH)$_{\rm 1D, LTE}$ to A(CH)$_{\rm 3D, LTE}$ abundances, and noted that 3D corrections to A(CI)$_{\rm 1D, NLTE}$ are likely insignificant, as indicated by \citet{Fabbian2009} and \citet{Dobrovolskas2013}. By requiring that A(CH)$_{\rm 3D, NLTE}$ = A(CI)$_{\rm 3D, NLTE}$, they found the best linear fit for ${\rm [Fe/H]}>-3$:
\begin{equation}
    {\rm A(CH)_{\rm 3D,NLTE}} = {\rm A(CH)_{\rm 3D,LTE}} + 0.483 + 0.240\,{\rm [Fe/H]_{1D,LTE}}
\end{equation}
For lower metallicities, where CH features and \ion{C}{I} lines are weak, the authors made the conservative assumption that A(CI)$_{\rm 3D, NLTE}$ = A(CH)$_{\rm 3D, LTE}$, and hence:
\begin{equation}
    {\rm A(CH)_{\rm 3D,NLTE}} = {\rm A(CH)_{\rm 3D,LTE}}.
\end{equation}  
The above equations imply that strong negative corrections should be applied to the observed A(CH)$_{\rm 1D,LTE}$ abundances for stars with [Fe/H]$\lesssim-1.4$, reaching $\sim -0.9\,$dex at $\rm[Fe/H]=-6$.

Recently, \citet{Popa2023} made the first attempt at computing 
the G band of the CH molecule in NLTE for a cool stellar atmosphere typical of red giants (\logg = 2.0, Teff = 4500 K). Their results showed that A(C)$_{\rm 1D, NLTE}$-A(C)$_{\rm 1D, LTE}$ corrections are consistently positive and increasing towards lower [Fe/H] and [C/Fe]. However, these findings are subject to significant uncertainties, since molecular collisional data are still poorly constrained \citep{Lind2024}. Furthermore, \citet{Popa2023} acknowledged that their analysis, based on 1D hydrostatic models, neglects time-dependent 3D phenomena such as convection and turbulence, that have been shown to lower CH abundances at least in LTE (see above). Therefore, it remains unclear whether accounting for 3D effects will overcompensate for the positive 1D\,NLTE - 1D\,LTE corrections,  resulting in the net negative corrections suggested by \citet{Norris19}.

\begin{figure*}
\begin{center}
\includegraphics[width=0.7\hsize]{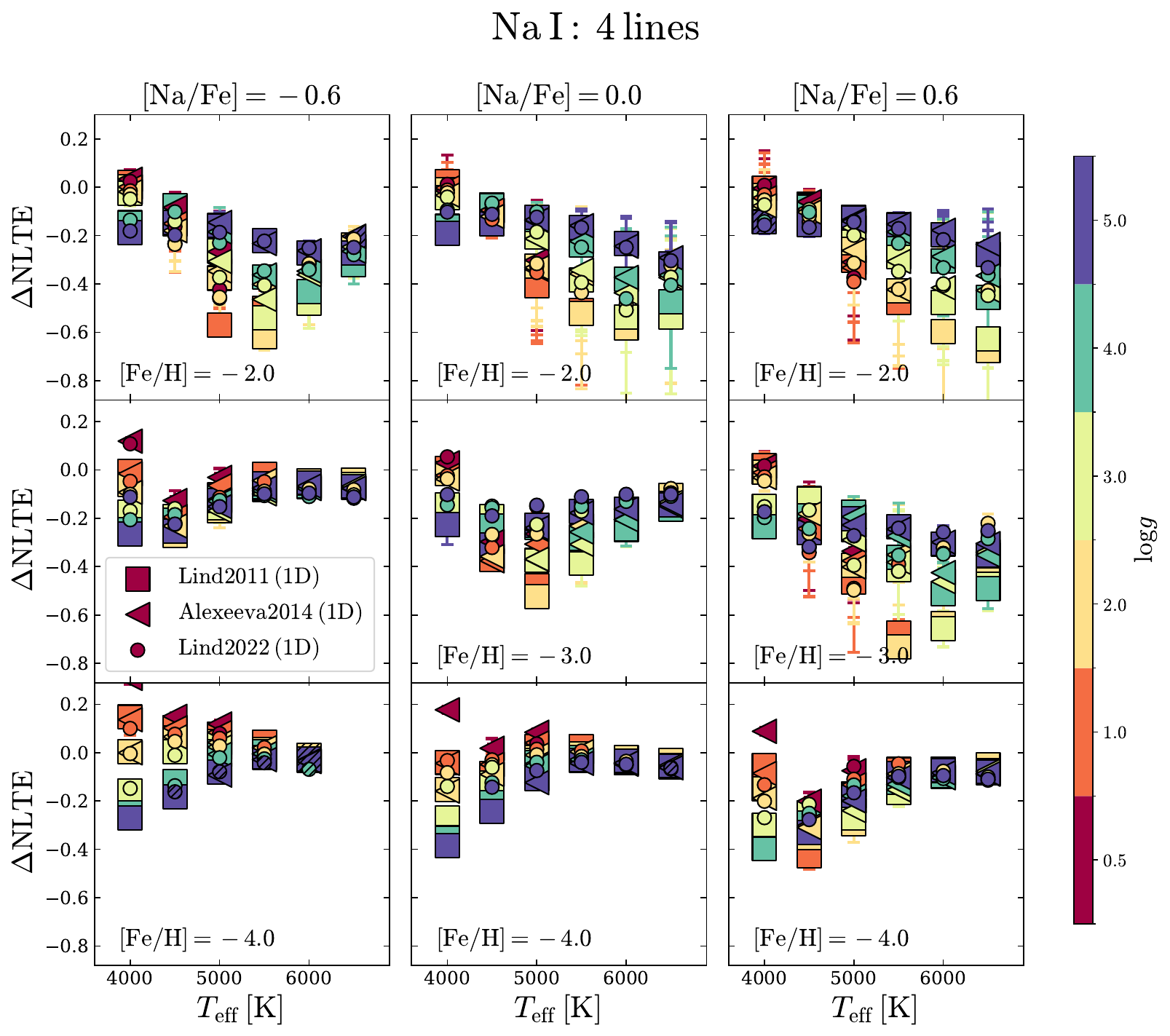} 
\caption{NLTE corrections for \ion{Na}{I}, colour-coded by $\logg$: squares are from \citet{Lind2011_Na1}, triangles from \citet{Alexeeva2014_Na1}, and circles from \citet{Lind2022}. Rows show three metallicities: $\rm [Fe/H]=-2$ (top), $\rm [Fe/H]=-3$ (middle), and $\rm [Fe/H]=-4$ (bottom). Columns show different [Na/Fe] LTE values: $\rm [Na/Fe]=-0.6$ (left), $\rm [Na/Fe]=0$ (middle) and $\rm [Na/Fe]=+0.6$ (right). Symbols are hatched diagonally in cases where only one line is available (i.e. ${\rm EW}>5\:{\rm m\AA}$).
Error bars represent the standard deviation of the NLTE corrections of the different \ion{Na}{I} lines at a given set of stellar parameters.}
\label{fig: Na1_common_lines}
\end{center}
\end{figure*} 

\subsection{Carbon I and oxygen I}
We include grids for the 3D\,NLTE corrections of \ion{C}{I} and \ion{O}{I} from \citet{Amarsi19}, in the provided range of $5000 \, {\rm K} \leq T_{\rm eff} \leq  6500 \,$K, $3\leq \logg \leq 5$, and 
$ -3\leq {\rm [Fe/H]} \leq 0$. In addition, we provide the grid for their 1D\,NLTE corrections for \ion{C}{I} and \ion{O}{I}, in the range $\rm 4000 \,K \leq T_{\rm eff} \leq7500\,$K, $0\leq \logg \leq5$, ${\rm -5 \leq [Fe/H] \leq 0}$ and ${\rm -0.4 \leq [X/Fe] \leq 1.2}$. We used three infrared lines for \ion{C}{I}, 909.5, 911.1 and 940.6\,nm, which are still visible in extremely metal-poor stars and the \ion{O}{I} triplet at 777\,nm. Since EWs were not provided, corrections were included for all three lines for the entire grid, both for \ion{C}{I} and \ion{O}{I}.

The comparison between the 1D\,NLTE and 3D\,NLTE results of \citet{Amarsi19} are shown in Figs~\ref{fig: C1_lines} and \ref{fig: O1_lines}. In addition, we computed the mean 1D\,NLTE corrections from \citet{Bergemann2021_O1} for the \ion{O}{I} 777\,nm triplet, which we find to be in general $\sim$0.1-0.3\,dex weaker than those of \citeauthor{Amarsi19} (\citeyear{Amarsi19}; see Fig.~\ref{fig: O1_lines}). Here we adopt the corrections from \citet{Amarsi19} as our fiducial \ion{O}{I} grid, as it has a larger range in [Fe/H] compared to that of \citet{Bergemann2021_O1}.

The work of \citet{Spite2013} calculated 3D\,NLTE corrections for \ion{C}{I} lines in two stars at ${\rm [Fe/H]}=-3.3$, $T_{\rm eff} \approx 6200\,$K, and $\logg=4.0$. They find corrections of $\Delta_{\rm 3D\,NLTE}=-0.45$\,dex for both stars, while the Amarsi corrections for these stars are less strong,  $\Delta_{\rm 3D\,NLTE}\lesssim-0.2$\,dex. Other works that have computed 1D\,NLTE corrections for \ion{C}{I} lines include \citet{Takeda2005a,Fabbian2006}, and \citet{Alexeeva2015}. The results of these three papers were compared in \citeauthor{Alexeeva2015} (\citeyear{Alexeeva2015}, their Fig.~5) for a test case of $T_{\rm eff}=6000\,$K and $\logg = 4$. Comparable results were found at $\rm[Fe/H]\gtrsim-1$, but significantly smaller corrections at $\rm[Fe/H]\lesssim-2$, compared to the older works. For this specific test case at low metallicities, the work of \citet{Amarsi19} is generally in agreement with that of \citet{Alexeeva2015}, within $\approx0.05$\,dex. None of these aforementioned works \citep{Takeda2005a, Fabbian2006, Alexeeva2015, Spite2013} provide full 1D NLTE \ion{C}{I} correction grids, and are therefore not included in our selected grids.

In regards to \ion{O}{I}, \citet{Takeda2003} calculated 1D\,NLTE corrections for a sample of Milky Way disk and halo stars (late-F through early-K types). They found a typical correction of $\Delta_{\rm 1D\,NLTE}=-0.1$\,dex for their stellar sample at $\rm[Fe/H]<-1$. This is comparable to the 1D\,NLTE results of \citet{Amarsi19} for similar stars, agreeing within $\approx0.05$\,dex. \citet{Sitnova2018_OI} found slightly lower corrections at $\rm[Fe/H]<-1$, $\Delta_{\rm 1D\,NLTE}\approx-0.05$\,dex, deviating by $\approx0.1$\,dex from \citet{Amarsi19}. The 1D\,NLTE corrections for \ion{O}{I} as calculated by \citet{Fabbian2009_OI}, on the other hand, are stronger than those of \citet{Amarsi19}, with differences up to $\approx0.4$\,dex at the lowest [Fe/H] (as seen by Fig.~7 in \citealt{Fabbian2009_OI}). As far as we are aware, the grids of \citet{Takeda2003}, \citet{Fabbian2009_OI} and \citet{Sitnova2018_OI} are not available publicly, and are therefore not included in this work.

\subsection{Molecular nitrogen (CN)}
In metal-poor stars, nitrogen is most commonly measured through the CN molecular bands, and less frequently through NH lines. Calculating NLTE grids for molecular bands is computationally expensive and complex. Furthermore, 3D\,effects are likely to be significant, since molecular lines are typically very sensitive to temperature. Finally, when N is measured through CN, the accurate abundance for C is also crucial, which is non-trivial to obtain, see the previous subsections. In the absence of a NLTE grid for CN, we include only [N/Fe]$_{\rm LTE}$ in our fiducial SAGA catalogue, marked specifically to indicate that this is a LTE result.

\subsection{Sodium I}
\label{sodium}
We computed NLTE corrections for the \ion{Na}{I} doublets at 5682/5688$\, \AA$ and 5889/5895$\, \AA$ employing the grids of \citet{Lind2011_Na1}, \citet{Alexeeva2014_Na1} and \citet{Lind2022} (see Fig.~\ref{fig: Na1_common_lines}). The \citet{Lind2022} corrections are available for $v_{\rm turb} = 1$ and 2 km/s, with the differences between them remaining below 0.13 dex across all lines and stellar parameters. We adopted the average of the two.

The two Na doublets are known to have significantly different NLTE corrections \citep[e.g.][]{Lind2022}. Therefore, applying a single mean correction across both could lead to inaccurate Na abundances, in cases where only one doublet is observed. To avoid this, we divided SAGA stars into three groups: those with Na abundances based on the resonance lines at 5889/5895$\, \AA$, those based on the subordinate lines at 5682/5688$\, \AA$ and those observed using both doublets. For each group, we applied the mean correction of the corresponding lines. For SAGA stars with no information on the Na lines used, we applied the mean corrections of all four lines, which are shown in Fig.~\ref{fig: Na1_common_lines}. All three versions of the grids are available in \texttt{NLiTE}. 

The mean corrections are predominantly negative, reaching $\sim -0.7\,$dex ($\sim -0.4\,$dex and $-1\,$dex for the subordinate and resonance lines, respectively), except at the lowest metallicities, where a slight, positive upturn is found. As seen in Fig.~\ref{fig: Na1_common_lines}, at $\rm[Fe/H]=-3$ and $\rm[Fe/H]=-2$, the corrections are generally more negative at higher temperatures and lower surface gravities. However, these trends appear to reverse at the lowest metallicity bin. 

Overall the agreement between the different works is quite good, differing by $\lesssim0.1$\,dex in most cases. Yet, in some parts of the parameter space, the differences between different grids can reach $\sim0.3$\,dex (see also Sec.~\ref{Differences between grids}). As our fiducial grid we choose that of \citet{Lind2022} since it covers the largest parameter space.

\subsection{Magnesium I}
\label{Magnesium}

\begin{figure*}
\begin{center}
\includegraphics[width=0.9\hsize]{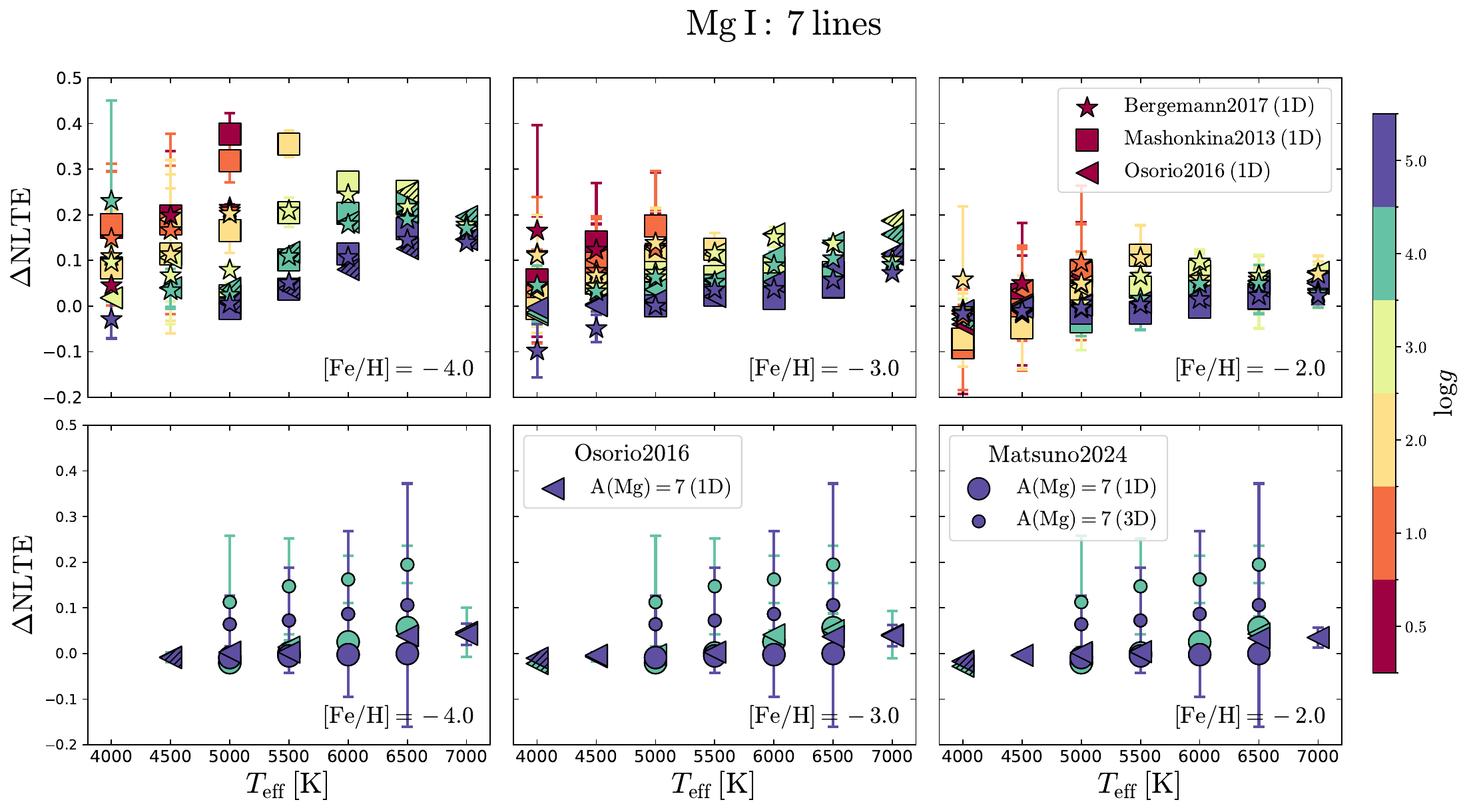} 
\caption{NLTE corrections for \ion{Mg}{I}, colour-coded by $\logg$: stars are from \citet{Bergemann2017_Mg1}, squares are from \citet{Mashonkina2013_Mg1}, triangles from \citet{Osorio2016_Mg1}, and circles from \citet{Matsuno2024}. The top row compares the corrections of \citet{Osorio2016_Mg1} assuming the same EWs as given by \citet{Bergemann2017_Mg1} and \citet{Mashonkina2013_Mg1} for each \ion{Mg}{I} line, at fixed $T_{\rm eff}$, $\logg$ and [Fe/H] values. The bottom row compares \citet{Osorio2016_Mg1} to the 1D and 3D NLTE corrections of \citet{Matsuno2024}, only available for dwarf stars, for A(Mg)=7. Error bars represent the standard deviation of different \ion{Mg}{I} lines (often smaller than the depicted symbols); hatched symbols indicate corrections that are based only on one line.}
\label{fig: Mg1_common_lines}
\end{center}
\end{figure*} 

We computed mean corrections for seven optical \ion{Mg}{I} lines, using the grids of \citet{Merle2011}, \citet{Mashonkina2013_Mg1}, \citet{Bergemann2017_Mg1}, \citet{Osorio2016_Mg1}, \citet{Lind2022} and  \citet{Matsuno2024} (see Fig.~\ref{fig: Mg1_common_lines}). 
The last three grids are given as a function of abundance A(Mg), while $\rm[Mg/Fe]=+0.4$ is adopted for metal-poor stars in the grids of \citet{Merle2011} and \citet{Mashonkina2013_Mg1}. By computing the \citet{Osorio2016_Mg1} corrections for the same EWs as given by \citet{Mashonkina2013_Mg1} and \citet{Bergemann2017_Mg1} at each $T_{\rm eff}$, $\logg$ and [Fe/H], we see that the three grids are in remarkable agreement (top panels of Fig.~\ref{fig: Mg1_common_lines}). The same is true for the corrections of \citet{Lind2022}, which are not displayed in Fig.~\ref{fig: Mg1_common_lines} to avoid overcrowding the figure. $\Delta$NLTE are predominantly positive and increase with decreasing $\logg$ and metallicity. The scatter in the corrections of the different lines is in most cases small, $\sigma<0.1$.

Recently, \citet{Matsuno2024} computed both 1D- and 3D-NLTE corrections for FG-type dwarfs. Their 1D corrections are similar to those of \citeauthor{Osorio2016_Mg1} (\citeyear{Osorio2016_Mg1}; bottom panels of Fig.~\ref{fig: Mg1_common_lines}), while those assuming 3D are typically higher by $0.1-0.2$\,dex (the difference $\Delta$NLTE$_{\rm 3D}$ - $\Delta$NLTE$_{\rm 1D}$ ranges between $-0.08$ and $+0.3$ approximately). We note that the 3D corrections of \citet{Matsuno2024} show a much stronger dependence on $v_{\rm turb}$ than their own 1D corrections, or those of \citeauthor{Osorio2016_Mg1} (\citeyear{Osorio2016_Mg1}; see Sec.~\ref{Microturbulence}). Specifically, the differences between $v_{\rm turb}=1\,$km/s $v_{\rm turb}=2\,$km/s reach up to 0.14\,dex in 3D, compared to just 0.016\,dex in 1D.
Similarly, their 3D corrections exhibit significantly larger line-to-line scatter, with the standard deviation reaching $\sigma=0.36\,$dex, in comparison to 0.07 dex for their 1D corrections and 0.045 dex for those of \citet{Osorio2016_Mg1}.

We adopt the grid from \citet{Osorio2016_Mg1} as default because it spans the broadest parameter range and includes an additional dependence on A(Mg).

\subsection{Aluminium I}
\label{aluminium}

\citet{Nordlander2017} and \citet{Lind2022} have provided grids of EWs for several Al lines, as a function of $T_{\rm eff}$, $\logg$, [Fe/H] and [Al/Fe], computed in LTE and NLTE. We derived the NLTE correction for each line by interpolating LTE equivalent widths onto the NLTE curves of growth. The \citet{Lind2022} corrections are given for $v_{\rm turb}=1$ and 2$\,$km/s; we adopted their average, noting that the differences between them do not exceed 0.08 dex when applied to all MP SAGA stars. They are based on 1D model atmospheres, while \citet{Nordlander2017} also provide values for temporally and spatially averaged $\langle {\rm 3D} \rangle$ hydrodynamical models.

\begin{figure*}
\begin{center}
\includegraphics[width=0.9\hsize]{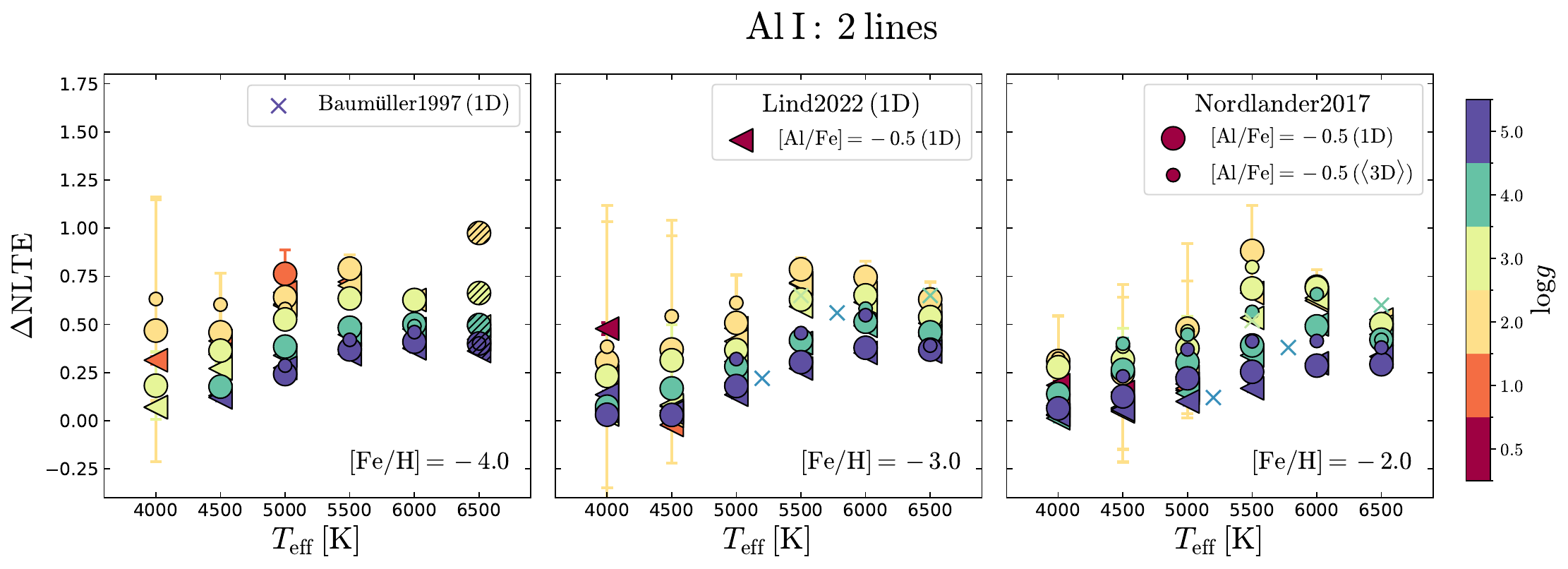} 
\caption{Mean NLTE corrections of the resonance \ion{Al}{I} lines at 3944 ${\rm \AA}$ and 3961 ${\rm \AA}$, colour-coded by $\logg$: triangles are from \citet{Lind2022} and circles from \cite{Nordlander2017}. Corrections for individual stars from \citet{Baumuller1997} are shown with crosses (X), for four test cases of \teff=5200, 5500, 5780, and 6500\,K; and \logg=4.50, 3.50, 4.44, and 4.00. Error bars represent the standard deviation of the corrections of the  two lines (often smaller than the depicted symbols). Hatched symbols indicate corrections that are only based on one line.}
\label{fig: Al_common_lines}
\end{center}
\end{figure*}

For each stellar parameter set, we computed the mean NLTE correction of the Al resonance lines at 3944\,${\rm \AA}$ and 3961\,${\rm \AA}$, which are used in the vast majority of MP stars. The resulting $\Delta {\rm NLTE}$ values are shown in Fig.~\ref{fig: Al_common_lines}, alongside 1D corrections for individual stars by \citet{Baumuller1997}. In addition, we computed mean corrections for the Al subordinate lines at 6696-8773\,${\rm \AA}$, which are often used at ${\rm [Fe/H] \gtrsim -1.5}$, where the resonance lines become too strong. Both sets of NLTE grids are provided in \texttt{NLiTE}. SAGA stars were categorized based on whether their Al abundances were derived from resonance or subordinate lines, and the corresponding corrections were applied accordingly. 

The resonance lines corrections are predominantly positive and tend to increase with decreasing $\logg$ and increasing $T_{\rm eff}$ (Fig.~\ref{fig: Al_common_lines}). The same trend is seen in the subordinate lines (not shown here) which however exhibit weaker corrections (see also Fig.~\ref{fig: NaAl_logg}).
The behavior of $\Delta$NLTE with [Al/Fe] is not monotonous but depends on the specific stellar parameters. We adopt the grid from \citet{Lind2022} as our reference, as it covers the largest [Fe/H] range.

\subsection{Silicon I}

In total, eight optical lines of \ion{Si}{I} were used for our average NLTE corrections, using the grids of \citet{Bergemann2013_Si} and \citet{Amarsi2017_Si} (see Fig.~\ref{fig: Si_common_lines}). The \citet{Amarsi2017_Si} corrections are provided for $v_{\rm turb}=1$ and 2$\,$km/s. We adopted the average of the two, noting that their difference remains below $0.05\,$dex for $99.4\%$ of MP SAGA stars (with a maximum difference of 0.22\,dex).

We find that the mean corrections are mostly positive and tend to increase with increasing \teff, decreasing \logg, and decreasing [Si/Fe]. The \citet{Amarsi2017_Si} corrections are significantly larger than those of \citet{Bergemann2013_Si} at high $T_{\rm eff}$, with the latter being in most cases close to zero at $\rm[Fe/H]\geq-3$. We adopt the corrections of \citet{Amarsi2017_Si} as default because they are given as a function of [Si/Fe], in addition to $T_{\rm eff}$, $\logg$ and [Fe/H]. 

 We note that, especially at lower metallicities, neutral silicon is often represented by only one line, both in observations and our calculations (diagonally hatched symbols in Fig.~\ref{fig: Si_common_lines}), since other lines become too weak for reliable detection. That is, primarily the \ion{Si}{I} 3906 ${\rm \AA}$ resonance line and less frequently the 4103 ${\rm \AA}$ line. Finally, we note that several studies have been conducted on the NLTE effects of Si lines in the infrared \citep[e.g.][]{Shi2012,Tan2016}, however, these are beyond the scope of this work since our focus is chemical abundance analysis done with optical spectra.

\subsection{Sulphur I}
\label{sulfur}

Several 1D\,NLTE grids of \ion{S}{I} are available in the literature, however, they typically cover a limited parameter space and do not always share common lines. Direct unified comparison between different grids is therefore not possible, nonetheless they are shown for reference in Fig.~\ref{fig: S_common_lines}.
We include the NLTE corrections for the \ion{S}{I} 8694\,\AA\ line from \citet{Takada-Hidai2002}. In addition we compute the average corrections for the \ion{S}{I} triplet at 9213, 9228 and 9238\,\AA\ from \citet{Skuladottir2015}, provided only for giant stars,  as well as the average corrections from \citet{Takeda2005}  for the 8695 and 9213\,\AA\ lines, and \citet{Korotin2008} for the 8695, 9213\,\AA, and the eighth  \ion{S}{I} multiplet at 6743–6757\,\AA, provided only for $2 \leq \logg \leq 4$. \citeauthor{Takada-Hidai2002}'s and \citeauthor{Takeda2005}'s corrections are given for $v_{\rm turb}=2\,$km/s. \citet{Skuladottir2015} adopt  $v_{\rm turb}=1.7\,$km/s, but note that their corrections are not sensitive to the adopted turbulence velocity.

All corrections are negative, with their absolute values increasing towards lower $\logg$ and higher \teff\ (Fig.~\ref{fig: S_common_lines}). However, there are significant variations among different lines. This is evident when comparing the corrections from \citet{Skuladottir2015} and \citet{Takada-Hidai2002} at $\rm[Fe/H] = -1$, or from the large scatter observed in the corrections from \cite{Korotin2008} at $\rm[Fe/H] = -2$.

The corrections for the eighth \ion{S}{I} multiplet, provided only by \citet{Korotin2008}, are the smallest, always $<-0.14$\,dex, but these lines become too weak to be observed at $\rm[Fe/H]<-1.5$. The corrections for the 8694\,\AA\ line are stronger, ranging from $\sim-0.4$ to 0\,dex, yet this line also becomes too weak at $\rm[Fe/H]<-2$.
For the same stellar parameters, the corrections for the 8694\,\AA\ line from different studies are generally consistent, with those of \citet{Takada-Hidai2002} being $\sim0.1\,$dex weaker (less negative) than the \citet{Takeda2005} and \citet{Korotin2008} corrections.

The first \ion{S}{I} multiplet (9212–9237\,\AA) contains the only observable lines at $\rm[Fe/H] < -2$, in our optical spectral range of interest ($3500-10\,000$\,\AA).
The corrections for these lines are significantly larger, reaching $<-1$\,dex. Among them, the 9213\,\AA\ line exhibits the strongest corrections but only by about 0.04 and 0.08\,dex compared to the 9228 and 9238\,\AA\ lines, respectively \citep{Skuladottir2015}.
For the same stellar parameters, there are large discrepancies in the corrections of the 9213\,\AA\ line between \citet{Takeda2005} and \citet{Korotin2008}, with  differences reaching 0.4\,dex.

In the case of sulphur, different studies use different lines, often with no overlap, making this element not very suitable for the method introduced here. Furthermore, no single grid fully covers the required range of stellar parameters, and includes all the commonly used \ion{S}{I} lines for metal-poor stars. Therefore we do not apply NLTE corrections to the limited number of \ion{S}{I} abundances available in SAGA. Instead we only present [S/Fe]$_{\rm LTE}$ in Table~\ref{table:Fiducial} (marked specifically). However, we provide all aforementioned NLTE correction grids in \texttt{NLiTE} for users with a specific set of S lines.

\subsection{Potassium I}

The NLTE corrections for \ion{K}{I} are based on the doublet at 7664 and 7698\,${\rm \AA}$, using the grid of \citet{Reggiani2019_K}.
This is given for $v_{\rm turb}$=1, 2 and 5\,km/s. We adopted the mean of the $v_{\rm turb}=1\,$km/s and $v_{\rm turb}=2\,$km/s  corrections for each \teff, $\logg$, [Fe/H] and A(K). We note that in 98.5$\%$ of cases the differences in the corrections between the two $v_{\rm turb}$ values are $<0.1$ dex (100$\%$ of cases with $\rm [Fe/H]<-2$). The largest difference reaches 0.29\,dex. In addition, we offer the corrections of \citet{Takeda2002}, which are provided only for the 7698\,${\rm \AA}$ line. 

The average NLTE corrections from \citet{Reggiani2019_K} are shown in Fig.~\ref{fig: K_common_lines}, for three values of A(K)=1.33, 2.33, and 3.33, while A(K)$_\odot=5.03$ according to \citet{Asplund2009}. For comparison, we plot the corrections of \citet{Andrievsky2010} for individual stars with $\rm 1.5<A(K)<3$ and those of \citet{Takeda2009} for stars with $0.07 \leq {\rm [K/Fe]} \leq 0.42$. 

The \citet{Reggiani2019_K} corrections are mainly negative, ranging from $-0.882$ to $+0.074$, and tend to increase (in absolute value) towards lower $\logg$ and $T_{\rm eff}$, and towards higher A(K). These trends appear also in the corrections of \citet{Takeda2002} (not shown in Fig.~\ref{fig: K_common_lines}), which range from $-1.279$ to $-0.072$ dex. 

The work of \citet{Neretina2020} also reports primarily negative NLTE corrections for \ion{K}{I} lines, in general agreement with previous studies. There are indications that the absolute value of the corrections can vary significantly, often by $>0.1$\,dex for the same stellar parameters, between the different works (Fig.~\ref{fig: K_common_lines}). However, since full grids are not available from \citet{Takeda2009,Andrievsky2010}; and \citet{Neretina2020} only a limited comparison can be made.

\subsection{Calcium I}

We computed average NLTE corrections for 25 optical \ion{Ca}{I} lines, using the grids of \citet{Spite2012_Ca} and \citet{Mashonkina2017_Ca1}. The results from these two sources are in fairly good agreement, typically within $\lesssim0.05$\,dex where the grids overlap (Fig.~\ref{fig: Ca1_common_lines}). Both show typically positive $\Delta$NLTE corrections that increase as $\logg$ decreases. Most corrections fall within the range $\Delta$NLTE$\in$[0, +0.4] and show only a small dependence on Ca: when [Ca/Fe] varies between 0 and +0.4\,dex, the NLTE corrections typically change by $\lesssim0.05$\,dex.

We did not include the corrections of \citet{Merle2011} here because they are available for only 10 lines in common with the sets of \citet{Spite2012_Ca} and \citet{Mashonkina2017_Ca1}. We choose as our fiducial grid the one of \citet{Mashonkina2017_Ca1} as it covers the widest parameter range.

\subsection{Scandium II}

Very few studies on the NLTE effects of Sc lines have been conducted. According to \citet{Zhang2014}, NLTE corrections for \ion{Sc}{II} are expected to be small ($-0.04$ to $+0.06$\,dex). As far as we are aware of, no NLTE grids for \ion{Sc}{II} exist in our targeted range of metallicity and stellar parameters. Therefore we cannot provide NLTE corrections for this element in \texttt{NLiTE}, but for convenience, we include [Sc/Fe]$_{\rm LTE}$ (marked specifically) in our fiducial SAGA catalogue.

\begin{figure*}
\begin{center}
\includegraphics[width=0.9\hsize]{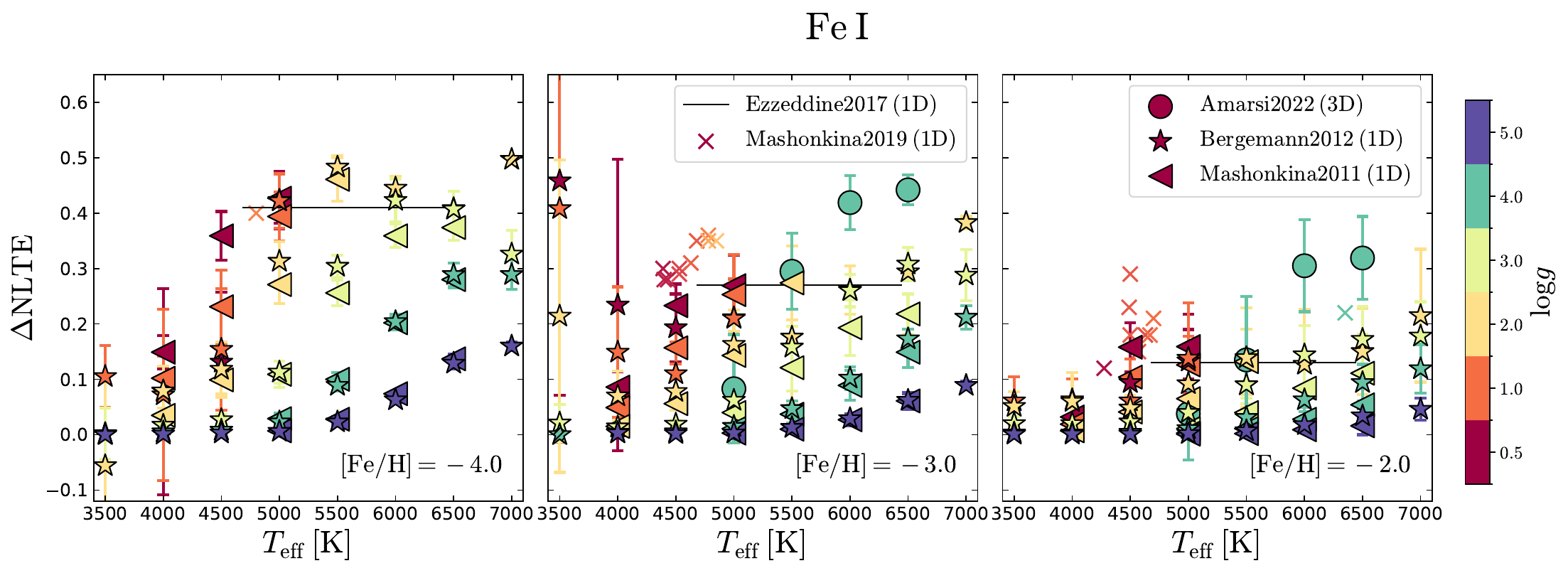}\\
\includegraphics[width=0.9\hsize]{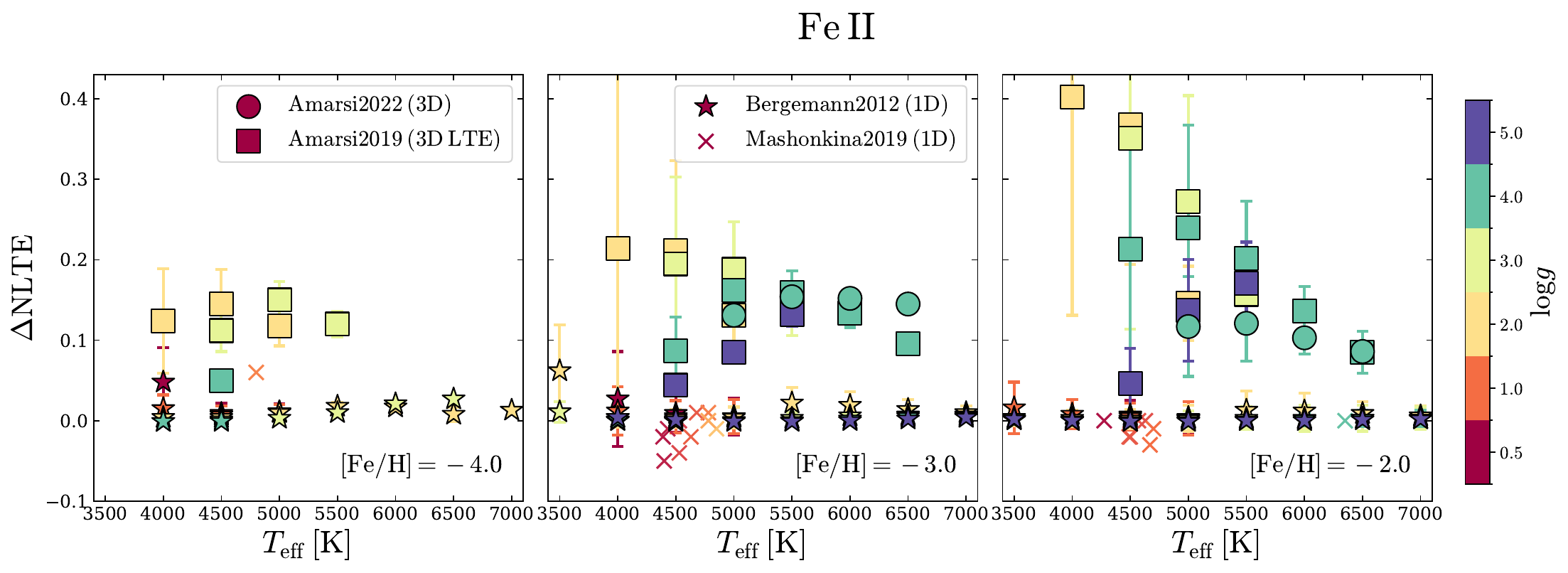} 
\caption{Corrections for \ion{Fe}{I} (top) and \ion{Fe}{II} (bottom), colour-coded by $\logg$: stars and triangles show 1D\,NLTE corrections from \citet{Bergemann2012_Fe1_Fe2} and \citet{Mashonkina2011_Fe1}, respectively, and circles show 3D\,NLTE corrections from \citeauthor{Amarsi2022} (\citeyear{Amarsi2022}; available only for dwarf stars). The 1D\,NLTE corrections for individual stars from \citet{Mashonkina2019_Fe} are shown with crosses (X). The $\Delta {\rm NLTE}$ relation of \citet{Ezzeddine2017} for \ion{Fe}{I} is shown with the black line (top row). Squares in the bottom row show the 3D\,LTE corrections for \ion{Fe}{II} from \citet{Amarsi19}. Error bars represent the standard deviation of different lines.}
\label{fig: Fe1_common_lines}
\end{center}
\end{figure*} 

\subsection{Titanium I and II}

We computed mean 1D NLTE corrections from 22 \ion{Ti}{I} and 41 \ion{Ti}{II} optical lines, using the grids of \citet{Bergemann2011_Ti1_Ti2} and \citeauthor{Sitnova2016_Ti2} (\citeyear{Sitnova2016_Ti2}; given for $\rm[\ion{Ti}{II}/Fe]=+0.3$).

As shown in Fig.~\ref{fig: Ti1_common_lines}, the corrections for both Ti ionization states are positive, with the ones for \ion{Ti}{I} being generally higher, and increasing with decreasing $\logg$. Recently, \citet{Mallinson2022} computed also 1D NLTE \ion{Ti}{I} and \ion{Ti}{II} corrections for 5 stars, including the Sun. Their \ion{Ti}{II} corrections are within $\lesssim0.1\:$dex of those from \citet{Sitnova2016_Ti2} for the same stars. On the other hand, the three metal-poor stars analysed in \citet{Mallinson2022} have NLTE corrections for \ion{Ti}{i} around $\approx$0.1\,dex lower, compared to similar stars in \citet{Bergemann2011_Ti1_Ti2}. 

For \ion{Ti}{I}, only \citet{Bergemann2011_Ti1_Ti2} provide a full grid. However, they note that while their NLTE model solves the discrepancy between the \ion{Ti}{I} and \ion{Ti}{II} lines in the Sun, it does not perform similarly well for metal-poor stars, but overestimates NLTE effects in the atmospheres of dwarfs and underestimates overionization for giants. Therefore, they stress that only \ion{Ti}{II} lines can be safely used for abundance analysis in MP stars. This is confirmed in the work of \citet{Sitnova2020} whose NLTE calculations were unable to restore the ionization balance of Ti in MP stars.

For Ti, we adopt the \citet{Bergemann2011_Ti1_Ti2} corrections as default as they span the widest parameter range.

\subsection{Vanadium I}

Unfortunately, no detailed NLTE study of vanadium exists in the literature. However, \citet{Ou2020} derived LTE abundances of \ion{V}{I} and \ion{V}{II} for a large sample of metal-poor stars. From their most reliable measurements they find an ionization imbalance of $\rm[\ion{V}{II}/\ion{V}{I}]=+0.25$ and argue that this is most likely due to NLTE effects on the lines of \ion{V}{I}. Given that \ion{V}{I} is most commonly used to measure vanadium in metal-poor stars, a typical $\rm \Delta NLTE\approx+0.25$ is expected. We provide [V/Fe]$_{\rm LTE}$ in our fiducial SAGA catalogue, and since the 1D\,NLTE corrections for \ion{V}{I} and \ion{Fe}{I} are expected to be of similar order, it is reasonable to expect $\rm [V/Fe]_{\rm NLTE}\approx[V/Fe]_{\rm LTE}$.

\subsection{Chromium I}

Mean NLTE corrections for 16 optical \ion{Cr}{I} lines were calculated from the grid of \citet{Bergemann2010_Cr} (see Fig.~\ref{fig: Cr1_common_lines}). The corrections are predominantly positive, with values reaching up to $\sim 1\,$dex, and tend to increase with increasing temperature and decreasing $\logg$. Overall, the NLTE corrections for the different lines are in good agreement, with $\sigma$ being much smaller than $\Delta{\rm NLTE}$ in all but a few cases at $\logg<1$ (see Secion~\ref{Scatter}).

\subsection{Manganese I}

We computed the average NLTE corrections based on 8 optical \ion{Mn}{I} lines from the grid of \citet{Bergemann19} (see Fig.~\ref{fig: Mn1_common_lines}). Similar to \ion{Cr}{I} (see above), the corrections are positive, increasing with increasing temperature and decreasing $\logg$ and can reach up to $\gtrsim +1 \,$dex. The corrections are consistent between the different \ion{Mn}{I} lines, except at $\logg<2.0$, where the standard deviation can reach as high as $\sigma>0.3$ (see Sec.~\ref{Scatter}).

\citet{Bergemann19} computed also 3D\,NLTE \ion{Mn}{I} corrections for selected metal poor models (${\rm [Fe/H]=-2}$ and ${\rm [Fe/H]=-1}$). They found that 3D NLTE abundances are typically higher than 1D NLTE abundances by up to 0.15 dex for metal-poor dwarf stars. The difference can be more pronounced for giants, reaching $\sim 0.3$\,dex for certain Mn lines. However, a full 3D\,NLTE grid for Mn is currently not available.

\subsection{Iron I}
\label{Iron I}

In total 82 \ion{Fe}{I} lines were included in the calculated 1D NLTE corrections, using the grids of \citet{Mashonkina2011_Fe1} and \citet{Bergemann2012_Fe1_Fe2} (see Fig.~\ref{fig: Fe1_common_lines}, top). The \ion{Fe}{I} corrections are always positive, and increase with decreasing $\logg$ and increasing $T_{\rm eff}$, reaching as high as $\sim+0.5 \,$dex in the most extreme cases. The \citet{Mashonkina2011_Fe1} and \citet{Bergemann2012_Fe1_Fe2} corrections are in good agreement for all $T_{\rm eff}$, $\logg$ and [Fe/H] values. We adopt their mean values at each stellar parameter set as our fiducial grid. 

\citet{Amarsi2022} provides 3D\,NLTE corrections, but only for FG type dwarfs ($T_{\rm eff}=5000-6500\,$K, $\logg = 4 -4.5\,$dex), and metallicities $\rm[Fe/H]>-3$. Their line list includes only 42 lines that overlap with our fiducial line list, from which we computed the average corrections.\footnote{Since \citet{Amarsi19} and \citet{Amarsi2022} do not provide EWs, we identify the Fe lines with $5\,\AA \leq {\rm EW} \leq 200\,\AA$ for each stellar atmosphere using data from \citet{Bergemann2012_Fe1_Fe2} and \citet{Mashonkina2011_Fe1}.} We note that the mean corrections from our fiducial grid \citep{Mashonkina2011_Fe1,Bergemann2012_Fe1_Fe2}  differ by less than 0.04 dex when computed for this subset of 42 lines compared to the full line list. 

We find that the 3D\,NLTE corrections of  \citet{Amarsi2022}  are systematically higher, by up to $\sim0.2-0.3\,$dex, compared to the 1D corrections of \citet{Bergemann2012} and \citet{Mashonkina2011_Fe1}, with the difference increasing with temperature (upper panels of Fig.~\ref{fig: Fe1_common_lines}). In addition, the scatter between different \ion{Fe}{I} lines is larger in the 3D corrections than in the 1D ones at fixed stellar parameters, yet remains always smaller than 0.12\,dex.

For comparison, we also show the 1D\,NLTE corrections for individual stars from \citet{Mashonkina2019_Fe} in Fig.~\ref{fig: Fe1_common_lines}. Those lie approximately 0.1-0.15 dex higher than those of \citet{Mashonkina2011_Fe1} for the same stellar parameters. In addition, we plot the $\Delta$NLTE-[Fe/H] relation from \citet{Ezzeddine2017}, found by fitting 1D\,NLTE Fe corrections of 22 stars. We note that \citet{Ezzeddine2017} only provide a fit to $\Delta$NLTE as a function of [Fe/H], but not as a function of \teff\ or \logg.

For all three grids, we have adopted corrections for $v_{\rm turb}=1.5\,$km/s, the peak of the microturbulence distribution of MP stars (Fig.~\ref{fig: SAGA_microturbulence}). We note that \citet{Bergemann2012} report that the value of the microturbulence has almost no influence on the size of the NLTE effects. On the other hand, 3D versus 1D corrections are sensitive to the $v_{\rm turb}$ adopted in 1D models. In particular, \citet{Amarsi2022} show that for saturated lines, for $v_{\rm turb}=0\,$km/s the estimated 3D\,NLTE versus 1D\,LTE corrections have similar absolute values as for $v_{\rm turb}=2\,$km/s but with opposite signs, making the difference between the two as high as 1 dex (see their Fig.~3). The difference between the $v_{\rm turb}=1\,$km/s and the $v_{\rm turb}=2\,$km/s is smaller, reaching up to $\sim0.5\,$dex at the lowest $T_{\rm eff}$ values.

\subsection{Iron II}

We provide mean corrections for 12 \ion{Fe}{II} optical lines, using the 1D\,NLTE grid of \citet{Bergemann2012_Fe1_Fe2} and the 3D\,LTE one of \citet{Amarsi19}; and a subset of 7 lines using the 3D\,NLTE grid of \citet{Amarsi2022} (see Fig.~\ref{fig: Fe1_common_lines}, bottom)\footnotemark[\value{footnote}].

The 1D\,NLTE corrections from \citet{Bergemann2012_Fe1_Fe2} are minimal, ranging from $<0.01$ to $\sim 0.04\,$dex. \citet{Mashonkina2019_Fe} report similar values with 1D\,NLTE corrections for individual stars varying between $-0.05$ and +0.06\,dex (Fig.~\ref{fig: Fe1_common_lines}). In contrast, the 3D\,NLTE corrections from \citet{Amarsi2022} are significantly larger, reaching $\sim 0.2\,$dex. Notice that the 3D\,LTE corrections from \citet{Amarsi19} are often even stronger, with the difference increasing at lower $T_{\rm eff}$ and higher [Fe/H]. This highlights the fact that 3D and NLTE effects often act in opposite directions \citep[e.g.][]{Asplund2005}.

As our fiducial grid, we choose \citet{Bergemann2012_Fe1_Fe2} since it covers the widest parameter range.

\subsection{Cobalt I}

Using the grid of \citet{Bergemann2010_Co}, we computed average NLTE corrections for 17 optical \ion{Co}{I} lines (Fig.~\ref{fig: Co1_lines}). The resulting corrections are mainly positive, and increase at higher $\logg$ and $T_{\rm eff}$ values, reaching as high as $\sim +1\,$dex. The NLTE corrections show significant scatter across the adopted lines, with $\sigma\gtrsim0.1$\,dex for about half the grid points, which is however small compared to the mean corrections (see Sec.~\ref{Scatter}).

\subsection{Nickel I}

\citet{Eitner2023} computed 1D\,NLTE Ni abundances for 264 stars from the {\it Gaia}-ESO survey and found that the slight sub-solar [Ni/Fe] trend observed at lower [Fe/H] in LTE is reversed under NLTE conditions, where at  ${\rm [Fe/H]} \lesssim-1$ stars exhibit slightly super-solar [Ni/Fe] ratios. The authors provide NLTE corrections only for 12 model atmospheres with $T_{\rm eff}=5750\,$K and $\logg=4.5$; $T_{\rm eff}=6500\,$K and $\logg=4.5$; and $T_{\rm eff}=5000\,$K and $\logg=3.0$, for the metallicities $\rm [Fe/H] = -3.0, -2.0, -1.0, 0.0$. The corrections are positive and show an increasing trend towards decreasing metallicity, at fixed $T_{\rm eff}$ and $\logg$, reaching 0.2-0.3 dex at $\rm[Fe/H]=-3$. 

Observationally, scatter in [Ni/Fe] is generally low and agrees well between different stellar types \citep[e.g.][]{Cayrel2004,Roederer14}. Furthermore, different galaxies agree quite well in their [Ni/Fe] ratios at $\rm [Fe/H]<-3$ \citep[e.g.][]{Skuladottir2024b}, arguing against very different NLTE corrections of Ni and Fe. Because of similarities in the structure of the Fe and Ni atoms, it has been argued that likely $\rm \Delta NLTE(Ni)\approx \Delta NLTE(Fe)$ \citep[e.g.][]{Skuladottir2021}. Because of the lack of a more complete grid of NLTE corrections for Ni, we therefore adopt here the conservative approach of providing only [Ni/Fe]$_{\rm LTE}$, and argue based on the observational evidence that it is reasonable to expect $\rm[Ni/Fe]_{\rm NLTE}\approx[Ni/Fe]_{\rm LTE}$. Finally, we note that although the results of \citet{Eitner2023} are not used for our fiducial catalogue, their grid is available for the community through our \texttt{NLiTE} tool, making it very easy to correct a large sample of stars with Ni abundance measurements.

\subsection{Copper I}
\label{copper}

Currently, there is no available NLTE grid for neutral copper. To examine whether we can construct a simple relation between $\Delta$NLTE and different stellar parameters, we compiled available 1D corrections for 37 individual metal poor stars ($-4.2<{\rm [Fe/H]}<-1$) from the studies of \citet{Shi2018_Cu}, \citet{Andrievsky2018_Cu} and \citet{Xu2022_Cu}.

\begin{figure}
\begin{center}
\includegraphics[width=0.95\hsize]{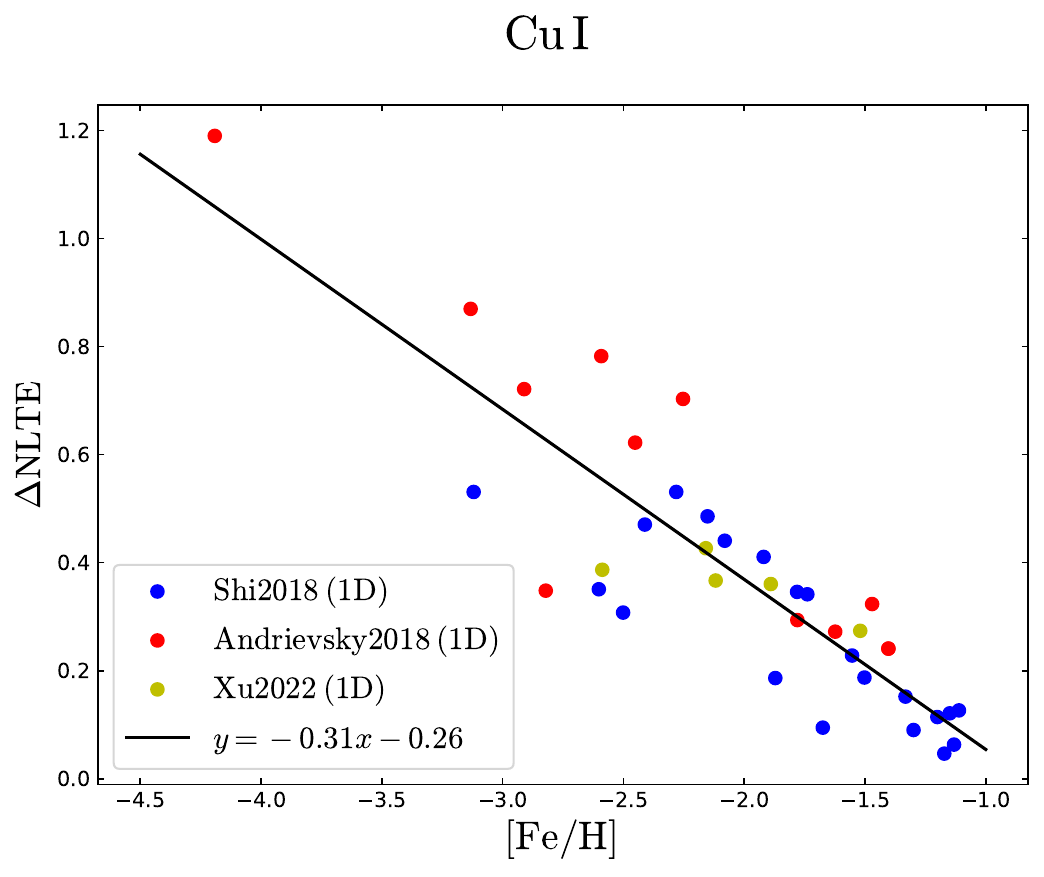} 
\caption{NLTE corrections for \ion{Cu}{I} for individual stars by \citeauthor{Shi2018_Cu} (\citeyear{Shi2018_Cu}; blue dots), \citeauthor{Andrievsky2018_Cu} (\citeyear{Andrievsky2018_Cu}; red dots), and \citeauthor{Xu2022_Cu} (\citeyear{Xu2022_Cu}; yellow dots) as a function of [Fe/H]. The black line represents the linear least-squares fit to the data.}
\label{fig: Cu1_lines}
\end{center}
\end{figure}

Figure~2 from \citet{Shi2018_Cu}, reveals that there is no clear trend between the \ion{Cu}{I} corrections and $T_{\rm eff}$ or $\logg$, at least for their sample of 28 stars. There is, however, a clear decreasing trend with metallicity that is also exhibited in the works of  \citet{Andrievsky2018_Cu} and \citet{Xu2022_Cu} (see Fig.~\ref{fig: Cu1_lines}).

The derived corrections by the three studies are somewhat different; we note that five of the stars shown in Fig.~\ref{fig: Cu1_lines} are shared between two or all three of the works. In particular \citet{Andrievsky2018_Cu} obtained higher corrections than \citet{Shi2018_Cu}, while the corrections of \citet{Xu2022_Cu} fall in between. These differences likely stem from variations in how inelastic collisions processes with hydrogen are modelled (see discussion in \citealp{Xu2022_Cu}). 

Here, we adopt the mean $\Delta$NLTE-[Fe/H] relation based on the three studies, described by the linear least-squares fit to the data:
\begin{equation}
    \Delta {\rm NLTE} = -0.26 -0.31\,\cdot {\rm [Fe/H]}
\end{equation}
(RMS=0.12), which is shown with a black line in Fig.~\ref{fig: Cu1_lines}.

\subsection{Zinc I}

We computed average 1D\,NLTE corrections for three \ion{Zn}{I} lines, at 4680, 4722 and 4810 ${\rm \AA}$, which are most commonly used in MP studies \citep{Sitnova2022_Zn1_Zn2}. The corrections are given as a function of $T_{\rm eff}$, $\logg$ and [Fe/H], assuming $\rm[Zn/Fe]=0$ (Fig.~\ref{fig: Zn_lines}).
They are given for [Fe/H] in the 
range $-5$ to 0. However, at $\rm [Fe/H]<-3.5$, \citet{Sitnova2022_Zn1_Zn2} provide corrections and measurable EWs only for the UV lines, which are excluded from our study. 

At $\rm[Fe/H]\geq-3.5$ we find that $\Delta$NLTE for \ion{Zn}{I} is primarily positive, $0\,\text{dex}<\Delta\text{NLTE}\lesssim0.3$\,dex, showing an increasing trend with temperature and decreasing $\logg$. Overall the 1D\,NLTE corrections of the different adopted lines are in very good agreement, with $\sigma\lesssim0.035$\,dex. 

The 1D\,NLTE corrections for \ion{Zn}{I} were also calculated by \citet{Takeda2005} who found typical values of $0\,\text{dex}\lesssim\Delta\text{NLTE}\lesssim0.1$\,dex for low-gravity giants, $\logg<3$, and similar results for higher gravities at $\rm[Fe/H]<-1$, $-0.05\,\text{dex}\lesssim\Delta\text{NLTE}\lesssim0.1$\,dex. The $\Delta$NLTE from \citep{Takeda2005} are thus somewhat less strong in comparison to our fiducial grid from \citet{Sitnova2022_Zn1_Zn2}.

\renewcommand{\arraystretch}{1.5}
\begin{table*}[htb]
\centering
\footnotesize
\caption{Fiducial NLTE-corrected SAGA catalogue}
\tabcolsep=0.27cm
\begin{tabular}{c c c c c c c }
\hline
\hline
Name & Reference & A(Li) & ${\rm [C/Fe]}$ & ${\rm [N/Fe]_{LTE}}$  & ${\rm [O/Fe]}$ &... \\ 
\hline
HD122563 & I.U.Roederer+,AJ, 147,136, 2014 &  <-0.15 & -0.48 & 0.65 & <0.43 & ... \\

BD+44$\_$493 & H.Ito+,ApJ, 773,33, 2013 &   1.09$\pm$0.11 & 0.63 & 0.40 & <2.15 & ... \\

BD+55$\_$1362 & R.G.Gratton+,AAP, 354,169, 2000 & 0.97 & -0.27 & 0.02 &0.54 & ... \\ 
.. & ...&  ... & ... & ... & ... & ...\\

\hline
\end{tabular}

\noindent\parbox{0.78\textwidth}{
    \vspace{2.5mm} 
    \small {\bf Notes:} This catalogue contains a single entry per star, prioritizing the one with the highest number of measured elements. Only stars with $T_{\rm eff}$, $\logg$ and [\ion{Fe}{I}/H] and/or [\ion{Fe}{II}/H] measurements are included (2014 stars in total). C has been corrected for evolutionary effects following \citet{Placco2014}. Abundance ratios are provided in NLTE except for [N/Fe], [S/Fe], [Sc/Fe], [V/Fe], and [Ni/Fe].}
\noindent\parbox{0.78\textwidth}{
    \vspace{2.mm} 
    \small (Full table available at the CDS.)}    
\vspace{0.8cm}
\label{table:Fiducial}
\end{table*}

\renewcommand{\arraystretch}{1.5}
\begin{table*}[htb]
\centering
\footnotesize
\tabcolsep=0.09cm
\caption{Extended catalogue including full information on individual NLTE corrections for all SAGA entries}
\tabcolsep=0.12cm
\begin{tabular}{c c c c c c c c c c c} 
\hline
\hline
Name & Reference & Ion & ${\rm [X/H]_{LTE}}$ & ${\rm OC_{X}}$  &   ${\rm OW_{W}}$       & ${\rm d[X/H]_{NLTE}}$   & ${\rm in\,grid_{X}}$ & ${\rm std{\scriptstyle d[X/H]_{NLTE}}}$ & ${\rm Nl{\scriptstyle d[X/H]_{NLTE}}}$ & ${\rm C_{cor,P}}$ \\ 
\hline
HD122563 & I.U.Roederer+,AJ,147,136,2014 & \ion{Li}{I} &  <-3.41 & 1 & 0 & 0.0 & - & - & - & -\\ 
HD122563 & I.U.Roederer+,AJ,147,136,2014 & \ion{CH} &  -3.37 & 0 & 0 & -0.648 & 1 & - & - & 0.77 \\ 
HD122563 & I.U.Roederer+,AJ,147,136,2014 & \ion{N}{I} &  -2.32 & - & - & - & - & - & - & - \\ 
... & ... & ... & ...& ... & ... & ... & ... & ... & ... & ...\\
HD122563 & I.U.Roederer+,AJ,147,136,2014 & \ion{Fe}{I} &  -2.97$\pm$0.15 & 0 & 0 & 0.20 & 1 & 0.049 & 72 & - \\ 
... & ... & ... & ...& ... & ... & ... & ... & ... & ... & ...\\
HD122563 & D.Prakapavicius+,AAP,599,128,2017 & \ion{O} &  -2.28$\pm$0.16 & - & 1 & - & - & - & - & - \\ 
HD122563 & D.Prakapavicius+,AAP,599,128,2017 & \ion{Fe}{I} &  -2.6 & 1 & 0 & 0.0 & - & - & - & - \\ 

BD+44$\_$493 & H.Ito+,ApJ,773,33,2013 & \ion{Li}{I} & -2.26$\pm$0.11 & 0 & 0 & 0.088 & 1 & 0.0 & 1 & -\\
BD+44$\_$493 & H.Ito+,ApJ,773,33,2013 & \ion{CH} & -2.48 & 0 & 0 & -0.564 & 1 & - & - & 0.0\\
BD+44$\_$493 & H.Ito+,ApJ,773,33,2013 & \ion{N}{I} & -3.43 & - & - & - & - & - & - & -\\
BD+44$\_$493 & H.Ito+,ApJ,773,33,2013 & \ion{O}{I} & <-1.49 & 0 & 0 & -0.032 & 0 & 0.0 & 3 & -\\
... & ... & ... & ...& ... & ... & ... & ... & ... & ...& ... \\
\hline
\end{tabular}

\noindent\parbox{0.98\textwidth}{
    \vspace{3mm} 
    \small {\bf Notes:} This catalogue contains all metal-poor entries in the SAGA database. For each NLTE-corrected chemical species X, the table provides: (1) the LTE abundance, ${\rm [X/H]_{LTE}}$; (2) the NLTE correction, ${\rm d[X/H]_{NLTE} \equiv [X/H]_{ NLTE}- [X/H]_{LTE}}$; (3) the flag ${\rm OC_X}$, set to 1 if the abundance was originally published corrected for NLTE effects (in which case ${\rm d[X/H]_{NLTE}=0}$); (4) the flag ${\rm OW_X}$, set to 1 if the abundance was derived using spectral lines not included in our line list (see Table~\ref{table: linelist}, in which case ${\rm d[X/H]_{NLTE}= NaN}$); (5) the flag ${\rm in\,grid_X}$, set to 1 if the star’s stellar parameters fall within the NLTE grid limits, and 0 otherwise; (6) the standard deviation of the average NLTE correction, ${\rm std{\scriptstyle d[X/H]_{NLTE}}}$; and (7) the number of lines used to compute the average correction at the closest grid point to the star, ${\rm Nl{\scriptstyle d[X/H]_{NLTE}}}$ (not given for CH and Cu). Additional columns include 'Line Type' indicating whether resonance or subordinate lines were used for the Na and Al corrections (not shown here) and '${\rm C_{cor,P}}$' providing the \citet{Placco2014} evolutionary carbon corrections.
    Abundance ratios for N, F, \ion{Si}{II}, P, S, \ion{Ca}{II}, Sc, V, \ion{Cr}{II}, \ion{Mn}{II}, Ni, \ion{Cu}{II}, \ion{Zn}{II}, and \ion{Sr}{I} are provided, but are not corrected for NLTE effects.
}
\noindent\parbox{0.98\textwidth}{
    \vspace{2.mm} 
    \small (Full table available at the CDS.)}    
\vspace{0.8cm}
\label{table: Complete}
\end{table*}

\renewcommand{\arraystretch}{1.5}
\begin{table*}[!ht]
\centering
\footnotesize
\caption{Coordinates and stellar parameters}
\begin{tabular}{c c c c c c c c c c c c c }
\hline
\hline
Name$^{\rm a}$ & Reference & R.A. & Decl. & $T_{\rm eff}$ & ${\rm log}g$ & ${\rm [Fe/H]^{\rm c}}$ & SAGA\,name$^{\rm b}$  \\ 
\hline
HD122563 & I.U.Roederer+,AJ,147,136,2014 & 14 02 31.80	 & +09 41 10.0 & 4500 & 0.55 & -2.97$\pm$0.15 & HD122563  \\

HD122563 & D.Prakapavicius+,AAP,599,128,2017 & 14 02 31.80	 & +09 41 10.0 & 4600 & 1.6 & -2.6 & HD122563  \\

BD+44$\_$493 & H.Ito+,ApJ, 773,33, 2013 & 02 26 49.70 &	+44 57 46.0 & 5430 & 3.4 & -3.83 &  BD+44$\_$493 \\ 

BD+55$\_$1362 & R.G.Gratton+,AAP, 354,169, 2000 & 10 04 43.18 & +54 20 43.4 & 5157 & 3.01 & -1.77 & HD87140 \\
... & ...& 	... & ...& ... & ... & ... & ...\\
\hline
\end{tabular}

\noindent\parbox{0.93\textwidth}{
    \vspace{3mm} 
    \small {\bf Notes:} This catalogue contains all metal-poor entries in the SAGA database.\\
    $^{\rm a}$ A unique name per star, in alphabetic order.\\
    $^{\rm b}$ Name in SAGA (multiple names may refer to the same object).\\
    $^{\rm c}$ [Fe/H] is given in LTE unless it was originally published in NLTE (see Table~\ref{table: Complete}).} 
\noindent\parbox{0.93\textwidth}{
    \vspace{2.mm} 
    \small (Full table available at the CDS.)}     
\vspace{0.4cm}
\label{table: Parameters}
\end{table*}

\subsection{Strontium II}

Mean corrections for the two resonance \ion{Sr}{II} lines, at 4077 and 4215\,${\rm \AA}$, were calculated using the grids of \citeauthor{Bergemann2012_Sr} (\citeyear{Bergemann2012_Sr}; provided for $\vt=1 \kms$) and \citet{Mashonkina2022_Sr} (see Fig.~\ref{fig: Sr2_lines}). The two studies are in overall good agreement where they are both defined (within 0.15 dex), and the two lines have very similar corrections. The corrections vary in the range $\rm -0.2 \lesssim \Delta NLTE \lesssim +0.4\,$dex and tend to increase with decreasing [Sr/Fe] and [Fe/H].

We adopt the corrections of \citet{Mashonkina2022_Sr} as default, as they span the broadest parameter range.

\subsection{Barium II}

We calculated average corrections for the five \ion{Ba}{II} optical lines (4554, 4934, 5854, 6142, and 6497\,${\rm \AA}$) that are most commonly used in studies of MP stars, utilizing the grids of \citet{Mashonkina2019_Ba2} and \citet{Gallagher2020}\footnote{https://www.chetec-infra.eu/3dnlte/abundance-corrections/barium/} (see Fig.~\ref{fig: Ba2_lines}). Unlike the common practice, \citet{Gallagher2020} provide their corrections as a function of the 3D\,NLTE abundance and not the 1D\,LTE abundance. Therefore, we derived $\Delta {\rm NLTE}$ by interpolating the LTE equivalent widths onto the 1D and 3D\,NLTE curves of growth. In addition, we computed the mean corrections of \citet{Korotin2015_Ba2}, although these do not include the 4934\,${\rm \AA}$ line and are thus not included in Fig.~\ref{fig: Ba2_lines} (but see Fig.~\ref{fig: BaFe1_logg}).

The 1D\,NLTE corrections of \citet{Mashonkina2019_Ba2} and \citet{Gallagher2020} are in agreement within $<0.1\,$dex.
Depending on the stellar parameters, the 3D\,NLTE corrections of \citet{Gallagher2020} differ from their 1D\,NLTE corrections by $-0.36$ to +0.21 dex, typically being higher than the 1D~corrections at low \teff, and lower at high \teff. The \citet{Korotin2015_Ba2} corrections are generally higher/lower than the \citet{Mashonkina2019_Ba2} at low/high $T_{\rm eff}$, with a maximum difference reaching $\sim1.5 \,$dex. Similarly to \citet{Mashonkina2019_Ba2}, in all four sets, $\Delta$NLTE tend to decrease with increasing [Ba/Fe]. 

\citet{Andrievsky2009} computed 1D\,NLTE corrections for \ion{Ba}{II} in 41 VMP stars ($-4.19 \leq {\rm [Fe/H]} \leq -2.07$) based on the 4554,~5854 and 6497\,${\rm \AA}$ lines. Their corrections range from $-0.34$\,dex to $+0.41$\,dex and are on average $\sim 0.1\,$dex higher than those of \citeauthor{Mashonkina2019_Ba2} (\citeyear{Mashonkina2019_Ba2}, averaged over the five optical lines) for the same stellar parameters. The difference between the two sets does not show a trend with $T_{\rm eff}$, $\logg$, [Fe/H] or [\ion{Ba}{II}/H].

We choose the \citet{Mashonkina2019_Ba2} corrections as our fiducial grid, since they span the widest parameter range.

\subsection{Europium II}

Mean corrections for 6 \ion{Eu}{II} optical lines were calculated using the grid of \citep{Mashonkina2000_Eu2} (see Fig.~\ref{fig: Eu2_lines}). The corrections are always positive, and tend to increase with decreasing [Fe/H] and [\ion{Eu}{II}/Fe], reaching a maximum of $\sim+0.35 \,$dex. The scatter between the different lines is always small, $\sigma\lesssim 0.1$.

For comparison, we also plot in Fig.~\ref{fig: Eu2_lines} the mean corrections from the limited grid provided by \citet{Mashonkina2014_Eu2}, which differ by less than 10$\%$ (and less than 0.05\,dex) from those of \citet{Mashonkina2000_Eu2} for the same stellar parameters.

\begin{figure}
\begin{center}
\includegraphics[width=0.95\hsize]{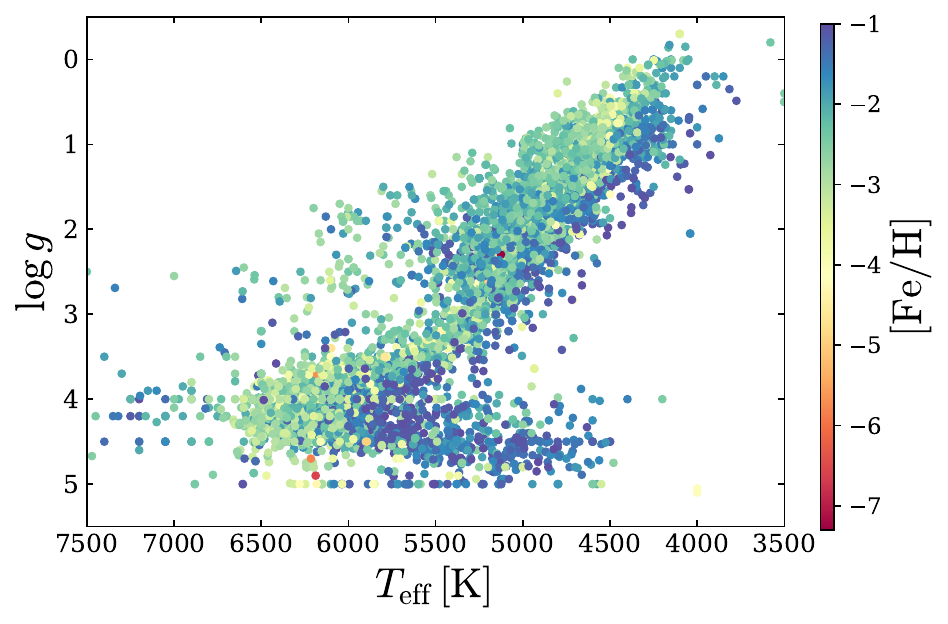} 
\caption{Kiel diagram of all metal-poor SAGA stars, colour-coded by metallicity [Fe/H].}
\label{fig: Kiel}
\end{center}
\end{figure}

\begin{figure}
\begin{center}
\includegraphics[width=0.95\hsize]{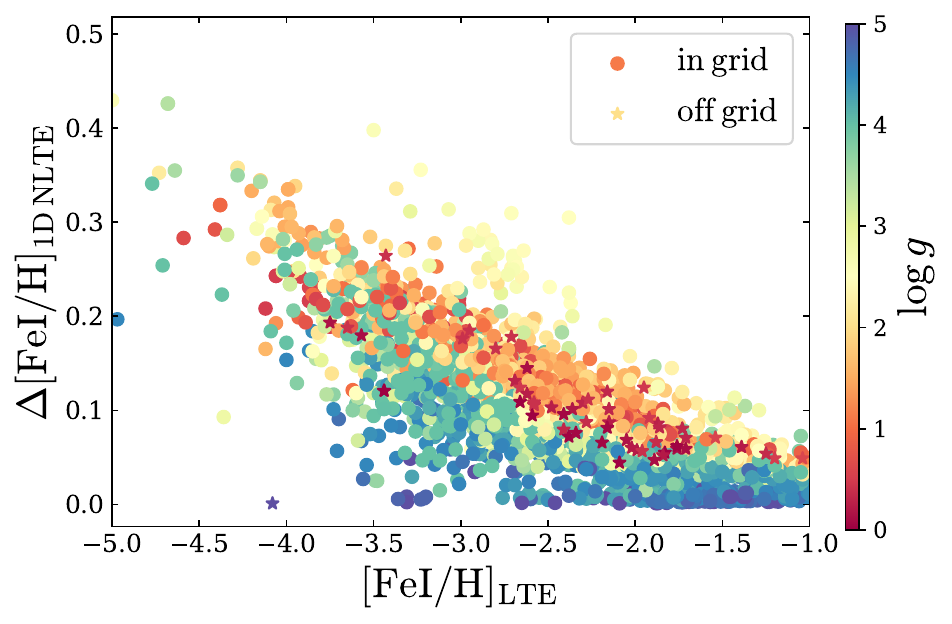} 
\caption{NLTE corrections for \ion{Fe}{I} in all metal-poor SAGA stars as a function of their [\ion{Fe}{I}/H]$_{\rm LTE}$, using the mean corrections from \citet{Mashonkina2011_Fe1} and \citet{Bergemann2012_Fe1_Fe2}. The data points are colour-coded by \logg. Circles represent stars whose parameters fall within the grid and star symbols mark those outside the grid boundaries.}
\label{fig: Fe1_logg1}
\end{center}
\end{figure}

\section{Correcting the SAGA database}
\label{sec:SAGA}

\subsection{Abundance tables}

We apply our NLTE interpolation code, \texttt{NLiTE}, to all metal-poor ($\rm[Fe/H]\leq-1$) Milky Way stars in the SAGA database (2023, April 10 version). The results are given in two online tables for all chemical species for the 7367 SAGA entries at $\rm[Fe/H]\leq-1$, corresponding to 3296 individual stars. The Kiel diagram of the sample, $\logg$ as a function of \teff, is shown in Fig.~\ref{fig: Kiel}. 

Table~\ref{table:Fiducial} provides our fiducial NLTE corrected SAGA database obtained by following the approach described in Sec.~\ref{Nlite}, using the fiducial grids listed in Table~\ref{table:Grids}. It includes [Fe/H]$_{\rm NLTE}$, the corrected [X/Fe]$_{\rm NLTE}$ abundance ratios and LTE abundance ratios for N, S, Sc, V and Ni. In case of multiple entries for the same star, in Table~\ref{table:Fiducial} we adopt the study which has the highest number of measured elements. The full details of our NLTE corrections for every individual element and SAGA entry (in some cases multiple analyses for the same star) are described in Table~\ref{table: Complete}. Table~\ref{table: Parameters} provides the coordinates and stellar atmospheric parameters of all entries.

\begin{figure*}
\begin{center}
    \includegraphics[width=0.49\hsize]{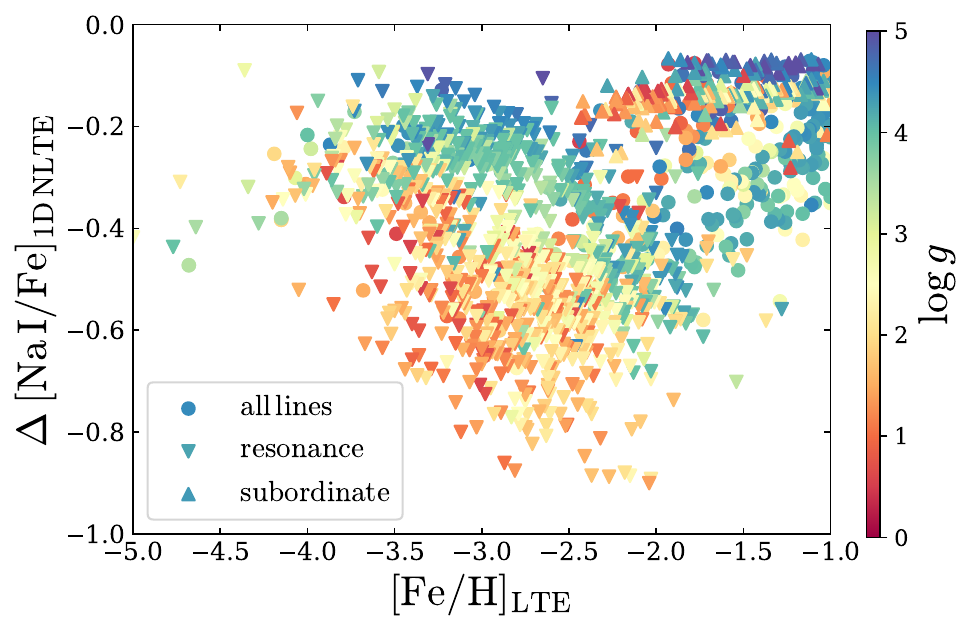} 
    \includegraphics[width=0.49\hsize]{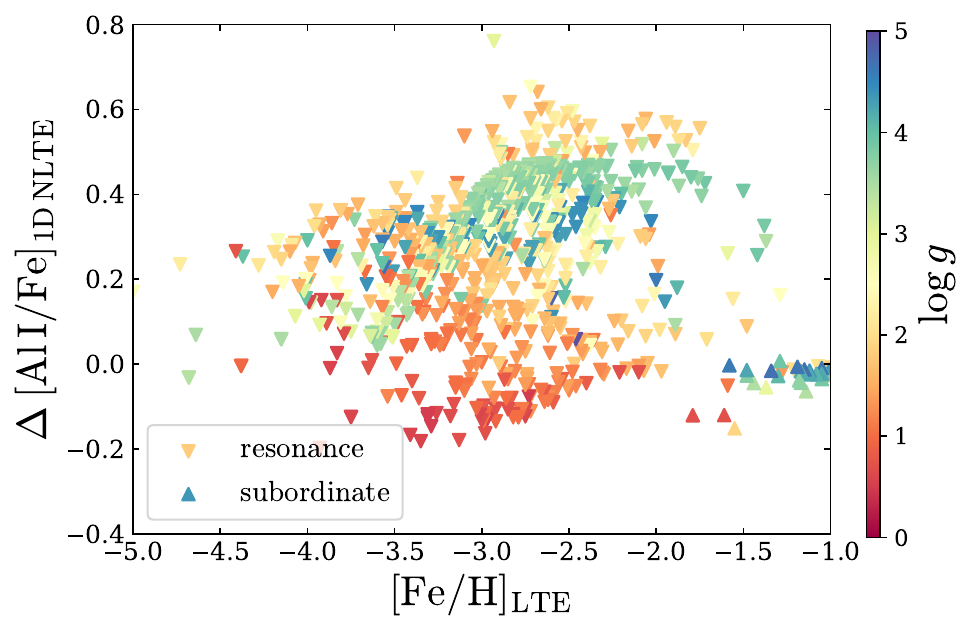} 
\caption{1D\,NLTE corrections for [Na/Fe] (left) and [Al/Fe] (right) as a function of their [Fe/H]$_{\rm LTE}$ for all MP SAGA stars, derived using the grids from \citet{Lind2022} for Na and Al, and the NLTE \ion{Fe}{I} in Fig.~\ref{fig: Fe1_logg1}. Stars with abundances determined from resonance or from subordinate lines are marked with downward and upward triangles, respectively.}
\label{fig: NaAl_logg}
\end{center}
\end{figure*} 

\begin{figure*}
\begin{center}
    \includegraphics[width=0.49\hsize]{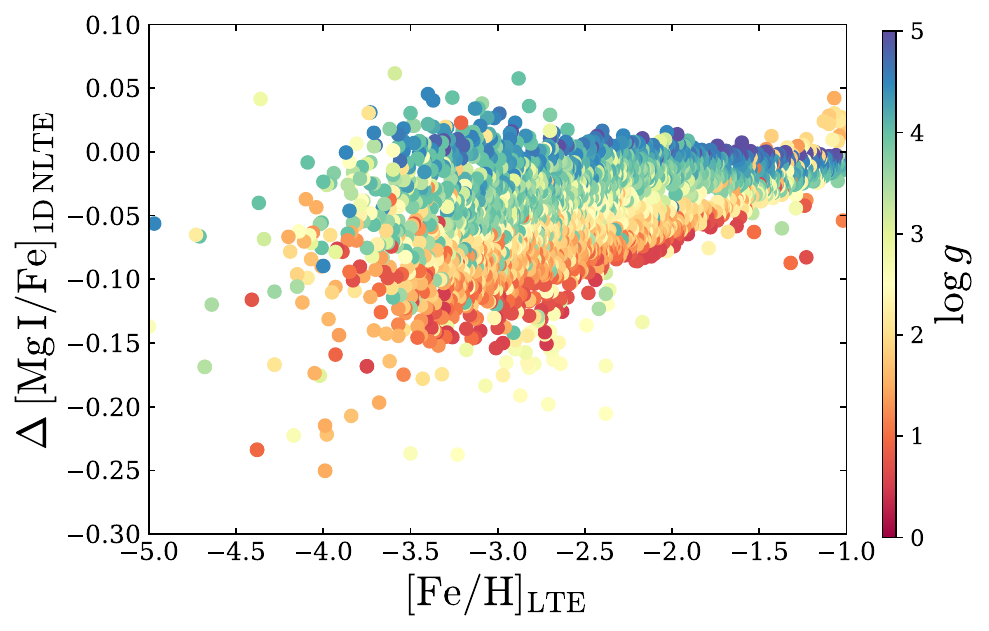} 
    \includegraphics[width=0.49\hsize]{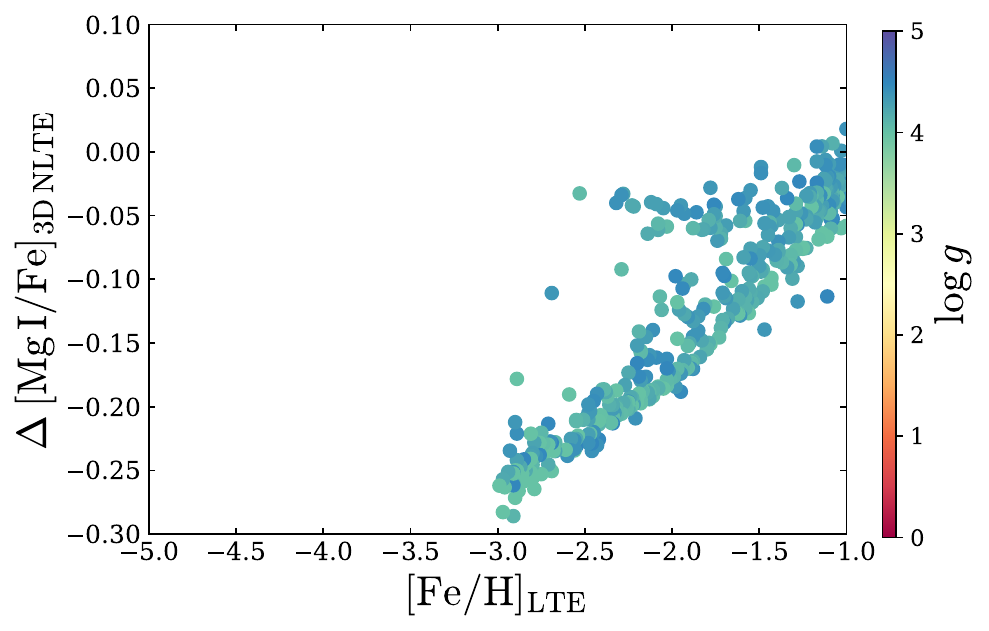} 
\caption{NLTE corrections for [Mg/Fe], colour-coded by \logg, for all MP SAGA stars. \textit{Left:} 1D\,NLTE corrections from \citet{Osorio2016_Mg1} for \ion{Mg}{I}, and \ion{Fe}{I} from Fig.~\ref{fig: Fe1_logg1}. \textit{Right:} 3D\,NLTE corrections from \citet{Matsuno2024} for \ion{Mg}{I}, and \citet{Amarsi2022} for \ion{Fe}{I}, defined only for dwarf stars.}
\label{fig: MgFe_logg}
\end{center}
\end{figure*}

The SAGA database is most appropriate for our study as it preferentially compiles 1D\,LTE abundances. However, occasionally these are not available, and the included abundances are already corrected for NLTE effects, most commonly in the cases of Li, Na, O and Al.
When possible, we retrieve the original 1D\,LTE abundances: for example \citet{Cohen2013} and \citet{Bandyopadhyay2018} apply a constant correction $\rm \Delta NLTE=+0.6\,$dex for Al, independent of stellar parameters, based on the results of \citet{Baumuller1997}. In other cases, it is not straightforward to obtain the 1D\,LTE abundances, and then they are marked as 'Originally Corrected', or  $\rm (OC)=1$ in the extended Table~\ref{table: Complete}, and we set $\Delta$NLTE=0.

Before applying our corrections, we make sure to exclude all entries with abundances derived through lines that are not included in our line list. In particular, we exclude all abundances derived through infrared or far-UV lines, carbon abundances based on the C$_2$ molecule, oxygen abundances based on the forbidden [OI] line or OH or CO features, iron abundances derived from the \ion{Ca}{II} H and K lines, as well as all abundances that involve lines from different ionization states (e.g. Sr abundances derived through \ion{Sr}{I} lines, Mn abundances derived through \ion{Mn}{II} lines, Ti abundances derived through both \ion{Ti}{I} and \ion{Ti}{II} lines, e.t.c.). These abundances are omitted from our fiducial NLTE-corrected Table~\ref{table:Fiducial} but remain available in the extended Table~\ref{table: Complete}, where they are flagged as `Other Wavelengths' (OW) = 1, with $\Delta$NLTE set to NaN.

\begin{figure*}
\begin{center}
    \includegraphics[width=0.49\hsize]{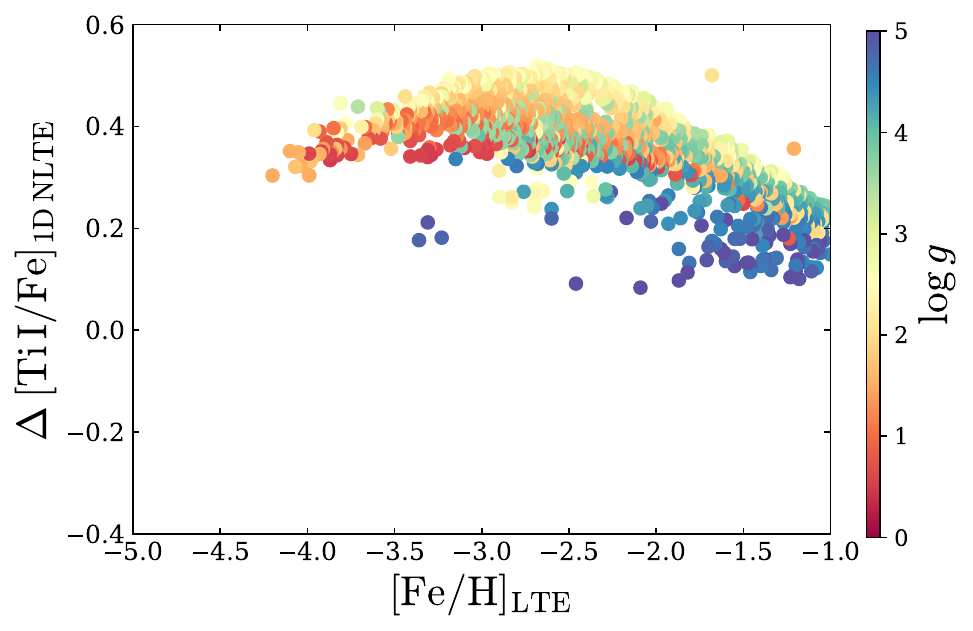} 
    \includegraphics[width=0.49\hsize]{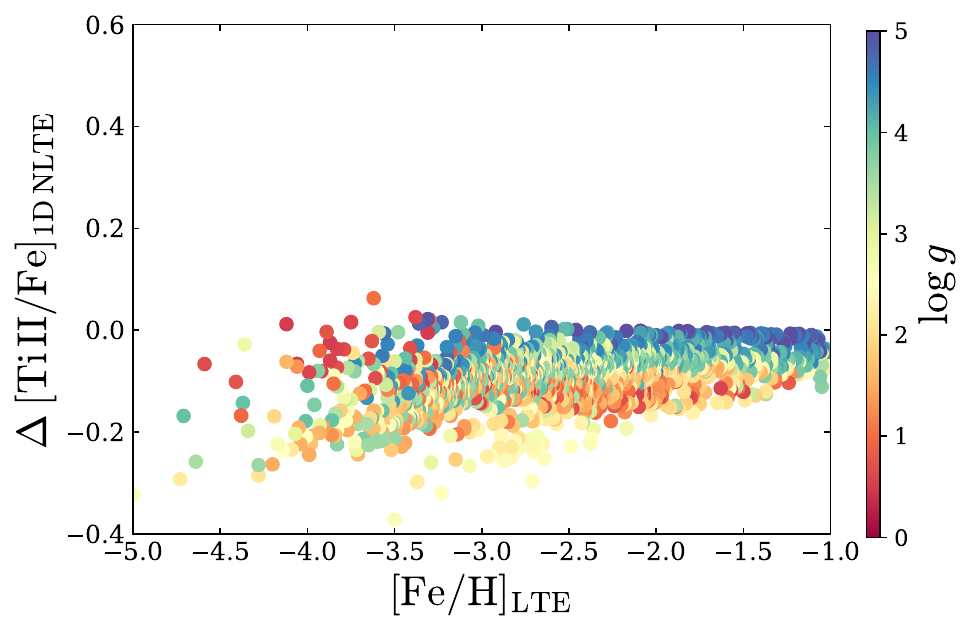} 
\caption{NLTE corrections for [\ion{Ti}{I}/Fe] (left) and [\ion{Ti}{II}/Fe] (right) for all MP SAGA stars based on \citet{Bergemann2011_Ti1_Ti2}, and the NLTE \ion{Fe}{I} from Fig.~\ref{fig: Fe1_logg1}.}
\label{fig: TiFe1_logg}
\end{center}
\end{figure*} 

\begin{figure}
\begin{center}
    \includegraphics[width=0.95\hsize]{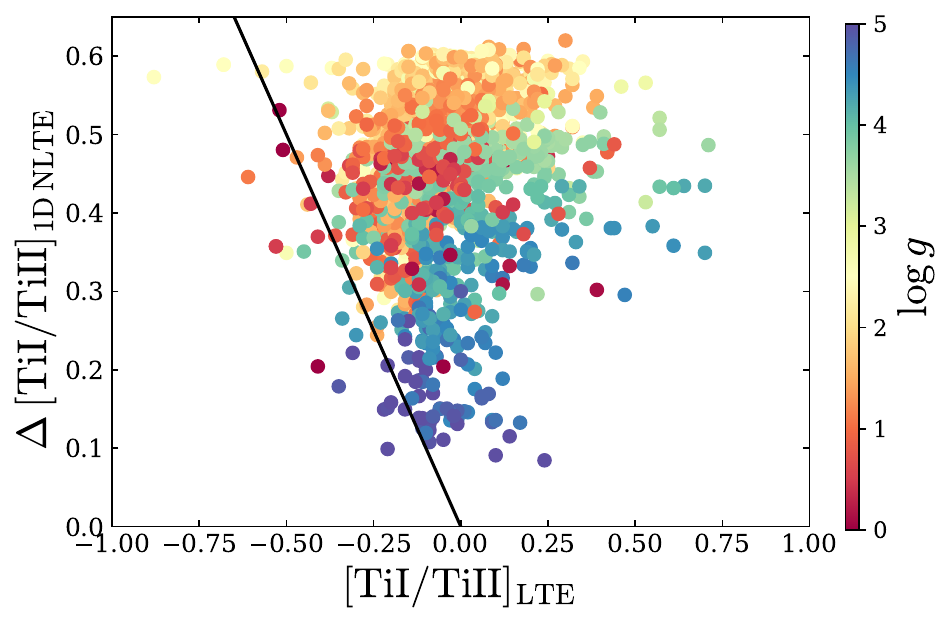} 
\caption{NLTE corrections for [\ion{Ti}{I}/\ion{Ti}{II}] for all MP SAGA stars, as a function of [\ion{Ti}{I}/\ion{Ti}{II}]$_{\rm LTE}$. The black solid line represents $\Delta [\ion{Ti}{I}/\ion{Ti}{II}]_{\rm NLTE} = - [\ion{Ti}{I}/\ion{Ti}{II}]_{\rm LTE}$, for which applying the NLTE corrections would result in ionization balance, [\ion{Ti}{I}/H]$_{\rm NLTE}$ =  [\ion{Ti}{Ii}/H]$_{\rm NLTE}$.}
\label{fig: TiTi}
\end{center}
\end{figure}

\begin{figure*}
\begin{center}
    \includegraphics[width=0.49\hsize]{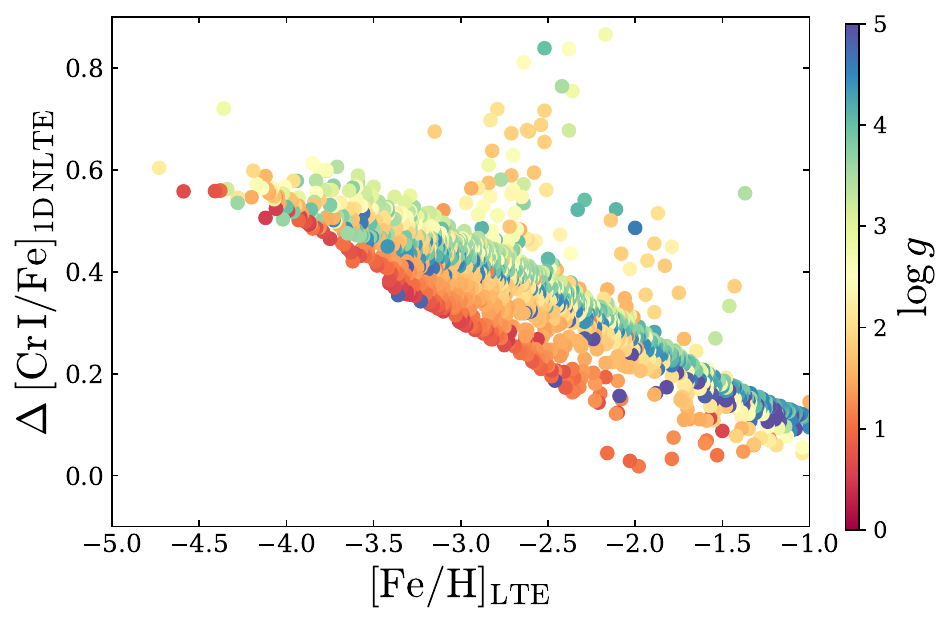} 
    \includegraphics[width=0.49\hsize]{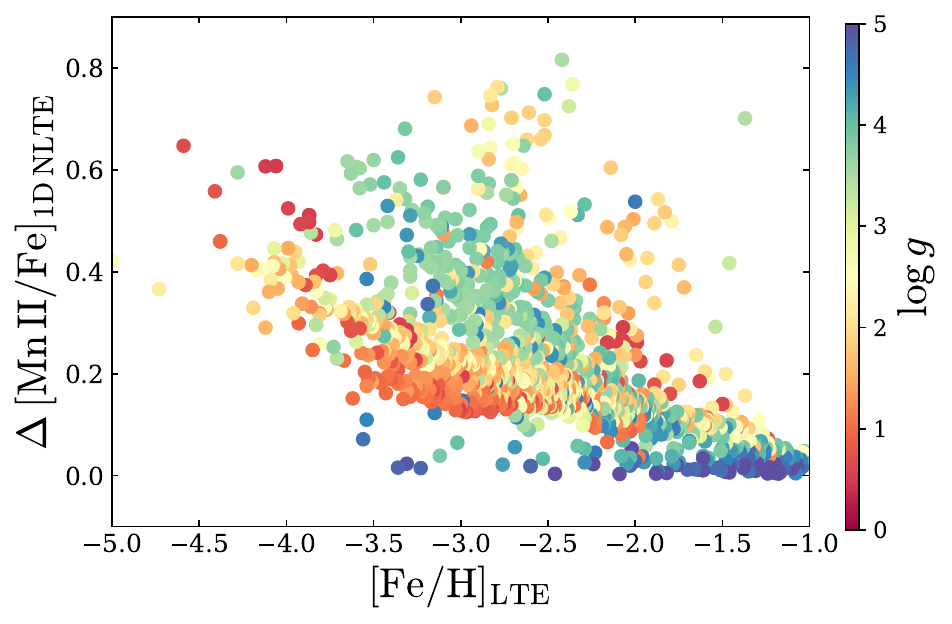} 
\caption{NLTE corrections for [\ion{Cr}{I}/Fe] (left) and [\ion{Mn}{I}/Fe] (right) for all MP SAGA stars, as a function of [\ion{Fe}{I}/H]$_{\rm LTE}$, using the grids of \citet{Bergemann2010_Cr} for \ion{Cr}{I}, \citet{Bergemann19} for \ion{Mn}{I}, and the NLTE \ion{Fe}{I} from Fig.~\ref{fig: Fe1_logg1}}
\label{fig: CrMn_logg}
\end{center}
\end{figure*} 

\subsection{The applied corrections}
\label{sec:appcorr}

Here we discuss the results of the NLTE corrections of the SAGA database for selected elements: Fe, Na, Mg, Al, Ti, Cr, and Mn. More details about other elements, as well as the relevant figures can be found in Appendix~\ref{app: applied_corr}.

The 1D\,NLTE corrections for \ion{Fe}{I} applied to all SAGA entries with $\rm [Fe/H]\leq-1$ are displayed in Fig.~\ref{fig: Fe1_logg1}. The corrections are based on the average of the \citet{Bergemann2012_Fe1_Fe2} and \citet{Mashonkina2011_Fe1} grids, which have a maximum discrepancy of 0.16\,dex. For stars with parameters outside of the available grid, we select the NLTE correction of the nearest boundary point, as described in Sec.~\ref{sec:code}.
The anti-correlation of $\Delta {\rm [Fe/H]_{1D \: NLTE}}$ with [Fe/H] is clearly visible, and the corrections are in general weakest for the highest \logg. 

Fig.~\ref{fig: NaAl_logg} shows $\Delta {\rm [Na/Fe]_{1D \: NLTE}}$ and $\Delta {\rm [Al/Fe]_{1D \: NLTE}}$ as a function of [Fe/H]$_{\rm LTE}$. These corrections show a more complex trend with [Fe/H] and \logg, than seen in Fig.~\ref{fig: Fe1_logg1}.
The [Na/Fe] corrections are always negative. At $\rm [Fe/H] \lesssim -2$, where Na abundances are primarily derived from the resonance doublet at 5889/5895$, \AA$, the corrections become stronger with decreasing $\logg$. At higher metallicities the corrections magnitude depends on the lines used, with the weakest corrections (typically > -0.2 dex) observed for stars measured through the subordinate lines at 5682/5688$, \AA$.

For aluminium, no clear trend with $\logg$ is observed. At $\rm [Fe/H] \gtrsim -1.5$, most Al abundances are based on the subordinate lines, which admit weak corrections. At lower metallicities the corrections are generally stronger, mostly positive and tend to increase with metallicity, driven by the correlation between $\Delta {\rm [Fe/H]_{1D \: NLTE}}$ and ${\rm [Fe/H]_{LTE}}$ (Fig.~\ref{fig: Fe1_logg1}). However, several giant stars display negative corrections. This occurs because although Al corrections remain positive in this regime, they are relatively small, whereas the Fe corrections are still significant. Figure~\ref{fig: NaAl_logg} therefore shows that it is not always easy to apply a simple relation to estimate the NLTE effects for a given set of stellar parameters. 

Figure~\ref{fig: MgFe_logg} displays the [Mg/Fe] corrections based on the 1D\,NLTE grid from \citeauthor{Osorio2016_Mg1} (\citeyear{Osorio2016_Mg1}; left) and the 3D-NLTE grid from \citeauthor{Matsuno2024} (\citeyear{Matsuno2024}; right panel). The 1D corrections exhibit a strong dependence on $\logg$ and are typically smaller than 0.15\,dex. The Mg 3D corrections of \citet{Matsuno2024}, available only for FG-type dwarfs, are also positive and exceed those of \citet{Osorio2016_Mg1} by $\sim0.06\,$dex on average, with a maximum difference of $\sim0.14\,$dex. However, the 3D \ion{Fe}{I} corrections of \citet{Amarsi2022} for the same stars exceed their 1D counterparts by a larger margin. As a result, the [\ion{Mg}{I}/\ion{Fe}{I}] 3D corrections for dwarfs are significantly more negative than their corresponding 1D values. Thus we see that 3D\,NLTE effects on Fe can have a quite significant impact on the [X/Fe]$_{\rm 3D\,NLTE}$ trend with [Fe/H], even if the element of interest, is not strongly affected by 3D effects.

The Ti ionization imbalance is a decades long problem in the abundance analysis of metal-poor stars. The 1D\,NLTE corrections for [\ion{Ti}{I}/Fe] and [\ion{Ti}{II}/Fe] are shown in Fig.~\ref{fig: TiFe1_logg}. We find that the corrections for neutral titanium are strong and positive, ranging from $\sim+0.1$ to $+0.5$ dex, while those for \ion{Ti}{II} are weaker and mostly negative, reaching down to about $-0.3$\,dex. Notably, these differences in the corrections do not align with the observed differences in the measured \ion{Ti}{I} and \ion{Ti}{II} abundances (Fig.~\ref{fig: TiTi}). For about 30$\%$ of the SAGA MP stars with both abundances available, [\ion{Ti}{I}/\ion{Ti}{II}] is already positive in LTE. Even in cases where the difference is negative in LTE, the difference in the corrections $\Delta [\ion{Ti}{I}/\ion{Ti}{II}]_{\rm NLTE}$ is significantly higher, meaning that the ionization imbalance would worsen after applying NLTE corrections. 

This issue has been previously highlighted by \citet{Sitnova2016_Ti2}, who noted that for more than half of the metal-poor stars they analysed, the agreement between \ion{Ti}{I} and \ion{Ti}{II} was better in LTE than in NLTE. Similarly, \citet{Bergemann2011_Ti1_Ti2} found that the ionization balance for three of the four MP stars they studied worsened in NLTE, and strongly advised against using \ion{Ti}{I} abundances in studies of Galactic chemical evolution. More recently, \citet{Mallinson2022} observed that while 1D NLTE corrections improved the ionization balance in some VMP giants, they disrupted it in other MP stars. The authors concluded that consistent 3D NLTE modelling is essential for resolving these discrepancies and advancing the field.

Finally we show the corrections for [\ion{Cr}{I}/Fe] and [\ion{Mn}{I}/Fe] in Fig.~\ref{fig: CrMn_logg}. Both are positive and reaching up to $\sim+0.8 \,$dex. The [\ion{Cr}{I}/Fe] corrections show a pronounced dependence on metallicity, which is also seen to a lesser degree in [\ion{Mn}{I}/Fe]. For both Cr and Mn, the corrections increase with decreasing $\logg$ at fixed $T_{\rm eff}$ and metallicity (see Figs~\ref{fig: Cr1_common_lines} and \ref{fig: Mn1_common_lines}). This trend is reversed here due to the same dependence of the Fe corrections. In addition, the trends for [\ion{Mn}{I}/Fe] appear more blurred out due to their strong dependence on effective temperature. These strong 1D\,NLTE corrections for [Cr/Fe] and [Mn/Fe] result in artificial trends with [Fe/H] in the LTE assumption, which if not corrected can lead to very wrong conclusions about the nucleosynthesis of these elements (see Sec.~\ref{Iron-peak elements Cr and Mn}).

Figs.~\ref{fig: Fe1_logg1}-\ref{fig: CrMn_logg} highlight our results when correcting a database with a reality-based distribution of stellar parameters. These results show the diversity in the 1D\,NLTE corrections for different elements, and their not always straightforward dependence on stellar parameters. Furthermore, we see indications of potentially significant effects when taking the full 3D structure of the stellar atmosphere into account, but unfortunately studies of full 3D\,NLTE grids are still very limited. With \texttt{NLiTE} we are able to efficiently provide large databases of MP stars with more accurate 1D\,NLTE abundances using the state-of-the-art grids available. 

\begin{figure}
\begin{center}
\includegraphics[width=0.9\hsize]{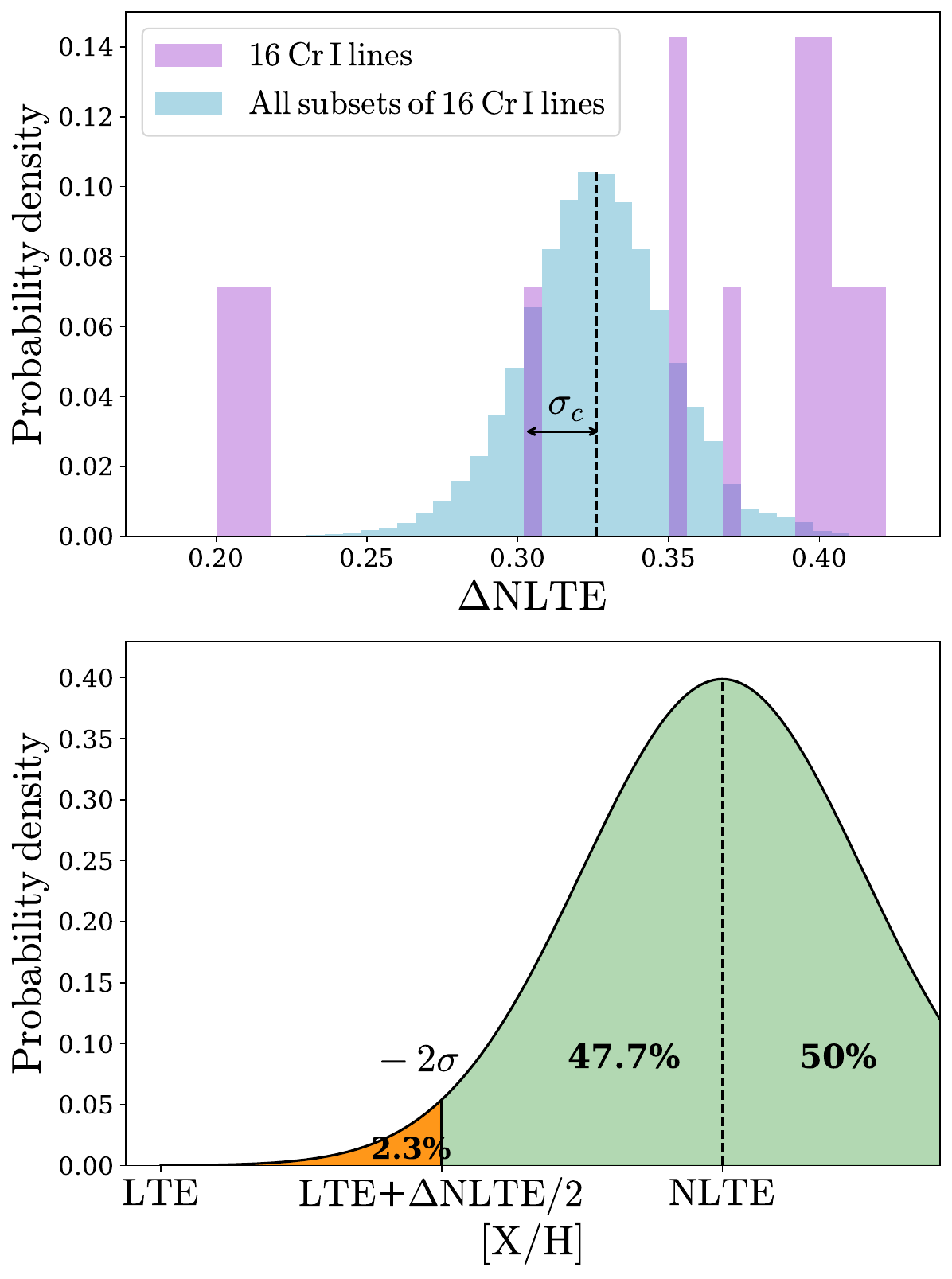} 
\caption{\textit{Top:} Normalized distribution of the NLTE corrections for 16 individual \ion{Cr}{I} lines and the mean NLTE corrections of all their possible subsets.
\textit{Bottom: }Illustration of our criterion to assess the reliability of average NLTE corrections. If the line-by-line NLTE value falls within the risky area (orange), applying the average NLTE correction leads to a larger offset compared to adopting the LTE value. The percentages shown correspond to the case where $2\sigma = |\Delta{\rm NLTE}|/2$.}
\label{fig: Gaussian}
\end{center}
\end{figure}

\begin{figure}
\begin{center}
    \includegraphics[width=0.95\hsize]{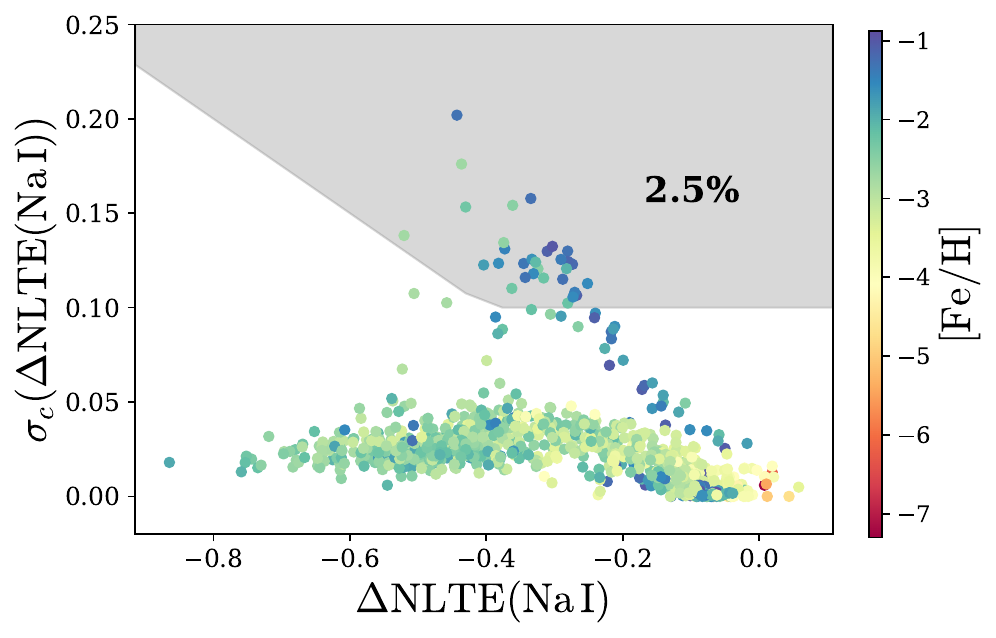} \\
    \includegraphics[width=0.95\hsize]{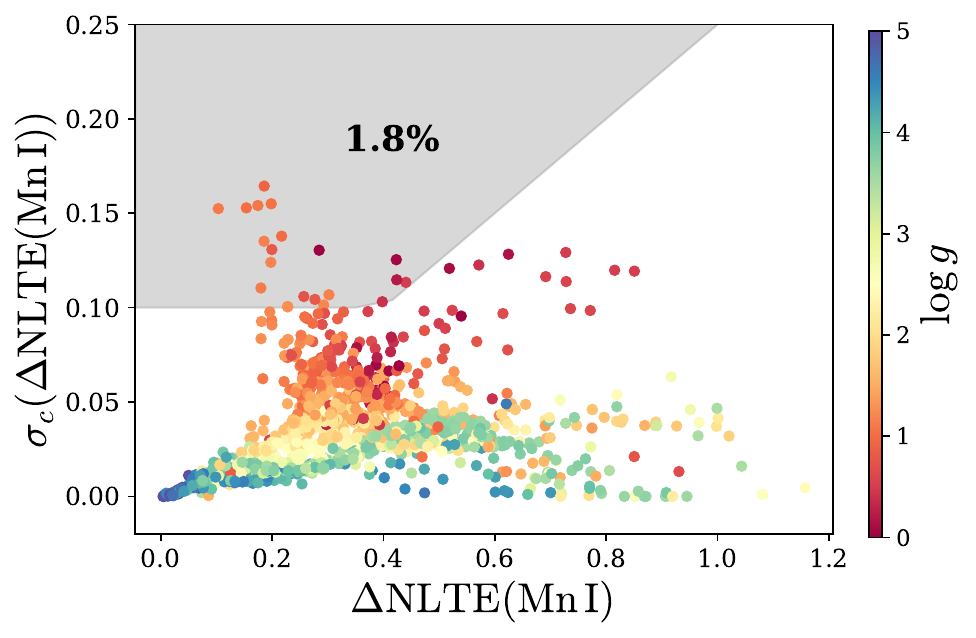} \\
\caption{Scatter of mean NLTE corrections from all subsets of lines as a function of $\Delta$NLTE, colour-coded by [Fe/H] (Na) or $\log g$ (Mn). Grey areas signify higher uncertainties on the average corrections, i.e. when $\sigma_c>0.1$\,dex and $\sigma_c>|\rm\Delta NLTE|/4$.}
\label{fig: scatter}
\end{center}
\end{figure}

\subsection{Scatter between lines}
\label{Scatter}

A key uncertainty in our approach is how well our average NLTE corrections represent those done line-by-line, i.e. corresponding to the actual (often unknown) line list used for individual stars in the SAGA database.
In other words, our average corrections might be shifted in stars where only a portion of our fiducial line list is used, and this might affect global abundance trends. To evaluate the statistical robustness of our corrections, we compute, for each SAGA star and chemical species, the standard deviation of the NLTE corrections across different lines, and provide it in Table~\ref{table: Complete}, along with the number of lines used.\footnote{The number of lines given in Table~\ref{table: Complete} refer to those at the closest NLTE grid point, since the number of detectable lines, i.e. lines with $5 \AA \leq {\rm EW} \leq 200 \AA$ can vary across neighboring grid points.} The same information is also available while using the \texttt{NLiTE} tool for individual elements.

Table~\ref{table: scatter} contains our analysis on the robustness of our method. Let $N$ be the number of detectable lines for a chemical species at given stellar atmosphere and $\sigma \equiv \sigma(\Delta{\rm NLTE})$ the standard deviation of their individual NLTE corrections. 
If only $n$ lines were used for an observation, the number of possible line combinations (subsets) is:
\begin{equation}
    N_n = \frac{N!}{n!(N-n)!}.
\end{equation}
The standard deviation of the mean NLTE corrections of all these subsets is:
\begin{equation}
    \sigma_n = \sigma\sqrt{\frac{1}{n} \bigg(1-\frac{n-1}{N-1}\bigg)}.
\end{equation}

If we consider all possible subsets with $n=1,..,N$,  the overall combined standard deviation is:
\begin{equation}
    \sigma_c = \sqrt{\frac{\sum_{n=1}^{N}{N_n \sigma_n}}{\sum_{n=1}^{N} {N_n}}}.
\end{equation}

Furthermore, when $N$ is large, the central limit theorem ensures that the distribution of the mean NLTE corrections from all subsets approaches a normal distribution. An example is shown at the top panel of Fig.~\ref{fig: Gaussian}, where the NLTE corrections of 16 individual \ion{Cr}{I} lines, for a star with $T_{\rm eff}=5000\,$K, $\logg=2$ and ${\rm [Fe/H]}=-2$, are compared to the distribution of the mean corrections of all possible subsets of these lines, which closely follows a Gaussian profile with standard deviation $\sigma_c$.

In such cases, the reliability of our average corrections can be assessed using standard statistical theory. Specifically, in stellar atmospheres where $2 \sigma_c  < |\Delta{\rm NLTE}|/2$, there is a $>97.7\%$ probability (green shaded area in Fig.~\ref{fig: Gaussian}) that the average correction $\Delta{\rm NLTE}$ improves the abundance estimate, i.e. brings it closer to the true value. Thus, if $\sigma_c <|\Delta{\rm NLTE}|/4$, then applying the correction is preferable to not doing so $>97.7\%$ of the time.
This analysis does not hold when $N$ is small, as the mean corrections from possible subsets may not approximate a Gaussian. Nevertheless, the threshold $\sigma_c<|\Delta{\rm NLTE}|/4$, still indicates a relatively narrow distribution.
We adopt this threshold to assess the reliability of our corrections, and the fraction of stars exceeding this limit for each element is listed in Table~\ref{table: scatter}.

When $\Delta{\rm NLTE}$ is small, $\sigma_c$ may surpass this limit despite being negligible in absolute terms. To account for this, we consider all corrections with $\sigma_c<0.1\,$dex to be acceptable, as they lie within the intrinsic uncertainties of NLTE models (see Sec.~\ref{Differences between grids}) and are unlikely to impact the inferred abundance trends significantly.

Fig.~\ref{fig: scatter} displays $\sigma_c$ as a function of $\Delta {\rm NLTE}$, for all individual SAGA MP stars for \ion{Na}{I} and \ion{Mn}{I}, which are the worst cases in Table~\ref{table: scatter}. The grey shaded area in each panel represents cases where $\sigma_c>0.1$\,dex and $\sigma_c>|\Delta {\rm NLTE}|/4$, i.e. areas with high scatter, where applying average NLTE corrections could potentially shift LTE abundances away from their true values. For \ion{Na}{I}, only $2.5\%$ of stars are in this risky area. As described in Sec.~\ref{sodium}, for \ion{Na}{I} (as for \ion{Al}{I}; see Sec.~\ref{aluminium}) separate corrections have been applied to stars observed through resonance and subordinate lines, since these admit very different NLTE corrections. Stars with the largest scatter typically correspond to higher [Fe/H], where both types of lines are detectable and used in abundance calculations (or cases where information on the Na lines are not provided).
Instead for \ion{Mn}{I}, the largest scatter is seen at low $\logg\lesssim1.5$. 

The scatter analysis for the remaining elements shows that the average corrections provide a robust representation of NLTE effects for the vast majority of stars. However, this analysis is not particularly well suited in all cases, e.g. for multiplets with very different $\Delta$NLTE values. In such cases, the abundances are much more likely to be measured from one multiplet (in part due to wavelength coverage of the observed spectra), rather than selecting one line from each multiplet. Different subsets of lines therefore have very different probabilities. This issue has the most significant effects in the case of Na, Al and S, and these elements have therefore been treated specially in this work (Sec.~\ref{sodium}, \ref{aluminium}, and \ref{sulfur}), and for these elements the users of \texttt{NLiTE} are advised to carefully select from the available grids those that correspond best to their line list.

\renewcommand{\arraystretch}{1.5}
\begin{table}[t]
\centering
\caption{\textbf{Line-to-line scatter of $\Delta$NLTE in SAGA MP stars.}}
\footnotesize
\tabcolsep=0.075cm
\begin{tabular}{c c c c c c c}
\hline
\hline
Species & ${\rm N_\star}$ &
$\mean{\sigma_{\rm c}}$  & 
$\sigma_{\rm c, \: max}$ 

& $\sigma_{\rm c}  \, {\scriptscriptstyle (>0.1)}$ 

& $\sigma_{\rm c} \, {\scriptscriptstyle \big(> \frac{|\Delta{\rm NLTE}|}{4}\big)}$ 

& $\sigma_{\rm c} \, {\scriptscriptstyle (>0.1)} \bigcap \sigma_{\rm c}\, {\scriptscriptstyle \big( >\frac{|\Delta{\rm NLTE}|}{4}\big)}$

\\
&  & \multicolumn{1}{c}{(dex)} & \multicolumn{1}{c}{(dex)} &  \multicolumn{1}{c}{($\%$)} & \multicolumn{1}{c}{($\%$)} & \multicolumn{1}{c}{($\%$)}  \\
\hline
\ion{C}{I} & 20 & 0.011  & 0.030 
 & 0.0 & 0.0 & 0.0 \\ 
\ion{O}{I} & 379 &   0.003 & 0.035  & 0.0 & 0.0 & 0.0  \\ 

\ion{Na}{I} & 1306 &  0.024  &  0.202  & 2.6 & 6.0 & 2.5 \\ 

\ion{Mg}{I} & 2092 & 0.027  &  0.078  & 0.0 & 71.5 & 0.0 \\ 

\ion{Al}{I} & 745 & 0.009  & 0.065 & 0.0 & 1.5 & 0.0 \\

\ion{Si}{I} & 1288 &   0.011  &  0.088 & 0.0 & 38.3 & 0.0 \\ 

\ion{K}{I} & 311 &  0.018  &  0.042  & 0.0 & 0.6 & 0.0\\ 

\ion{Ca}{I} & 2007 & 0.018  &  0.106  & 0.1 & 31.8 & 0.0 \\ 

\ion{Ti}{I} & 1270 &  0.019  &   0.069 & 0.0 & 0.1 & 0.0 \\ 

\ion{Ti}{II} & 1730 & 0.007  &   0.085 & 0.0 & 40.7 & 0.0 \\ 

\ion{Cr}{I} & 1372 &   0.018  &  0.092  & 0.0 & 0.5 & 0.0 \\ 

\ion{Mn}{I} & 1158 &  0.036 &  0.164  & 2.5 & 6.0 & 1.8 \\ 

\ion{Fe}{I} & 2067 &  0.005  &  0.074  & 0.0 & 1.3 & 0.0 \\

\ion{Fe}{II} & 1946 & 0.004  &  0.038  & 0.0 & 97.0 & 0.0 \\ 

\ion{Co}{I} & 1137 &  0.053 &   0.195  & 2.9 & 0.2 & 0.2 \\ 

\ion{Zn}{I} & 947 &  0.003  &   0.016  & 0.0 & 1.6 & 0.0 \\ 

\ion{Sr}{II} & 1385 & 0.012  &  0.030 & 0.0 & 28.6 & 0.0 \\ 

\ion{Ba}{II} & 1648 & 0.027  &  0.095  & 0.0 & 62.0 & 0.0 \\ 

\ion{Eu}{II} & 1026 & 0.018  &  0.062  & 0.0 & 8.5 & 0.0  \\ 
\hline
\end{tabular}

\noindent\parbox{0.49\textwidth}{
    \vspace{3mm} 
    \footnotesize {\bf Notes:} For each chemical species, we list: the number of individual MP SAGA stars (from Table~\ref{table:Fiducial}) with abundance measurements, ${\rm N_\star}$, the mean and maximum combined standard deviation of NLTE corrections across all possible line subsets, $\mean{\sigma_c}$ and $\sigma_{\rm c,max}$, the fraction of stars with  $\sigma_c>0.1\,$dex, the fraction of stars with $\sigma_c >{|\Delta{\rm NLTE}|/4}$ and the union of the two (see text for details).}

\label{table: scatter}
\end{table}

\subsection{Differences between grids}
\label{Differences between grids}

\begin{figure}
\begin{center}
    \includegraphics[width=0.92\hsize]{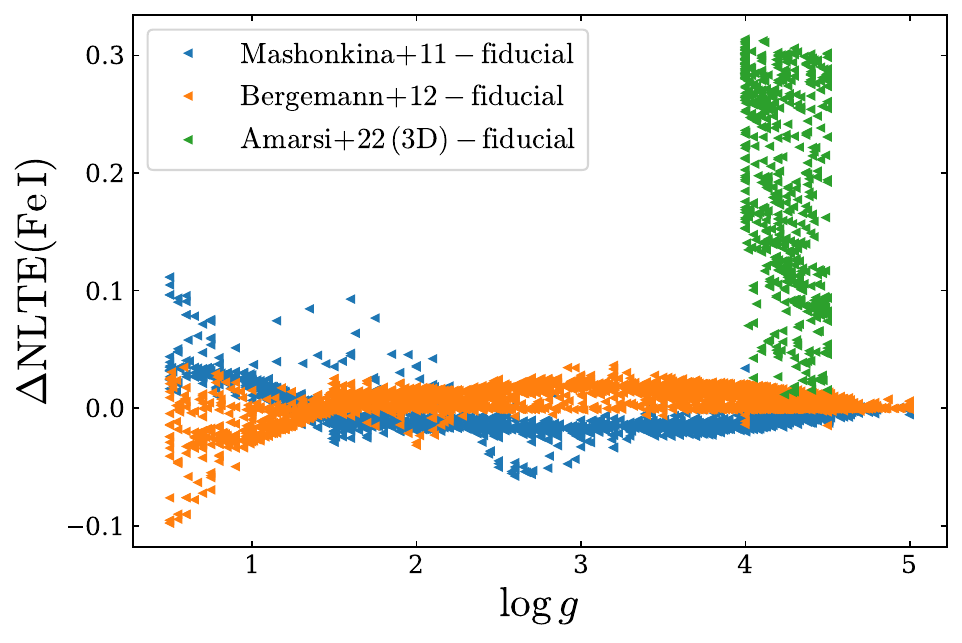} \\
    \includegraphics[width=0.92\hsize]{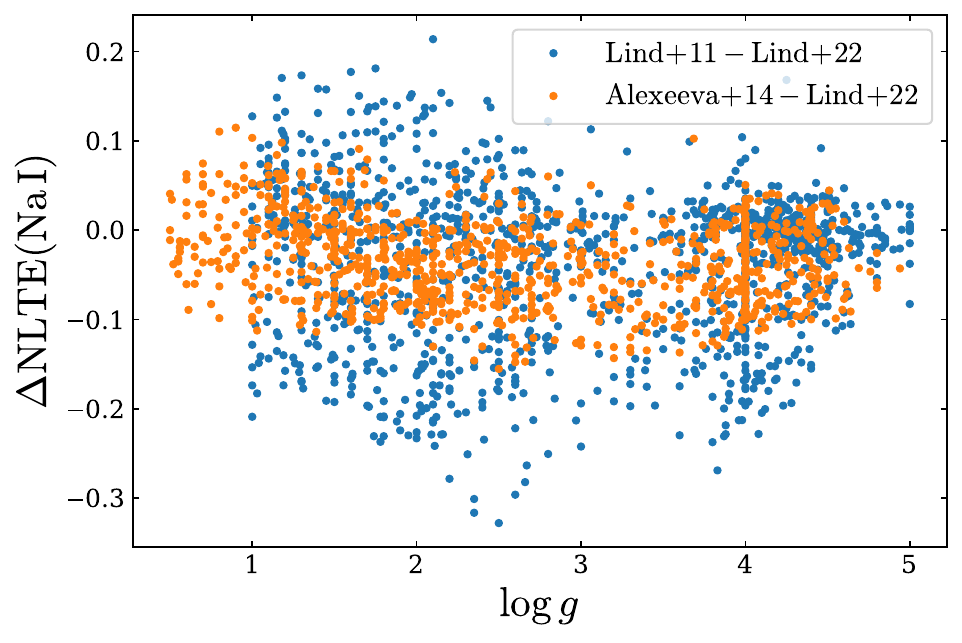} \\
    \includegraphics[width=0.92\hsize]{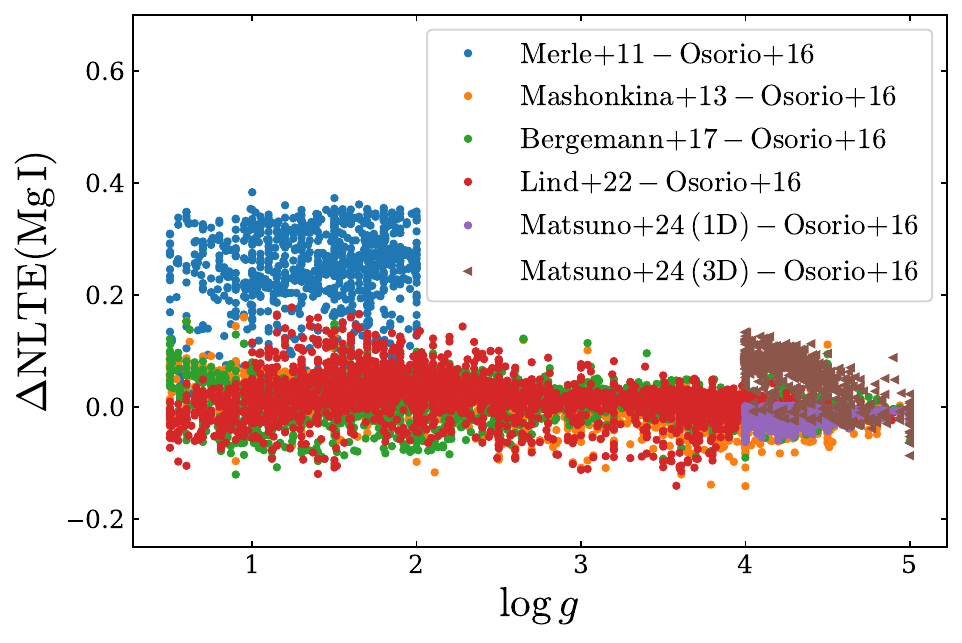} \\
    \includegraphics[width=0.92\hsize]{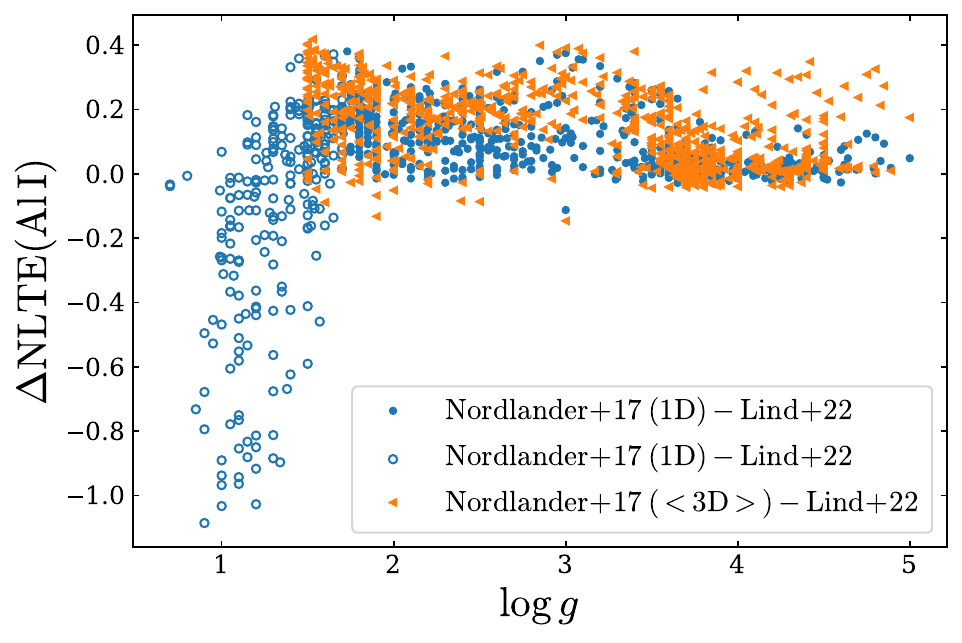}
\caption{Differences between the NLTE corrections for MP SAGA stars derived using all available grids, from top to bottom, for \ion{Fe}{I}, \ion{Na}{I}, \ion{Mg}{I} and \ion{Al}{I}. The fiducial \ion{Fe}{I} grid represents the mean corrections from \citet{Mashonkina2011_Fe1} and \citet{Bergemann2012_Fe1_Fe2}. Notice that the range on the y-axes is not always the same.}
\label{fig: diff}
\end{center}
\end{figure}

Fig.~\ref{fig: diff} examines the difference in the NLTE corrections for \ion{Fe}{I}, \ion{Na}{I}, \ion{Mg}{I} and \ion{Al}{I} when different grids are applied to MP SAGA stars. 

We find that the 1D\,NLTE corrections for \ion{Fe}{I} from \citet{Mashonkina2011_Fe1} and \citet{Bergemann2012_Fe1_Fe2} are in excellent agreement, differing by less than 0.05 dex for $\sim95\%$ of the stars. In contrast, the 3D\,NLTE corrections of \citet{Amarsi2022}, available only for dwarf stars with $4\leq \logg \leq 4.5$, are systematically higher with differences increasing from $<0.05\,$dex at $T_{\rm eff}=5000$\,K to 0.25-0.3\,dex at $T_{\rm eff}=6500\,$K (see also Sec.~\ref{Iron I}).

The NLTE corrections for Na are also in generally good agreement, with $73\%$ and $92\%$ of stars in the corrections of \citet{Lind2011_Na1} and \citet{Alexeeva2014_Na1}, respectively, differing by less than 0.1\,dex from those of our default grid from \citet{Lind2022}.
The NLTE corrections for Na are predominantly negative across all three grids considered. We note that cases where the corrections differ by more than 0.1\,dex between the grids do not show a clear trend with $\logg$, $T_{\rm eff}$, metallicity or lines used.

For \ion{Mg}{I}, our default grid from \citet{Osorio2016_Mg1}, is highly consistent with the 1D\,NLTE grids from \citet{Mashonkina2013_Mg1}, \citet{Bergemann2017_Mg1}, \citet{Lind2022}, and \citet{Matsuno2024}, with over 95$\%$ of stars exhibiting differences of less than 0.1 dex. An exception is the 1D\,NLTE grid from \citet{Merle2011}, which shows stronger positive corrections by $\sim0.2\,$dex, on average. The 3D\,NLTE corrections from \citet{Matsuno2024} are also systematically higher than those from \citet{Osorio2016_Mg1} with the difference being about 0.1\,dex at $\logg=4$ and decreasing with increasing $\logg$.

Bigger discrepancies are observed when considering Al.
Here, only 50-60$\%$ of stars have corrections that differ less than 0.1\,dex between the \citet{Nordlander2017} 1D and $\langle {\rm 3D} \rangle$ grids, and our fiducial grid from \citet{Lind2022}. 
At $\logg \geq 1.7$ the 1D corrections from \citet{Nordlander2017} are mostly positive and on average 0.08 dex higher than those from \citet{Lind2022}. For stars with $\logg<1.7$, the mean absolute difference between the two grids increases to approximately $0.25\,$dex. There, the corrections from \citet{Nordlander2017} are often strong and negative, while those from \citet{Lind2022} remain positive. In this low-$\logg$ regime, marked with open circles in Fig.~\ref{fig: diff}, the 1D \citet{Nordlander2017} corrections appear to have the wrong sign, likely due to issues in the grid or the underlying calculations, and  should not be considered reliable (Thomas Nordlander, private communication). The $\langle {\rm 3D} \rangle \,$NLTE corrections of \citet{Nordlander2017} are always positive and exceed those of \citet{Lind2022} by $\sim0.15\,$dex, on average.

The comparison of the corrections derived using different grids for \ion{Li}{I}, \ion{Si}{I}, \ion{K}{I}, \ion{Ca}{I}, \ion{Ti}{I}, \ion{Sr}{II} and \ion{Ba}{II} is discussed in Appendix~\ref{app: applied_corr}.

\subsection{The \ion{Fe}{i}-\ion{Fe}{II} ionization balance}

\begin{figure*}
\begin{center}
\includegraphics[width=0.95\hsize]{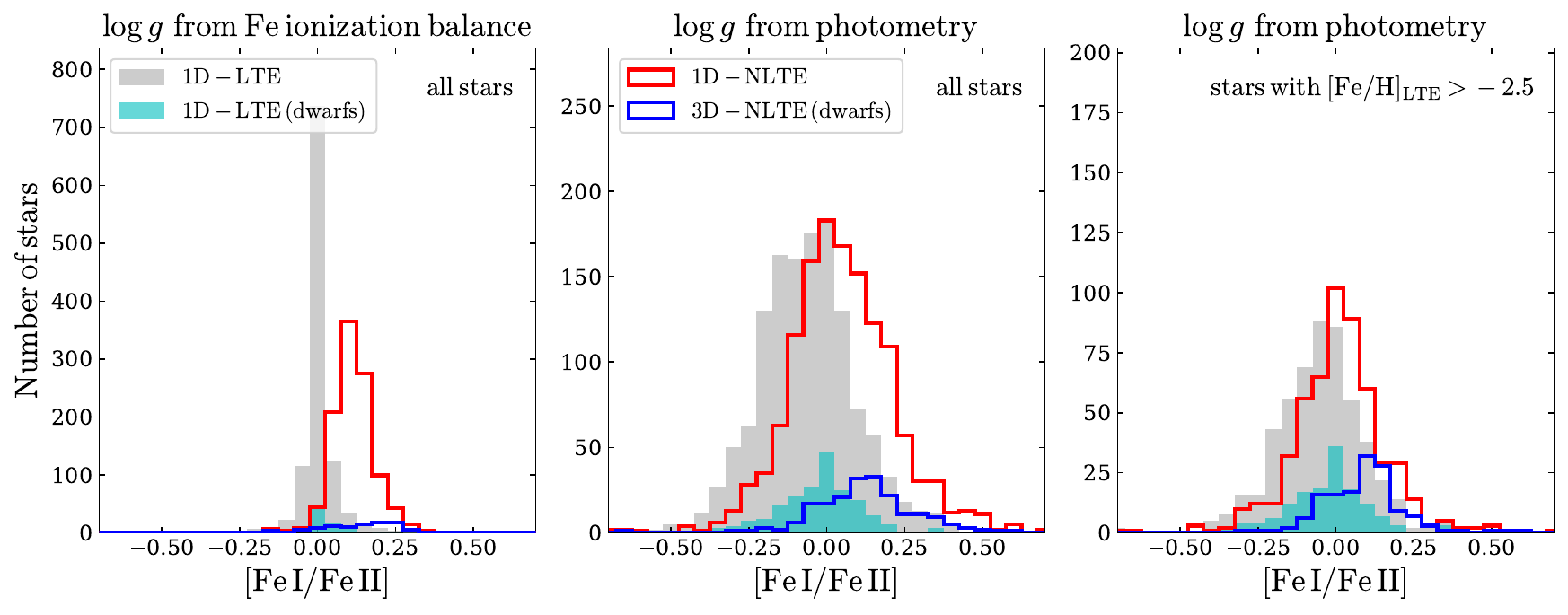} 
\caption{Ionization balance of Fe for all MP SAGA stars for which $\logg$ has been derived using the standard spectroscopic method (left), and  theoretical isochrones or stellar parallaxes for all MP stars (middle), and only stars with ${\rm [Fe/H]_{ LTE}} > -2.5$ (right). 
Shaded histograms are the original LTE distributions of all MP stars (grey) and of dwarf stars (cyan; $T_{\rm eff}=[5000,6500]$, $\logg=[4,4.5]$). Red and blue show the corresponding distributions after applying 1D\,NLTE corrections for all stars \citep{Bergemann2012_Fe1_Fe2} and 3D\,NLTE corrections for the dwarf sample \citep{Amarsi2022}, respectively.}
\label{fig: Fe1_Fe2}
\end{center}
\end{figure*} 

Stellar parameters are most commonly determined based on either the spectroscopic or the photometric method. In simplified terms, the spectroscopic method adjusts the stellar parameters so that all \ion{Fe}{I} and \ion{Fe}{II} lines give consistent Fe measurements. However, this is often done in LTE, neglecting NLTE effects that can strongly affect iron lines in a different way, e.g. \ion{Fe}{I} vs. \ion{Fe}{II} lines, especially at low metallicity \citep[see e.g.][]{Mucciarelli2020}. This can result in unreliable results, e.g. in surface gravities being underestimated by up to $\sim$0.5\,dex, which results in lower [Fe/H] abundances \citep[e.g.][]{Sitnova2015}. Alternatively, the stellar parameters can be based on photometry and distance estimates to the star, e.g. with parallaxes and isochrone fitting and/or other empirical relations \citep{Flower1996,Sitnova2015, Mashonkina2017,Mucciarelli2020}. These methods should be less sensitive to possible NLTE effects. 

Fig.~\ref{fig: Fe1_Fe2} shows the distribution of the ionization balance, [\ion{Fe}{I}/\ion{Fe}{II}], for the SAGA database.  We include only stars with parameters that lie within the respective grid limits (see Table~\ref{table:Grids}). For stars with surface gravities computed through the standard spectroscopic method (1077 entries), the original LTE [\ion{Fe}{I}/H] and [\ion{Fe}{II}/H] abundances were forced to agree with each other. Therefore, as expected, after applying the NLTE corrections the ionization balance becomes worse: the scatter increases and the peak moves at $+0.1\,$dex in 1D\,NLTE and at $+0.22\,$dex in 3D\,NLTE (available for 97 stars). This is a direct consequence of the fact that the NLTE corrections are systematically higher for \ion{Fe}{I} than for \ion{Fe}{II} (see Fig.~\ref{fig: Fe1_common_lines}).

On the other hand, when \logg\ is determined from photometry (1332 entries; middle panel of Fig.~\ref{fig: Fe1_Fe2}), in LTE the [\ion{Fe}{I}/H] is on average lower than [\ion{Fe}{II}/H] by $\sim$0.065\,dex. In this case, we would expect NLTE corrections to bring the distribution closer to equilibrium. Indeed, in 1D\,NLTE, the ionization balance shows modest improvement, with the mean offset reducing to +0.047\,dex. The 3D\,NLTE corrections (available for 199 stars), instead, appear to worsen the agreement, shifting the peak to +0.13\,dex and increasing the scatter from 0.14 to 0.16\,dex.

These findings align with previous studies suggesting that current models may overestimate NLTE departures for \ion{Fe}{I} at low [Fe/H]. For example, \citet{Mashonkina2019_Fe} report that at $\rm[Fe/H]<-3.5$, 1D\,NLTE \ion{Fe}{I} abundances exceed \ion{Fe}{II} abundances by up to +0.35 dex in five out of six cases. Similarly, in \citet{Bergemann2012_Fe1_Fe2} the ionization balance of their only EMP star analysed worsens in NLTE compared to LTE (though they note that the parallax of this star is highly uncertain), and \citet{Amarsi2022} find poor ionization balance for one of their two analysed VMP stars.

Indeed, if we consider only stars with $\rm[Fe/H]>-2.5$ (right panel of Fig.~\ref{fig: Fe1_Fe2}) the situation improves. In 1D\,NLTE, 61$\%$ of stars have $-0.1 < {\rm [\ion{Fe}{I}/\ion{Fe}{II}]}<+0.1$ compared to 52$\%$ in LTE, and the peak shifts from $-0.05$ to 0.0\,dex, while in 3D\,NLTE, the average offset decreases now to +0.09\,dex. For science cases that require high precision and accuracy, improvement of both the NLTE models and the chemical abundance analysis is advised.

\subsection{Impact on the SAGA MDF}
\label{sec:sagamdf}

\begin{figure}
\begin{center}
\includegraphics[width=0.95\hsize]{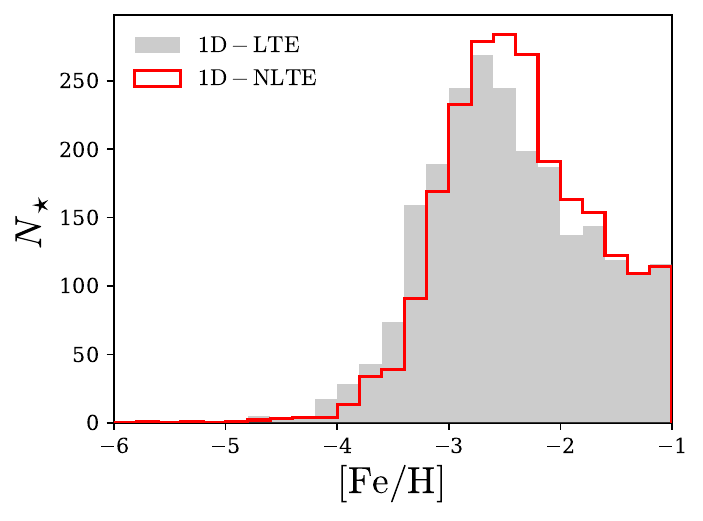} 
\caption{Metallicity distribution function of all MP SAGA stars in LTE (grey) and after applying our fiducial 1D\,NLTE corrections~(red).}
\label{fig: MDF}
\end{center}
\end{figure}

Fig.~\ref{fig: MDF} illustrates the impact on the SAGA metallicity distribution function (MDF) after applying the mean 1D-NLTE \ion{Fe}{I} corrections from \citet{Mashonkina2011_Fe1} and \citet{Bergemann2012_Fe1_Fe2}. Here, we include only individual stars; in cases where multiple entries from different surveys/authors exist for the same star, we keep the one with the highest number of measured abundances, as in Table~\ref{table:Fiducial}.
We find that the SAGA MDF becomes steeper and its peak shifts from $\rm[Fe/H]=-2.7$ to $\rm[Fe/H]=-2.5$. The number of EMP stars ($\rm[Fe/H]\leq-3$) decreases by $31\%$, i.e. from 526 in LTE to 363 in NLTE; we note that here only stars with \ion{Fe}{I} and/or \ion{Fe}{II} measurements are included. We therefore recommend that careful studies of metal-poor MDFs take NLTE effects into account, which can be done easily with \texttt{NLiTE} for large databases.

\begin{figure*}
\begin{center}
\includegraphics[width=0.95\hsize]{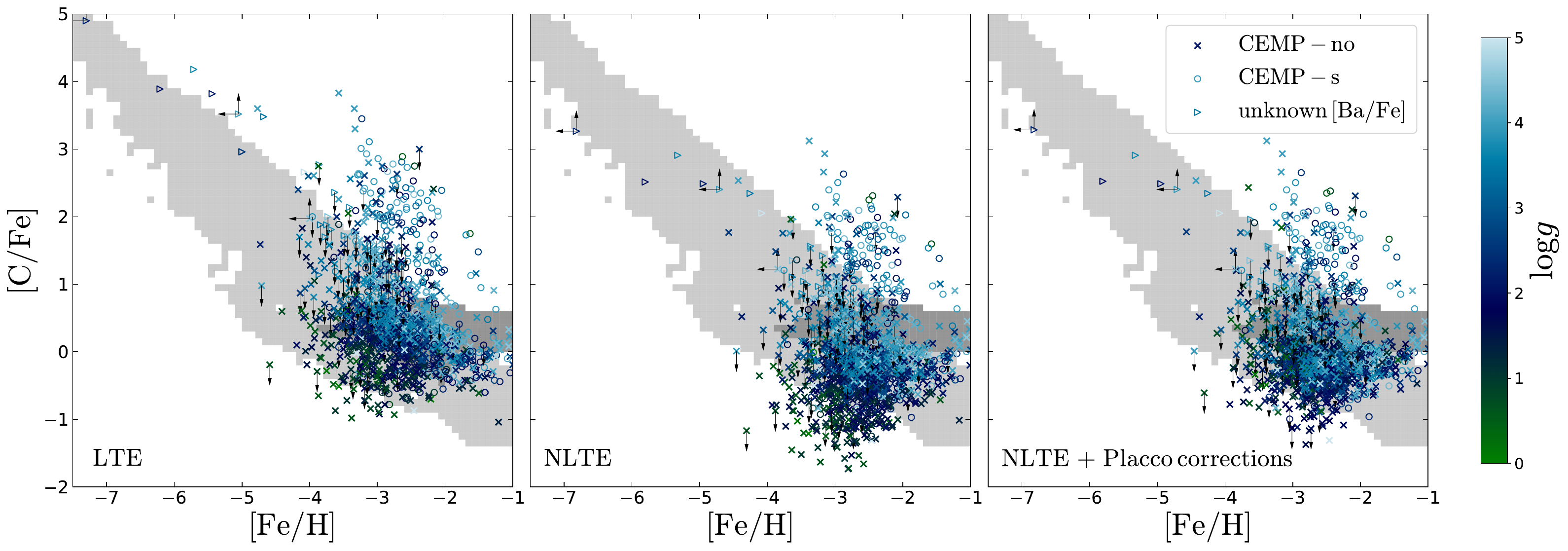} 
\caption{Metal-poor Galactic halo stars in the [C/Fe]--[Fe/H] diagram as predicted by our {\sc NEFERTITI} model: Pop~III descendants (in light grey) and Pop~I/II descendants (in dark grey). Data points show all SAGA MP stars colour-coded by \logg: in LTE (left); in NLTE (middle); and after applying both the NLTE and the C evolutionary corrections (right). Crosses (X) show CEMP-no stars ($\rm [Ba/Fe]<0$), circles show CEMP-s stars ($\rm[Ba/Fe]>0$) and triangles stars with unknown Ba. }
\label{fig: nlte_C}
\end{center}
\end{figure*} 

\begin{figure*}
\begin{center}
\includegraphics[width=0.95\hsize]{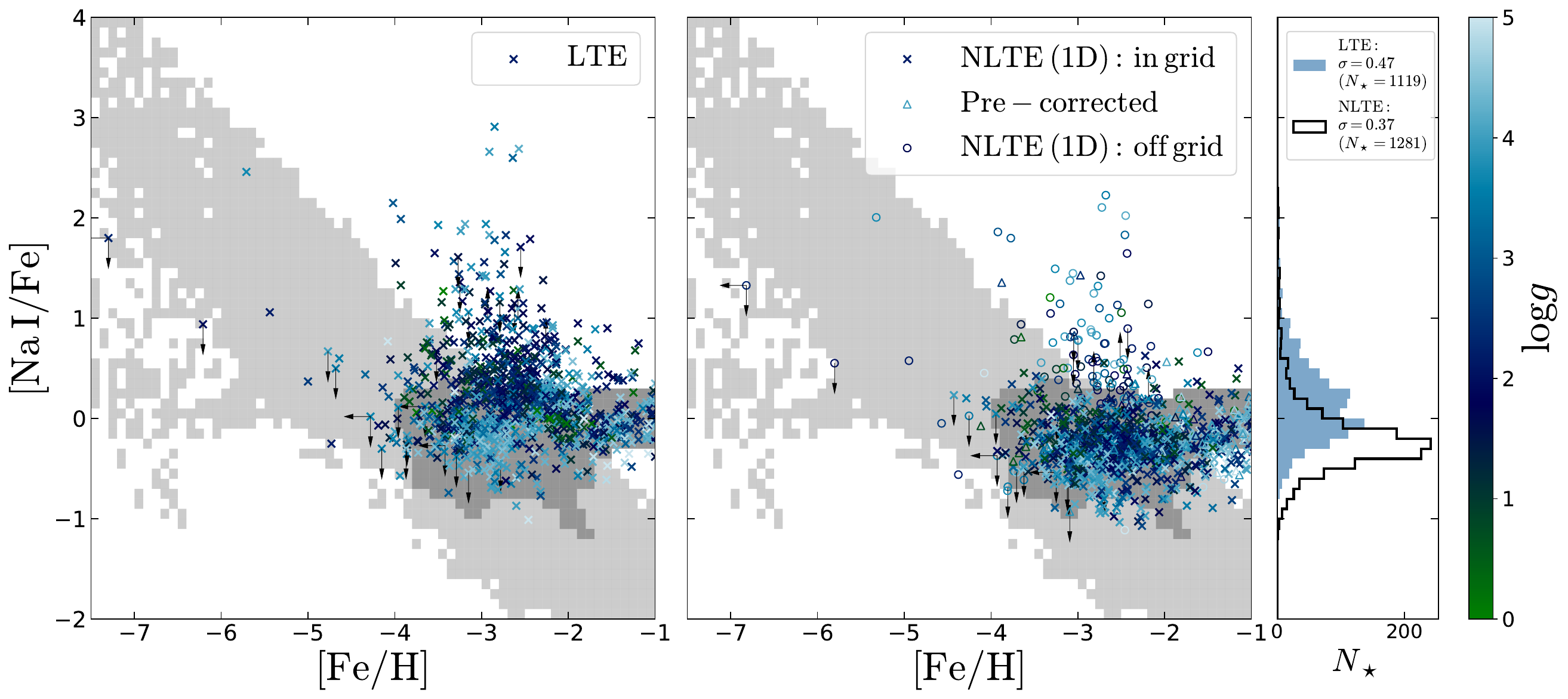}\\
\includegraphics[width=0.95\hsize]{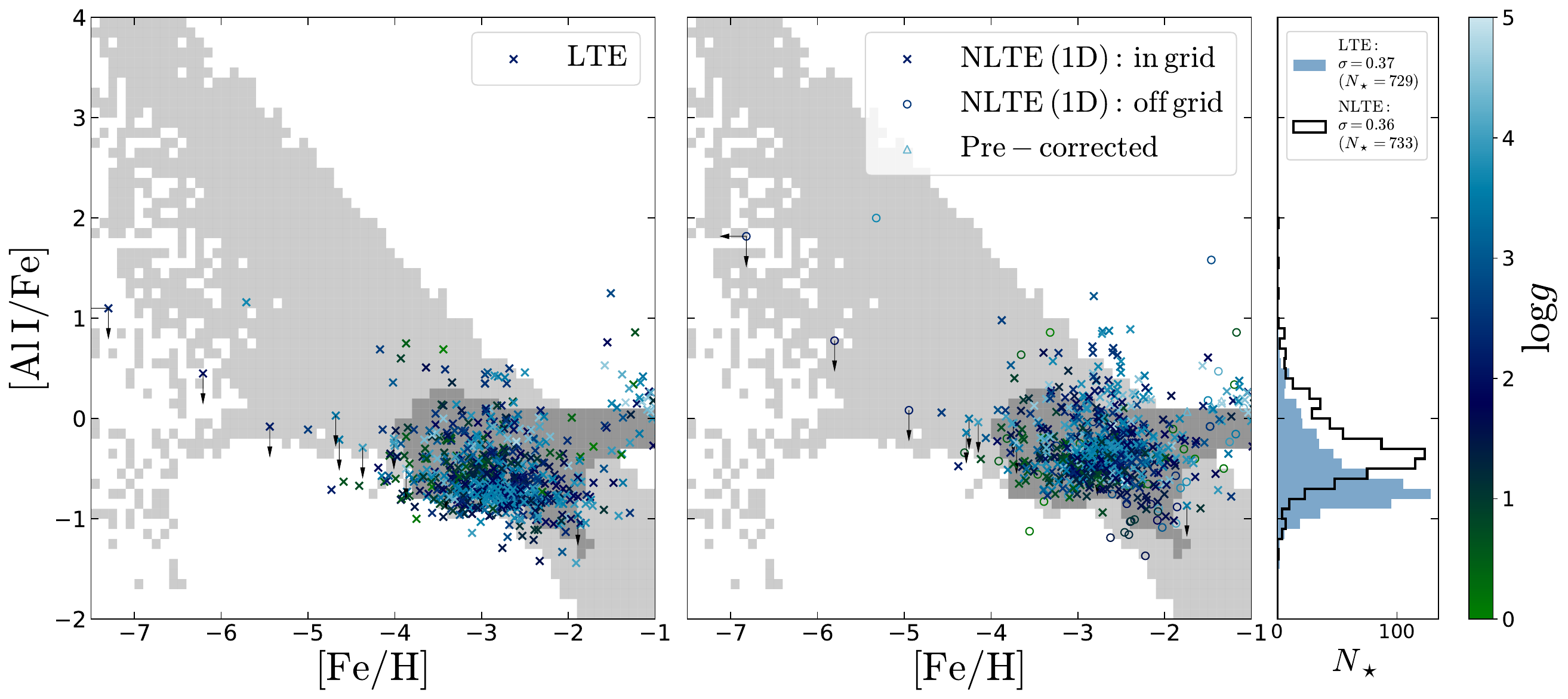}
\caption{Distribution of MP Galactic halo stars in the [Na/Fe]--[Fe/H] (top) and [Al/Fe]--[Fe/H] (bottom) diagrams, as predicted by the {\sc NEFERTITI} model: Pop~III descendants (in light grey), and Pop~I/II descendants (in dark grey). Data points represent all MP SAGA stars, colour-coded with $\logg$, in LTE (left) and NLTE (right). Stars with parameters within the grids are marked with crosses (X), while circles indicate stars outside these limits.
Triangles represent stars whose published abundances were already corrected for NLTE effects. The right marginal plots compare the SAGA [Na/Fe] and [Al/Fe] distributions in LTE (blue) and NLTE (black), excluding stars with upper or lower limits. The standard deviations of the distributions are noted at the top.}
\label{fig: nlte_NaAl}
\end{center}
\end{figure*}

\begin{figure*}
\begin{center}
\includegraphics[width=0.95\hsize]{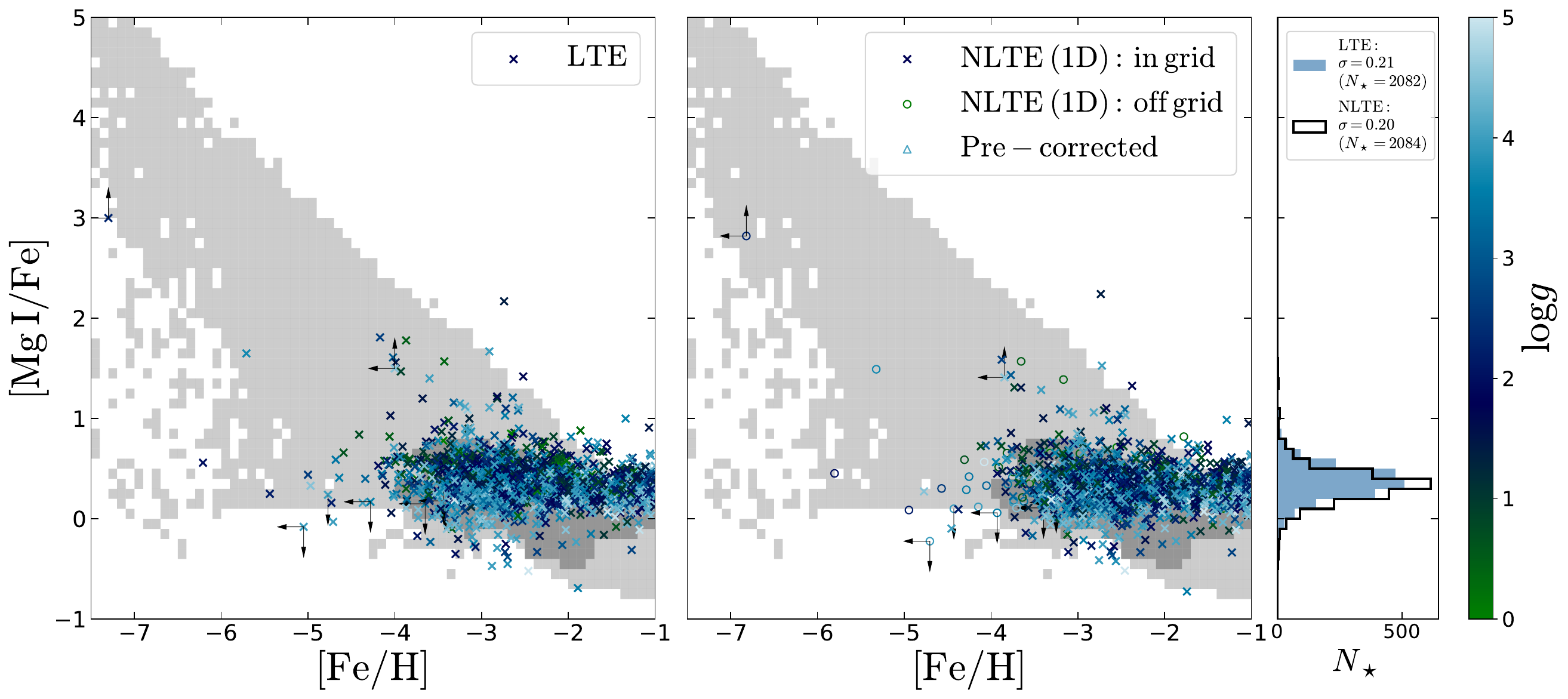}\\
\includegraphics[width=0.95\hsize]{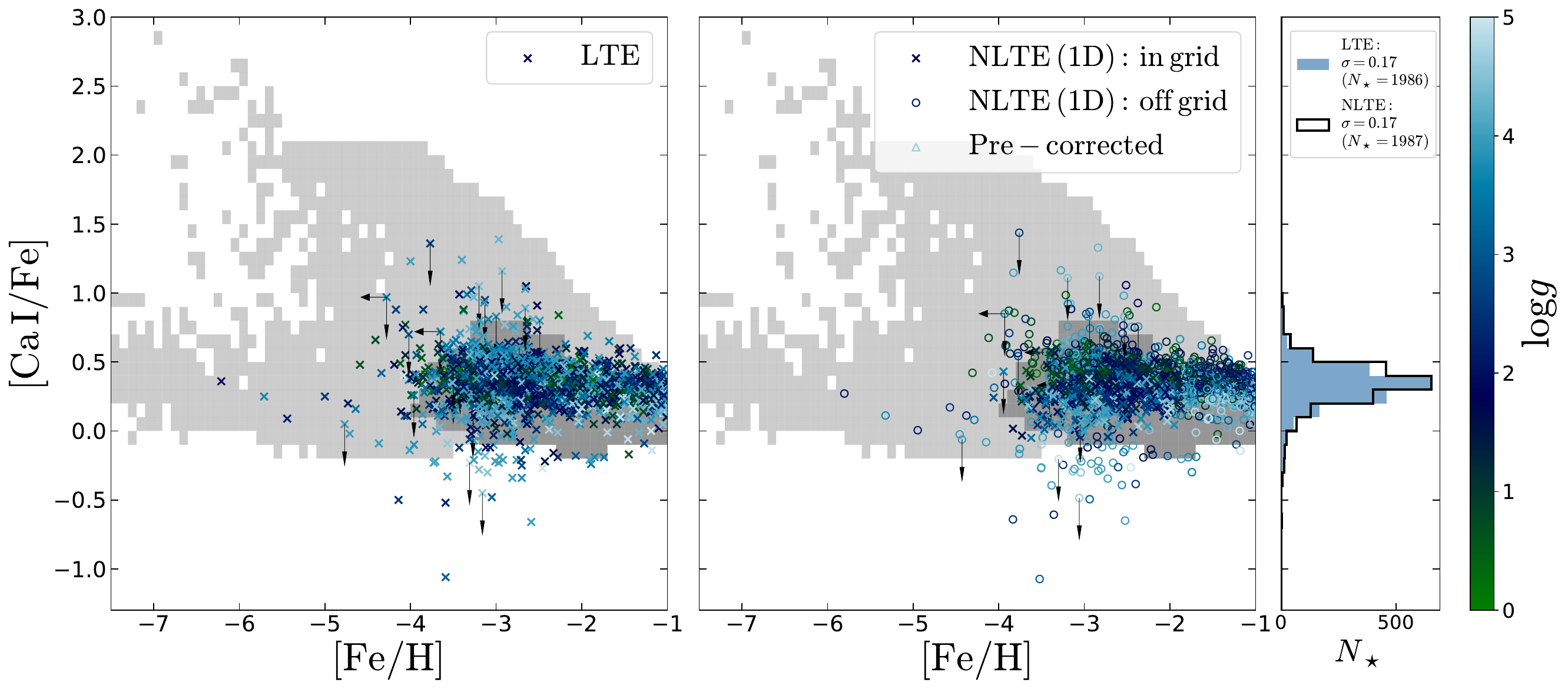}
\caption{Same as Fig.~\ref{fig: nlte_NaAl}, but for [Mg/Fe] (top) and [Ca/Fe] (bottom); observed abundances have been corrected using the grids of \citet{Osorio2016_Mg1} for Mg, \citet{Mashonkina2017_Ca1} for \ion{Ca}{I} and the mean corrections of \citet{Mashonkina2011_Fe1} and \citet{Bergemann2012_Fe1_Fe2} for \ion{Fe}{I}.}
\label{fig: nlte_MgCa}
\end{center}
\end{figure*}

\section{Comparison with the NEFERTITI model}
\label{sec:nefertiti}

The main goal of this work is to provide a large fully NLTE-corrected database of MP stars that can be contrasted against chemical evolution models in a meaningful way. In an upcoming publication (Koutsouridou et al. in prep.) we will go into detail on how combining these data with our models can help constrain the properties of the first stars and early chemical enrichment. Here instead, we focus more on the technical aspect of how adding the NLTE corrections affects the comparison with the predictions of our cosmological galaxy formation and chemical enrichment model {\sc NEFERTITI} (NEar FiEld cosmology: Re-Tracing Invisible TImes), described in detail in \citet{Koutsouridou2023}.

\subsection{Description of the NEFERTITI model}

{\sc NEFERTITI} is a state-of-the-art semi-analytical model that runs on halo merger trees derived from N-body cosmological simulations or Monte Carlo techniques. It builds upon previous
semi-analytical models for the Local Group formation \citep{Salvadori2007,Salvadori2010,Salvadori2015,Pagnini2023}.

The model follows the flow of baryons from the intergalactic medium (IGM) into the dark matter (DM) halos, the formation of stars and stellar evolution within each galaxy, and the return of mass and metals into the interstellar medium (ISM) and the IGM through stellar feedback. In addition, it traces the position and dynamics of baryons (gas and stars) through their associated DM particles and is therefore able to probe the spatial distribution of all stellar populations down to redshift $z=0$. Stars form in DM halos that surpass a minimum mass, which evolves through cosmic times to account for the effects of photodissociating and ionizing radiation \citep{Salvadori2009}. The IGM initially has a primordial composition enabling the formation of Pop~III stars in the first star-forming halos. Once the metallicity within a halo exceeds the critical value $Z_{\rm crit}=5.15 \times 10^{-5} \: {\rm Z_\odot}$ \citep{Caffau2011, deBen2017}, assuming $Z_\odot = 0.0134$ \citep{Asplund2009}, normal (Pop~II/I) stars form according to a Larson Initial Mass Function (IMF) with $m_\star=[10-100]\: M_\odot$ and a peak at $m_{\rm ch}=0.35 \: M_\odot$. The IMF of both Pop~III and Pop~II/I stars is stochastically sampled (as in \citealt{Rossi2021}) and the evolution of individual stars is followed in their proper timescales.

In this study, we employ NEFERTITI coupled with a cold dark matter (DM) N-body simulation of a Milky Way analogue (fully described in \citealt{Koutsouridou2023}), which successfully reproduces the present-day global properties of the Milky Way (metallicity and mass of both stars and gas) and the metallicity distribution function of the Galactic halo \citep{Bonifacio2021}. We adopt the yields of \citet{Heger2002, Heger2010} for Pop~III stars and those of \citet{Karakas2010} and \citet{Limongi2018} for Pop~II/I stars. Here, we assume a Larson IMF for Pop~III stars, with $m_\star=[0.8,1000]\: M_\odot$ and $m_{\rm ch}=10 \: M_\odot$, consistent with stellar archaeology observations \citep{Rossi2021, Pagnini2023, Koutsouridou2024}. We note that varying the Pop~III IMF or the energy distribution of Pop~III supernovae (SNe) can impact the overall extent of the predicted [X/Fe]–[Fe/H] distributions and influence the number of surviving stars in each bin. However, these variations have a minimal effect on the bulk of present-day metal-poor (MP) stars, which are dominated by normal Pop~II enrichment \citep{Koutsouridou2023}, represented by the dark grey regions in Figs.~\ref{fig: nlte_C}-\ref{fig: nlte_CrMn}.

\subsection{Comparison to the NLTE-SAGA database}

\begin{figure*}
\begin{center}
\includegraphics[width=0.95\hsize]{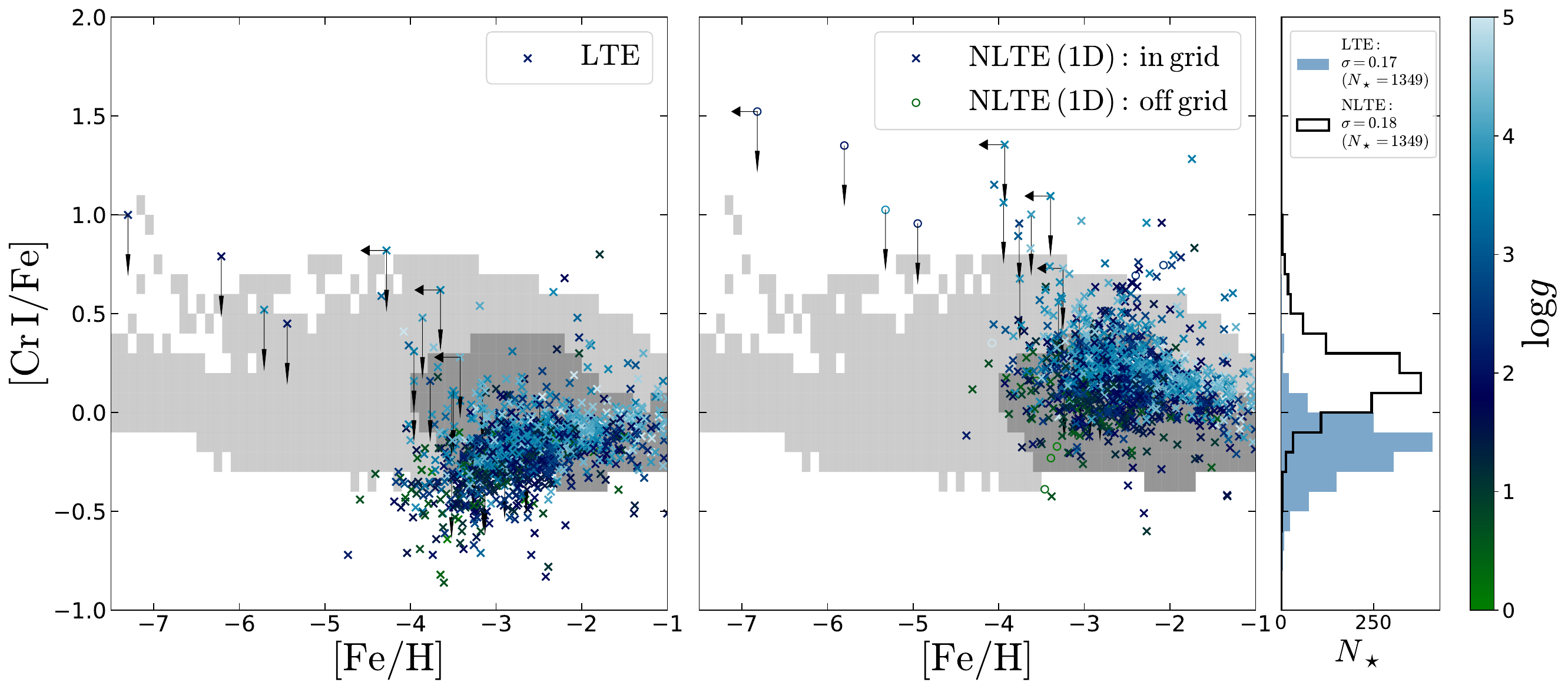}\\
\includegraphics[width=0.95\hsize]{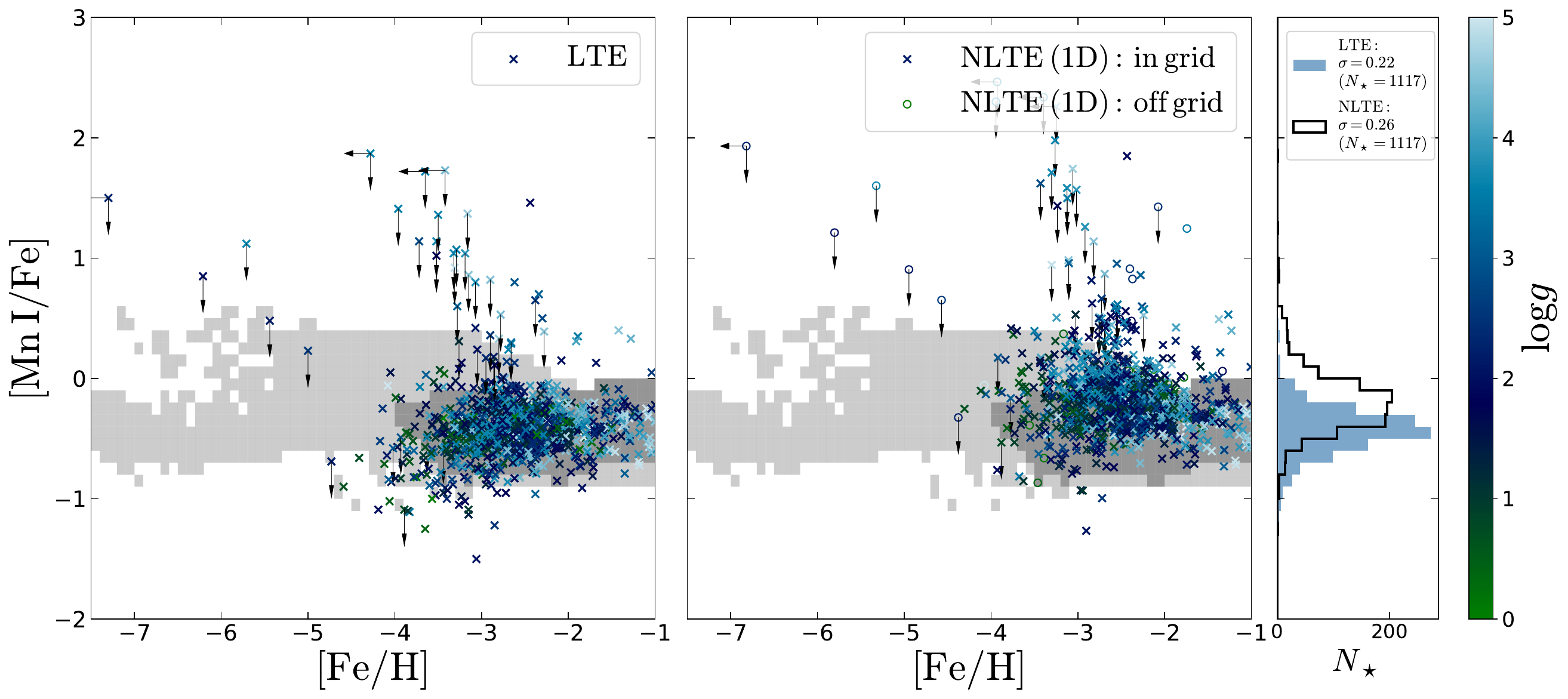}
\caption{Same as Fig.~\ref{fig: nlte_NaAl}, but for [Cr/Fe] (top) and [Mn/Fe] (bottom).}
\label{fig: nlte_CrMn}
\end{center}
\end{figure*}

In the following subsections, we compare the NLTE-SAGA database with our NEFERTITI model. Figures~\ref{fig: nlte_C}-\ref{fig: nlte_CrMn} display the comparison for selected elements, both in LTE and NLTE. We include predictions for the chemical abundances of both the descendants of Pop~III stars (light grey, low-density regions) and those of Pop~I and II stars (dark grey, high-density regions), which represent the bulk of the stellar population \citep{Koutsouridou2023}. In all of the following discussion our fiducial \ion{Fe}{I} abundance corrections are adopted (see~Fig.~\ref{fig: Fe1_logg1}). Discussion on additional elements can be found in Appendix~\ref{app: nefertiti}.

\subsubsection{Carbon predictions}
The elemental abundance that is perhaps the most indicative of pollution from Pop~III stars is carbon. Figure~\ref{fig: nlte_C} displays the [C/Fe]-[Fe/H] distribution of all individual stars with CH measurements (left) and after applying the CH 3D\,NLTE corrections of \citet{Norris19} and the average \ion{Fe}{I} NLTE corrections from \citet{Mashonkina2011_Fe1} and \citet{Bergemann2012_Fe1_Fe2} (middle panel). In the right panel we have additionally applied the evolutionary corrections for carbon from \citet{Placco2014}.\footnote{\url{https://vplacco.pythonanywhere.com/}} Grey shaded areas, identical across all panels, depict the distribution of all stars surviving in the Galactic halo at $z=0$ as predicted by our model. The stars residing in the light grey area have received $\geq50\%$ of their metals from Pop~III stars, while the dark grey regions correspond to stars whose metals come mainly  ($>50\%$) from Pop~I/II stars.

In \citet{Koutsouridou2023}, we showed that our model failed to reproduce the UMP (${\rm [Fe/H]}\leq-4$) stars with the highest ${\rm A(C) \gtrsim 6-6.5}$, a problem faced by several other studies (e.g. \citealp{Cooke2014, Magg2020, Komiya2020, Chiaki2020, Jeon2021, Rizzuti2025}). In this region, the enrichment of CEMP stars is predicted to be dominated by Pop~III SNe \citep[e.g.][]{Koutsouridou2023}. \citet{Komiya2020} showed that even assuming an extremely inefficient mixing of SN yields does only partially reproduce these stars while introducing inconsistencies in the metallicity distribution function.
\citet{Rossi2023}, using a model calibrated on the
UFD Böotes I, proposed that these stars might form through enrichment by Pop~II asymptotic giant branch (AGB) stellar winds after a powerful SN expels all gas from a halo, along with the iron-rich signature of Pop~II SNe. Here, we find that accounting for NLTE effects resolves this discrepancy: all UMP stars now align with the predictions of our model, even after correcting for evolutionary effects (middle and right panels of Fig.~\ref{fig: nlte_C}). 

At $\rm [Fe/H]>-4$, there are still several stars that lie above our predicted distribution. Most of them are CEMP-$s$ stars, i.e.~stars enriched in $s$-process elements (${\rm [Ba/Fe]}\geq 0$, marked with circles in Fig.~\ref{fig: nlte_C}) that are believed to have acquired their carbon by mass transfer from an AGB companion (e.g.  \citealt{Aoki2007,Starkenburg_2014, Hansen2016b}). Since our model does not currently account for binary systems, it is expected that CEMP-s stars are not reproduced.

However, a small number of CEMP-no stars (X symbols), whose abundances are thought to reflect their birth environment, also fall above our predictions at ${\rm [Fe/H]} > -4$. Additionally, NLTE corrections shift some stars to lower carbon abundances, causing them to drop below our distribution. However, the bulk of the [C/Fe] measurements do fall within our predictions, giving a generally good agreement (right panel).

As seen in the left panel of Fig.~\ref{fig: nlte_C}, the SAGA stars exhibit an increasing [C/Fe] with $\logg$ at fixed metallicity, a pattern that persists in NLTE (middle panel), since the empirical corrections of \citet{Norris19} are independent of $\logg$. Although this trend weakens after applying the Placco corrections, it does not fully disappear. Therefore, it is possible that once more physically motivated corrections for CH become available, the scatter at fixed [Fe/H] could further decrease, improving even more the alignment with our theoretical predictions.

\subsubsection{Light elements Na and Al}
The [Na/Fe]--[Fe/H] distribution of SAGA stars in LTE shows significant scatter, particularly in the range $-4\leq {\rm [Fe/H]}\leq -2$, where many stars exceed our model predictions by up to $\sim3 \,$dex (Fig.~\ref{fig: nlte_NaAl}; top). Such high [Na/Fe] ratios are produced only by Pop~III SNe; however, their iron yields are too small by two-three orders of magnitude to account for these observations. Correcting for NLTE effects \citep{Lind2022} improves significantly the agreement with our predictions. The peak of the [Na/Fe] distribution shifts from $\rm[Na/Fe] = -0.05$ to $\rm[Na/Fe] = -0.25$, and the scatter is notably reduced\footnote{The number of MP SAGA stars with LTE [Na/Fe] measurements plotted on the left of Fig.~\ref{fig: nlte_NaAl} is smaller than the number on the right, because 162 stars have [Na/Fe] abundances that have already been corrected for NLTE effects in the original studies.} from $\sigma=0.47$ to $\sigma=0.37$. As a result, the number of stars lying above our predicted distribution decreases by $75\%$. In addition, most of these stars are outside the limits of our adopted NLTE grid (marked as circles), $\rm[Na/Fe]_{\rm LTE}>+0.8$. Therefore, it is possible that even stronger negative corrections may apply to these stars, which could bring them into closer agreement with our model.

The NLTE corrections for [Al/Fe] are also strong, drastically shifting the peak of the observed distribution from $\rm[Al/Fe]=-0.75$ to $\rm[Al/Fe]=-0.35$ as shown in the rightmost bottom panel of Fig.~\ref{fig: nlte_NaAl}. Although the scatter decreases only slightly, the predominantly positive corrections improve the agreement with the model predictions. Fewer stars now fall below the model distribution, though some move above it under NLTE. We note that when adopting the 1D\,NLTE corrections from \citet{Nordlander2017}, instead of the \citet{Lind2022} ones, the scatter increases significantly, from $\sigma=0.37$ to  $\sigma=0.51$ with several stars in the range $-4\leq {\rm [Fe/H]}\leq-2$ and $\logg<1.5$, acquiring strong negative corrections, and falling below our predicted distribution, to values as low as  ${\rm [Al/Fe]} \simeq -2$ (see Sec.~\ref{Differences between grids}). In contrast, when using the $\rm \langle 3D \rangle$ corrections from \citet{Nordlander2017} the [Al/Fe] corrections are always positive, thus, in this case no stars occupy the region at $\rm[Al/Fe]<-1$, and the scatter reduces to $\sigma=0.35$.

\subsubsection{Alpha-elements Mg and Ca}

Fig.~\ref{fig: nlte_MgCa} displays the SAGA [Mg/Fe]--[Fe/H] (top) and [Ca/Fe]--[Fe/H] (bottom) distributions in comparison to our model predictions. NLTE corrections have a minimal impact on the observed distributions. For [Mg/Fe], they slightly reduce the scatter from 0.21 to 0.20, while for [Ca/Fe], the distribution remains virtually unchanged.
In both cases, the observed distributions align well with our theoretical ones, with a few outliers.

\subsubsection{Iron-peak elements Cr and Mn}
\label{Iron-peak elements Cr and Mn}

Fig.~\ref{fig: nlte_CrMn}, displays the distribution of SAGA stars in the [Cr/Fe]--[Fe/H] (top) and [Mn/Fe]--[Fe/H] (bottom) diagrams, using the 1D\,NLTE grids of \citet{Bergemann2010_Cr} for Cr and \citet{Bergemann19} for Mn.

In LTE, both of these distributions show a downward trend with decreasing [Fe/H] which disappears after applying the NLTE corrections. The NLTE [Mn/Fe]--[Fe/H] mean relation becomes flat while the [Cr/Fe]--[Fe/H] shows a slightly upward trend, with [Cr/Fe] increasing with decreasing [Fe/H]. The peaks of both distributions significantly shift upwards: from $-0.15$ to $+0.15$ for [Cr/Fe] and from $-0.45$ to $-0.15$ for [Mn/Fe]. NLTE corrections do not reduce the scatter in these cases.

Applying NLTE corrections reduces the number of stars with $\rm [Cr/Fe] \leq -0.4$, which are not reproduced by our model. However, the corrected [Cr/Fe] distribution now tends to lie on the high side of model predictions, similar to other Fe-peak elements such as Co and Zn  (see Fig.~\ref{fig: nlteCuZn} in Appendix~\ref{app: nefertiti}). Noticeably, the discrepancy pertains the bulk of the stellar population, and thus it is likely linked to the production mechanism of Fe-peak elements from Pop~II/I stars, or possibly to inaccuracies in the NLTE corrections or lack of consideration for 3D\,effects. Since the corrections for these elements can be large, up to $\rm|\Delta NLTE|\approx1$ (see Figs.~\ref{fig: Cr1_common_lines} and \ref{fig: Mn1_common_lines}), some uncertainty is expected. Although beyond the scope of this paper, it is interesting to note that a contribution of Pop~II hypernovae with $\rm [Fe/H]\approx-1$ might partially alleviate the problem, producing super-solar values of [Cr,Zn/Fe] (and [Al/Fe]) while leaving [Mn/Fe] unaltered \citep[e.g. see Fig.~9 from the review of][]{Nomoto2013}. Both in the case of Cr and Mn, the abundance trend is significantly altered when applying NLTE corrections. We therefore emphasize that it is misleading to discuss chemical evolution and nucleosynthesis of these elements without fully taking NLTE effects into account.

\section{Summary and conclusions}
\label{Discussion}

Metal-poor ($\rm[Fe/H]\leq-1$) stars provide a unique window into the early Universe, but accurately interpreting their chemical abundances requires accounting for non-LTE (NLTE) effects, which are often neglected in large stellar samples. For the first time, we have provided a fully NLTE-corrected catalogue of all MP stars in the SAGA database, reaching down to the lowest metallicities, $\rm[Fe/H]<-4$. In addition, we have developed \texttt{NLiTE},\footnote{\url{https://nlite.pythonanywhere.com/}} an online tool designed for the community to efficiently apply NLTE corrections to large stellar samples of MP stars. To complete this task we have undergone a detailed review of the currently available state-of-the-art 1D and 3D\,NLTE grids in the literature (Sec.~\ref{sec:grids}). Together with the \texttt{NLiTE} tool, this allowed us to do an extensive comparison of the available grids in order to understand the uncertainties of the NLTE calculations for individual elements.

The \texttt{NLiTE} tool (Sec.~\ref{Nlite}) provides corrections for 24 chemical species from Li to Eu, by interpolating within precomputed NLTE grids, which depend on the stellar atmospheric parameters: \teff, \logg, [Fe/H], and, when relevant, A(X) or [X/Fe]. These grids are based on publicly available NLTE datasets (Table~\ref{table:Grids}) that are originally linked to specific spectral lines. However, in the SAGA database the original selection of lines is not always available. The novelty of \texttt{NLiTE} is that it is averaged over a predetermined line list (Table~\ref{table: linelist}). The selected lines are those most frequently used to measure chemical abundances in MP stars, using optical spectra ($3\,500\,\AA\lesssim \lambda\lesssim10\,000\,\AA$). Our selection takes into account line detectability for each set of stellar parameters (see Sec.~\ref{Linelists}). We find that in most cases the scatter between the $\Delta$NLTE corrections of different line sets is small $\sigma_{\textsl{lines}}<0.1$\,dex, and/or small in comparison to the mean correction (Sec.~\ref{Scatter}).

A full NLTE-SAGA catalogue is provided for MP stars in the Milky Way, obtained using \texttt{NLiTE} (Tables~\ref{table:Fiducial} and \ref{table: Complete}). We find that the magnitudes and signs of NLTE corrections vary significantly depending on the chemical species and do not always show a straightforward dependence on stellar parameters (Sec.~\ref{sec:appcorr}).
The strongest [Element/Fe] NLTE corrections reaching $>0.5$\,dex are found for Na, Al, Si, K, Cr, Mn, and Co. Significant corrections are found for \ion{Fe}{I}, spanning from 0 to $+0.4$\,dex, that shift the SAGA metallicity distribution to higher [Fe/H] (Sec.~\ref{sec:sagamdf}), decreasing the number of ${\rm [Fe/H]\leq-3}$ stars by 31$\%$. Therefore we strongly encourage including NLTE effects when studying the metallicity distribution of the Galactic halo.

A detailed comparison between different 1D\,NLTE grids (Sec.~\ref{sec:grids}) reveals a generally good agreement for Li, Fe, Mg, K, Ca, \ion{Ti}{II}, Sr and Ba with corrections for over 90$\%$ of stars differing by less than 0.1\,dex and maximum discrepancies of $<0.2$\,dex (Sec.~\ref{sec:appcorr}). 
Larger discrepancies are found for Na and \ion{O}{I} (up to 0.30\,dex), Al (up to 0.55\,dex), and \ion{Si}{I} (up to 0.87\,dex). 
In general, quite significant discrepancies can be found when comparing 1D and 3D\,NLTE corrections, e.g. in the case of O, Mg, Al, Fe and Ba (Sec.~\ref{sec:grids}). This highlights that full 3D\,NLTE corrections are needed for highly accurate abundances. Unfortunately, there is still a lack of 3D\,NLTE grids that cover the full range of typical stellar parameters.

We compared the NLTE-SAGA catalogue to predictions from {\sc NEFERTITI}, our cosmological model for galaxy formation and evolution \citep{Koutsouridou2023}. Incorporating 3D\,NLTE effects resolves a key discrepancy for stars with ${\rm [Fe/H]}<-3.5$, which previously exhibited higher [C/Fe] than predicted by models (e.g. \citealp{deBen2017,  Komiya2020, Rossi2023, Koutsouridou2023}). After applying NLTE corrections, the observed carbon abundances align well with our theoretical predictions (Sec.~\ref{sec:nefertiti}). NLTE corrections also significantly improve the agreement between observed and theoretical Na abundances, reducing the number of stars with higher-than-expected [Na/Fe] by $75 \%$. Similarly, positive Al corrections bring the observed distribution closer to theoretical predictions.

The levels of Mg and Ca relative to Fe remain largely consistent with model predictions and exhibit minimal sensitivity to NLTE effects. For Si, Cr, and Mn, previously observed metallicity trends diminish, and their distributions are broadly in line with theoretical expectations. In contrast, K, Ti, Co, and Zn remain underproduced in the model compared to observations, a common issue in galactic chemical evolution studies (e.g.  \citealp{Romano2010, Zhao2016, Kobayashi2020, Rossi2024}). NLTE corrections mitigate but do not fully resolve this discrepancy for K, while they exacerbate it for \ion{Ti}{I}, Co, and Zn. Meanwhile, Cu remains within the predicted distribution both before and after applying NLTE corrections.

This study demonstrates the crucial role of NLTE corrections in refining stellar abundance measurements, improving consistency with chemical evolution models, and offering new insights into the nucleosynthetic history of the Milky Way.
By providing more accurate abundance measurements, \texttt{NLiTE} ensures that theoretical predictions can be tested against reliable data, preventing erroneous conclusions and enhancing our understanding of the first stars and early chemical evolution.

\section{Data availability}

Tables~\ref{table: linelist}, \ref{table:Fiducial}, \ref{table: Complete} and \ref{table: Parameters} are available in electronic form at the CDS via anonymous ftp to \url{https://cdsarc.cds.unistra.fr/} (130.79.128.5) or via \url{http://cdsweb.u-strasbg.fr/cgi-bin/qcat?J/A+A/}. The tables are also provided in .tsv format at \url{https://nlite.pythonanywhere.com/}.\\

\begin{acknowledgements}
This project has received funding from the European Research Council Executive Agency (ERCEA) under the European Union's Horizon Europe research and innovation program (acronym TREASURES, grant agreement No 101117455”). I.K. and S.S. acknowledge ERC support (grant agreement No. 804240). We thank the following experts for insightful discussion and suggestions that helped improve the paper: A.M.~Amarsi, M.~Bergemann, K.~Lind, L.~Mashonkina, and T.~Nordlander.
\end{acknowledgements}

\bibliographystyle{aa_url.bst}
\bibliography{nlte}

\appendix

\section{Adopted NLTE grids}
\label{Adopted NLTE grids}

Table~\ref{table:Grids} lists the NLTE grids that we have adopted from the literature for each chemical species.

\renewcommand{\arraystretch}{1.5}
\begin{table*}[t]
\centering
\footnotesize
\caption{The 1D and 3D\,NLTE grids included in NLiTE.}
\begin{tabular}{c c c c c c c c}
\hline
\hline
& & & \multicolumn{5}{c}{Grid range} \\\cline{4-8}
Species & Authors & Type & $T_{\rm eff}$ & ${\rm log}g$ & ${\rm [Fe/H]}$ & ${\rm [X/Fe]}$ & A(X) \\
\hline
\ion{Li}{I} & \citet{Lind2009_Li1} & 1D NLTE & [4000,8000] & [1,5]& $[-3,0]$ & - & $[-0.3,4.2]$\\
\ion{Li}{I} & \citet{Sbordone2010} & 1D/3D NLTE & [5500,6500]&[3.5,4.5] & $[-3,-2]$& -&$[1.2,3.3]$ \\
\ion{Li}{I} & \citet{Wang2021} & 3D NLTE$^{\star}$ & [4000,7000]&[1.5,5] & $[-4,0.5]$& -&$[-0.5,4]$ \\
\hline
CH & \citet{Norris19} & 3D NLTE$^{\star}$ & - & - & $[-6,-1]$ & - & -\\
\hline
\ion{C}{I} & \citet{Amarsi19} & \begin{tabular}{@{}c@{}} 1D NLTE$^{\star}$ \\ 3D NLTE \end{tabular} & \begin{tabular}{@{}c@{}} $[4000,7500]$ \\ $[5000,6500]$ \end{tabular}    & \begin{tabular}{@{}c@{}} $ [0,5]$ \\ $[3,5]$ \end{tabular}  & \begin{tabular}{@{}c@{}} $[-5,0]$ \\ $[-3,0]$ \end{tabular}  & $[-0.4,1.2]$ & -\\
\hline
\ion{O}{I} & \citet{Amarsi19} & \begin{tabular}{@{}c@{}} 1D NLTE$^{\star}$ \\ 3D NLTE \end{tabular} & \begin{tabular}{@{}c@{}} $[4000,7500]$ \\ $[5000,6500]$ \end{tabular}    & \begin{tabular}{@{}c@{}} $ [0,5]$ \\ $[3,5]$ \end{tabular}  & \begin{tabular}{@{}c@{}} $[-5,0]$ \\ $[-3,0]$ \end{tabular}  & $[-0.4,1.2]$ & -\\
\ion{O}{I} & \citet{Bergemann2021_O1} & 1D NLTE & [4000,7500] & [0.5,5] & $[-2.5,-1]$ & - & -\\
\hline
\ion{Na}{I} & \citet{Lind2011_Na1} & 1D NLTE & [4000,6500] & [1,5] & $[-5,-1]$ & $[-0.6,0.6]$ & -\\
\ion{Na}{I} & \citet{Alexeeva2014_Na1} & 1D NLTE & [4000,6500] & [0.5,5] & $[-5,-2]$ & $[-0.6,0.6]$ & -\\
\ion{Na}{I} & \citet{Lind2022} & 1D NLTE$^{\star}$ & [4000,7500] & [0,5] & $[-5,-0.5]$ & $[-1,0.8]$ & -\\
\hline
\ion{Mg}{I} & \citet{Merle2011} & 1D NLTE & [3500,5250] & [0.5,2] & $[-4,-1]$ & - & -\\
\ion{Mg}{I} & \citet{Mashonkina2013_Mg1} & 1D NLTE & [4000,6500] & [0.5,5] & $[-5,-2]$ & - & -\\
\ion{Mg}{I} & \citet{Osorio2016_Mg1} & 1D NLTE$^{\star}$ & [3500,7500] & [0.5,5] & $[-5,-1]$ & - & [2.6,8.6]\\
\ion{Mg}{I} & \citet{Bergemann2017_Mg1} & 1D NLTE & [3500,7500] & [0.5,5] & $[-5,-1]$ & - & -\\
\ion{Mg}{I} & \citet{Lind2022} & 1D NLTE & [4000,7500] & [0,5] & $[-5,-0.5]$ & $[-1,0.8]$ & -\\
\ion{Mg}{I} & \citet{Matsuno2024} & 1D/3D NLTE & [5000,6500] & [4,5] & - & - & [5,7.6]\\
\hline
\ion{Al}{I} & \citet{Nordlander2017} & \begin{tabular}{@{}c@{}} 1D NLTE \\ $\langle {\rm 3D} \rangle$ NLTE \end{tabular} & 
\begin{tabular}{@{}c@{}} $[4000,7500]$ \\ $[4000,7000]$ \end{tabular} & 
\begin{tabular}{@{}c@{}} $[0,5]$ \\ $[1.5,5]$ \end{tabular}
& $[-4,-0.5]$ & $[-2,1.9]$ & -\\ 
\ion{Al}{I} & \citet{Lind2022} & 1D NLTE$^{\star}$ & [4000,7500] & [0,5] & $[-5,-0.5]$ & $[-1,0.8]$ & -\\ 
\hline
\ion{Si}{I} & \citet{Bergemann2013_Si} & 1D NLTE & [3500,7500] & [0.5,5] & $[-5,-1]$ & - & -\\
\ion{Si}{I} & \citet{Amarsi2017_Si} & 1D NLTE$^{\star}$ & [4000,7500] & [1,5] & $[-5,-1]$ & $[-0.5,1.5]$ & -\\
\hline
\ion{S}{I} & \citet{Takada-Hidai2002} & 1D NLTE & [4500,6500] & [1,5] & $[-2,0]$ & $[-0.2,1.2]$ & -\\
\ion{S}{I} & \citet{Takeda2005} & 1D NLTE & [4500,6500] & [2,4] & $[-3,0]$ & - & -\\
\ion{S}{I} & \citet{Korotin2008} & 1D NLTE & [5000,6500] & [2,4] & $[-3,0]$ & - & -\\
\ion{S}{I} & \citet{Skuladottir2015} & 1D NLTE & [4000,4750] & [0,1.5] & $[-2.5,-1]$ & [0,0.6] & -\\
\hline
\ion{K}{I} & \citet{Takeda2002} & 1D NLTE & [4500,6500] & [1,5] & $[-3,0]$ & - & [2.55,6.54]\\
\ion{K}{I} & \citet{Reggiani2019_K} & 1D NLTE$^{\star}$ & [4000,8000] & [0,5] & $[-5,0.5]$ & - & [0.08,6.83]\\ 
\hline
\ion{Ca}{I} & \citet{Spite2012_Ca} & 1D NLTE & [4750,7000] & [1,4] & $[-3.5,-1]$ & [0,0.3] & -\\ 
\ion{Ca}{I} & \citet{Mashonkina2017_Ca1} & 1D NLTE$^{\star}$ & [4000,6500] & [0.5,5] & $[-5,-2]$ & [0,0.4] & -\\ 
\hline
\ion{Ti}{I} & \citet{Bergemann2011_Ti1_Ti2} & 1D NLTE$^{\star}$ & [3500,7500] & [0.5,5] & $[-5,-1]$ & - & -\\ 
\hline
\ion{Ti}{II} & \citet{Bergemann2011_Ti1_Ti2} & 1D NLTE$^{\star}$ & [3500,7500] & [0.5,5] & $[-5,-1]$ & - & -\\ 
\ion{Ti}{II} & \citet{Sitnova2016_Ti2} & 1D NLTE & [4000,6500] & [0.5,5] & $[-4,0]$ & - & -\\ 
\hline
\ion{Cr}{I} & \citet{Bergemann2010_Cr} & 1D NLTE$^{\star}$ & [3500,7500] & [0.5,5] & $[-5,-1]$ & - & -\\ 
\hline
\ion{Mn}{I} & \citet{Bergemann19} & 1D NLTE$^{\star}$ & [3500,7500] & [0.5,5] & $[-5,-1]$ & - & -\\
\hline
\ion{Fe}{I} & \begin{tabular}{@{}c@{}c@{}} Mean corrections from \\ \citet{Mashonkina2011_Fe1} \\ $\&$ \citet{Bergemann2012_Fe1_Fe2} \end{tabular}
 & 1D NLTE$^{\star}$ & [3500,7500] & [0.5,5] & $[-5,-1]$ & - & -\\ 
\ion{Fe}{I} & \citet{Mashonkina2011_Fe1} & 1D NLTE & [4000,6500] & [0.5,5] & $[-5,-2]$ & - & -\\ 
\ion{Fe}{I} & \citet{Bergemann2012_Fe1_Fe2} & 1D NLTE & [3500,7500] & [0.5,5] & $[-5,-1]$ & - & -\\ 
\ion{Fe}{I} & \citet{Amarsi2022} & 3D NLTE & [5000,6500] & [4,4.5] & $[-3,-1]$ & - & -\\ 
\hline

\end{tabular}
\label{table:Grids}
\end{table*}

 \begin{table*}
 \ContinuedFloat 
\centering
\footnotesize
\caption{(Continued)}
\begin{tabular}{c c c c c c c c}
\hline
\hline
& & & \multicolumn{5}{c}{Grid range} \\\cline{4-8}
Species & Authors & Type & $T_{\rm eff}$ & ${\rm log}g$ & ${\rm [Fe/H]}$ & ${\rm [X/Fe]}$ & A(X) \\
\hline
\ion{Fe}{II} & \citet{Bergemann2012_Fe1_Fe2} & 1D NLTE$^{\star}$ & [3500,7500] & [0.5,5] & $[-5,-1]$ & - & -\\ 
\ion{Fe}{II} & \citet{Amarsi19} & 3D LTE& [4000,6500] & [1.5,5] & $[-4,-1]$ & - & -\\ 
\ion{Fe}{II} & \citet{Amarsi2022} & 3D NLTE & [5000,6500] & [4,4.5] & $[-3,-1]$ & - & -\\ 
\hline
\ion{Co}{I} & \citet{Bergemann2010_Co} & 1D NLTE$^{\star}$ & [3500,7500] & [0.5,5] & $[-5,-1]$ & - & -\\ 
\hline
\ion{Ni}{I} & \citet{Eitner2023} & 1D NLTE$^{\star}$ & [5000,6500] & [3,4.5] & $[-3,0]$ & - & -\\ 
\hline
\ion{Cu}{I} & \begin{tabular}{@{}c@{}c@{}}\citet{Andrievsky2018_Cu} \\ $\&$ \citet{Shi2018_Cu} \\ $\&$ \citet{Xu2022_Cu} \end{tabular}
 & 1D NLTE$^{\star}$ & - & - & $[-4.2,-1]$ & - & -\\ 
\hline
\ion{Zn}{I} & \citet{Sitnova2022_Zn1_Zn2} & 1D NLTE$^{\star}$ & [4000,6500] & [0.5,5] & $[-3.5,0]$ & - & -\\ 
\hline
\ion{Sr}{II} & \citet{Mashonkina2022_Sr} & 1D NLTE$^{\star}$ & [4000,6500] & [0.5,5] & $[-5,-2]$ & $[-1.5,1]$ & -\\ 
\ion{Sr}{II} & \citet{Bergemann2012_Sr} & 1D NLTE & [4500,6400] & [2.2,4] & $[-3.9,-1]$ & $[-0.5,0.5]$ & -\\ 
\hline
\ion{Ba}{II} & \citet{Mashonkina2019_Ba2} & 1D NLTE$^{\star}$ & [4000,6500] & [0.5,5] & $[-5,-2]$ & $[-1.5,0.5]$ & -\\
\ion{Ba}{II} & \citet{Gallagher2020} & 1D/3D NLTE & [4000,6500] & [1.5,4.5] & $[-3,-1]$ & $[-2.5,2]$ & -\\ 
\ion{Ba}{II} & \citet{Korotin2015_Ba2} & 1D NLTE & [4000,6500] & [0.5,5] & $[-2,-1]$ & $[-0.2,0.4]$ & -\\ 
\hline
\ion{Eu}{II} & \citet{Mashonkina2000_Eu2} & 1D NLTE$^{\star}$ & [4000,5750] & [0.5,3] & $[-3,-2]$ & $[-0.5,1.5]$ & -\\ 
\hline
\end{tabular}
\noindent\parbox{0.85\textwidth}{
    \vspace{3mm} 
    \small {\bf Note:} $^{\star}$ denotes our fiducial NLTE grid for each element.}

\vspace{1cm}
\end{table*}

\section{Comparison of NLTE corrections} 
\label{app: NLTEcorr}

\begin{figure*}
\begin{center}
\includegraphics[width=0.9\hsize]{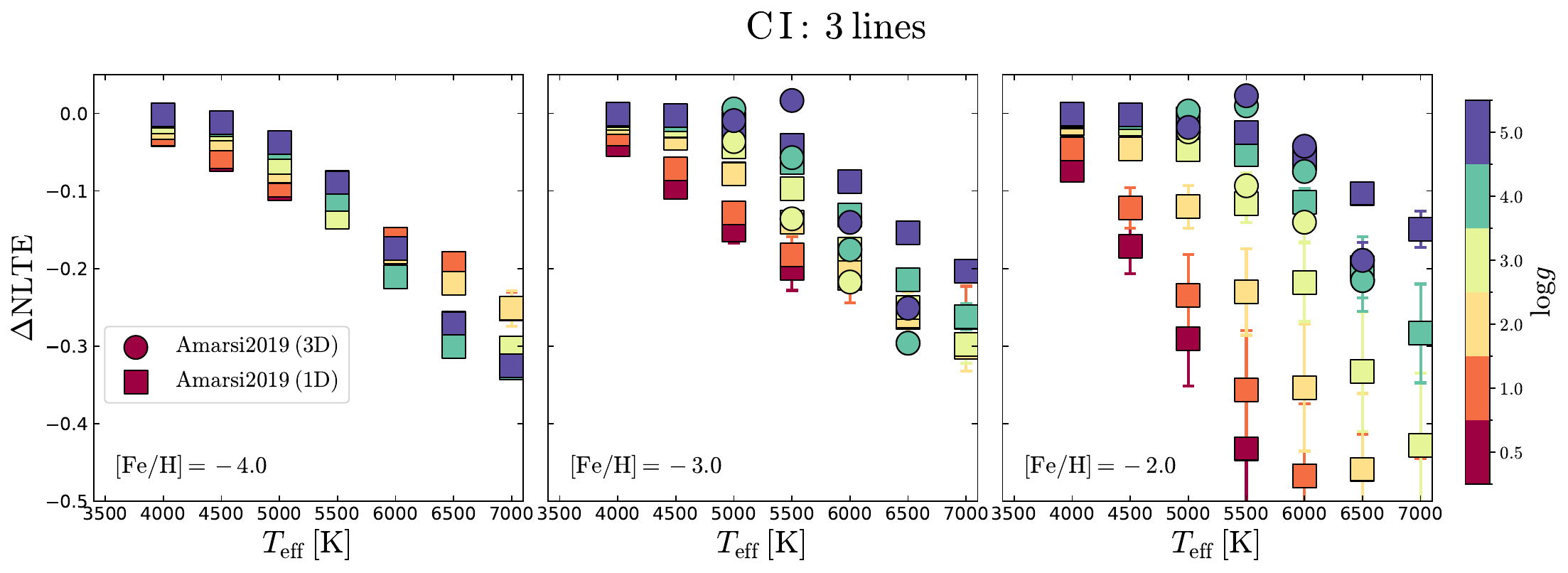} 
\caption{Comparison of 1D\,NLTE (squares) and 3D\,NLTE (circles) corrections for \ion{C}{I} from \citet{Amarsi19}, colour-coded by $\log g$. Three metallicities are shown, assuming $\rm[C/Fe]=+0.5$ in all cases: $\rm [Fe/H]=-4$ (left), $\rm [Fe/H]=-3$ (middle), and $\rm [Fe/H]=-2$ (right). Error bars show the standard deviation between the lines used. }
\label{fig: C1_lines}
\end{center}
\end{figure*}

\begin{figure*}
\begin{center}
\includegraphics[width=0.9\hsize]{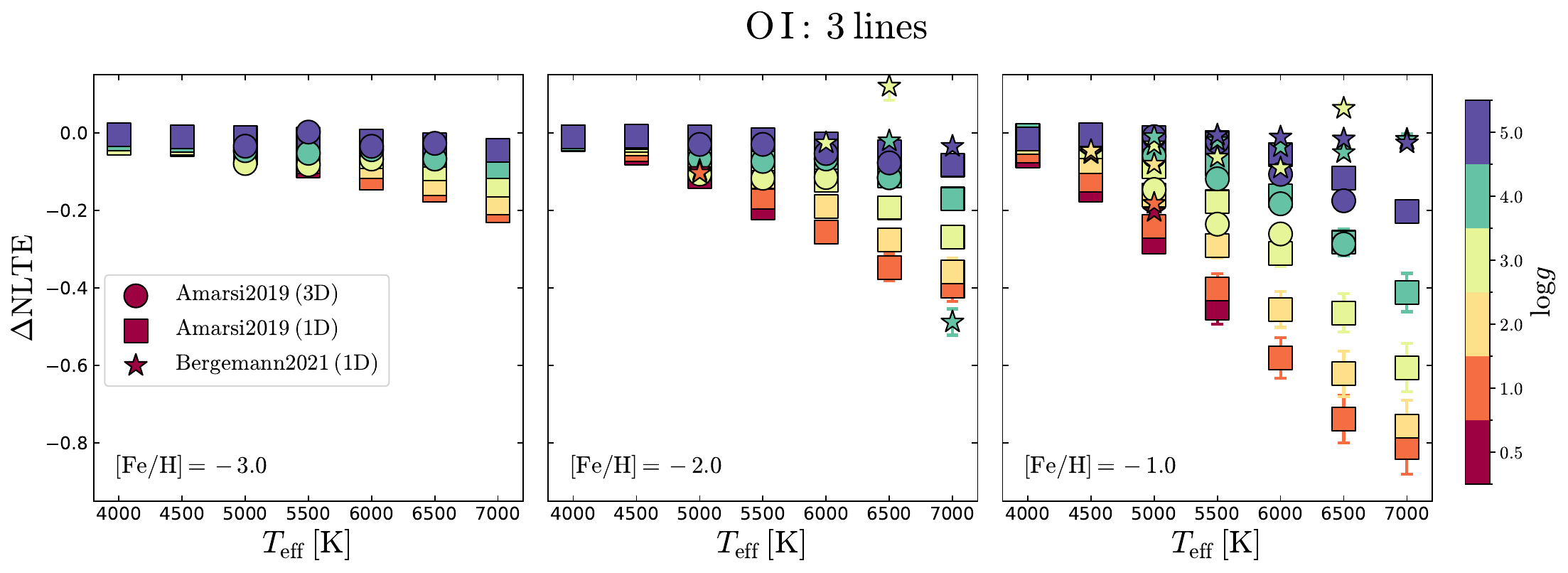} 
\caption{Same as Fig.~\ref{fig: C1_lines}, but for \ion{O}{I}, assuming $\rm[O/Fe]=+0.5$, for $\rm[Fe/H]=-3,-2$, and $-1$, from left to right. Star symbols show 1D\,NLTE corrections by \citet{Bergemann2021_O1}.}
\label{fig: O1_lines}
\end{center}
\end{figure*}

\begin{figure*}
\begin{center}
\includegraphics[width=0.9\hsize]{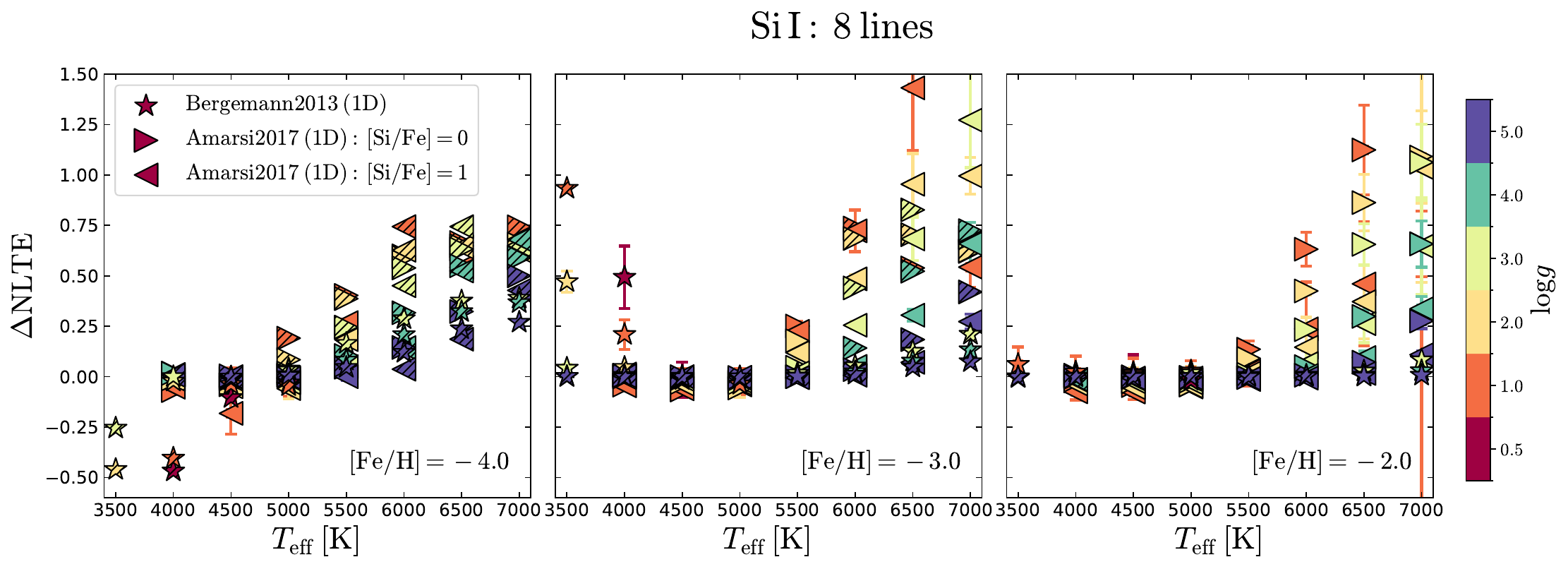} 
\caption{Same as Fig.~\ref{fig: C1_lines}, but for \ion{Si}{I}. Star symbols and triangles show the 1D\,NLTE corrections by \citet{Bergemann2013_Si} and \citet{Amarsi2017_Si}, respectively. Diagonally hatched symbols show cases where only one line is detectable (other lines are too weak, $\rm EW<5$\,m\AA), in this case the resonance \ion{Si}{I} line at 3960\AA.}
\label{fig: Si_common_lines}
\end{center}
\end{figure*}

\begin{figure*}
\begin{center}
\includegraphics[width=0.9\hsize]{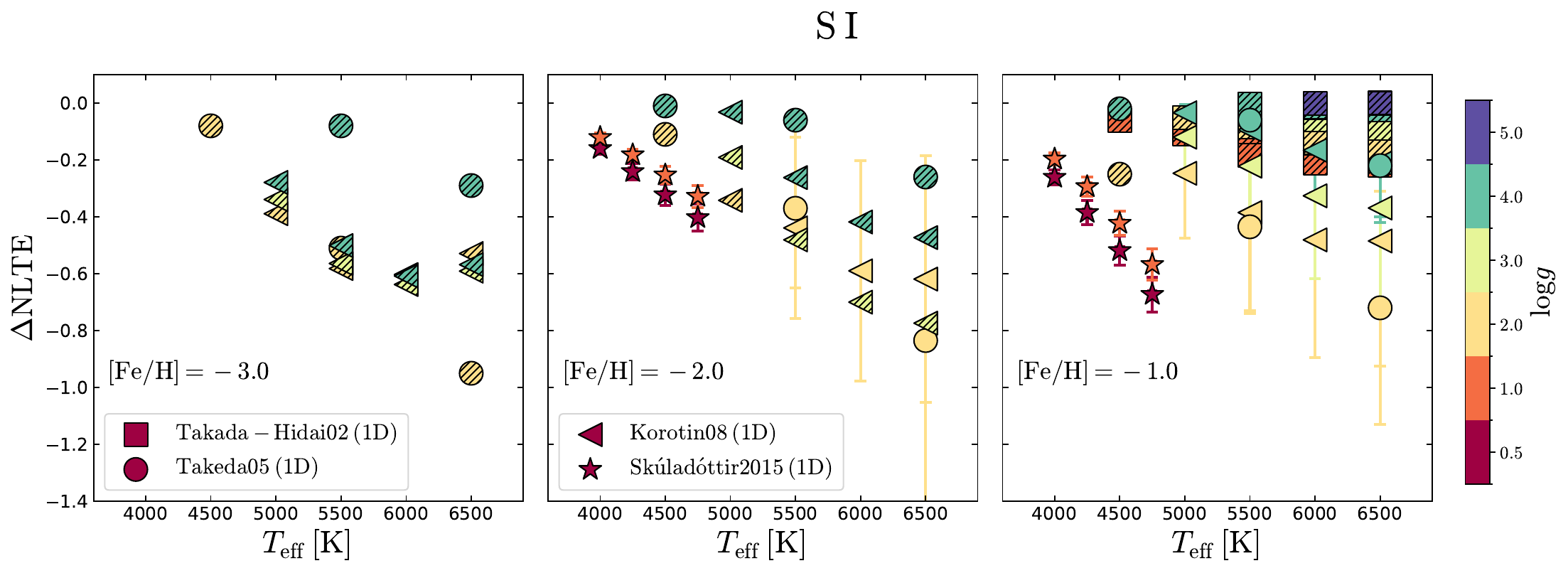} 
\caption{Same as Fig.~\ref{fig: Si_common_lines}, but for \ion{S}{I}. Squares show the corrections of \citet{Takada-Hidai2002} for the 8694\,\AA\ line; stars show the average corrections of the \ion{S}{I} triplet at 9213, 9228 and 9238\,\AA\ from \citet{Skuladottir2015}; circles show the average corrections of the 8694 and 9213\,\AA\ lines from \citet{Takeda2005}; and triangles the average corrections of the 8694, 9213 and the 8th \ion{S}{I} multiple at 6543-6557\,\AA\ from \citet{Korotin2008}. Diagonally hatched symbols represent cases where only one line is available, which is the 9213\,\AA\ line for the \citet{Takeda2005} and \citet{Korotin2008} grids at ${\rm [Fe/H] \leq -2}$; and the 8694\,\AA\ line for the \citet{Takada-Hidai2002} line at ${\rm [Fe/H]=-1}$.}
\label{fig: S_common_lines}
\end{center}
\end{figure*}

\begin{figure*}
\begin{center}
\includegraphics[width=0.9\hsize]{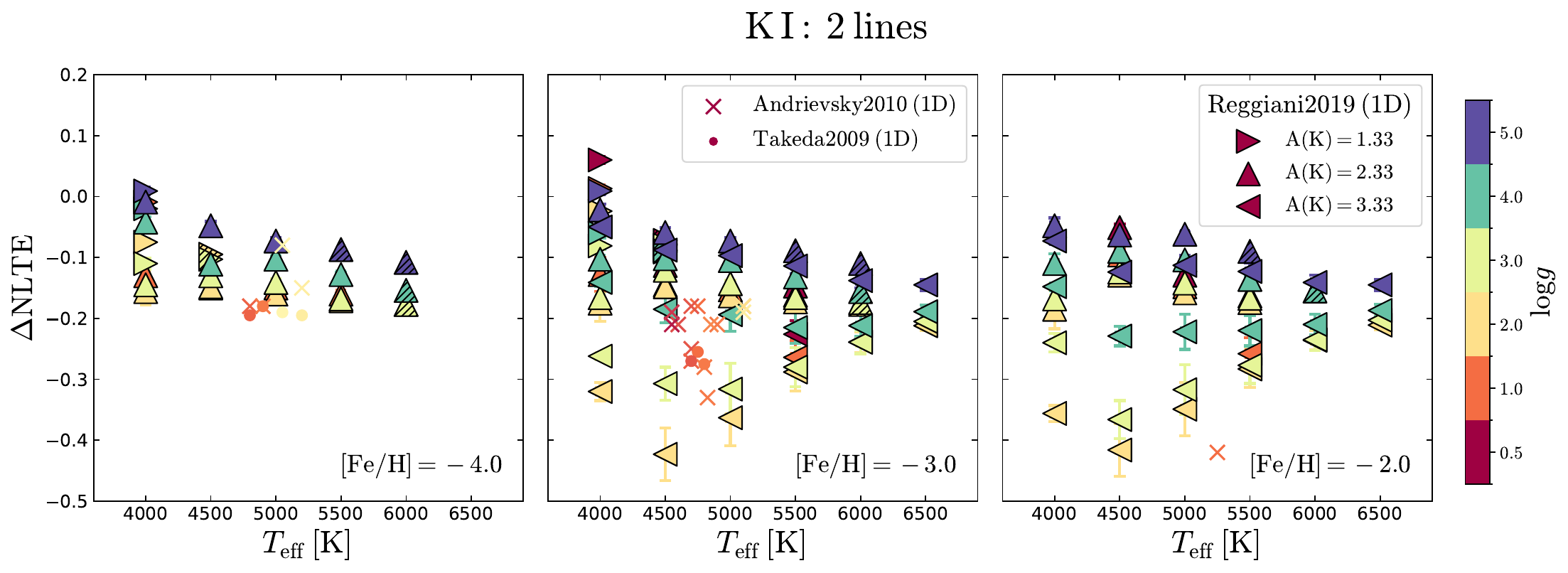} 
\caption{Same as Fig.~\ref{fig: Si_common_lines}, but for \ion{K}{I}. Triangles show 1D\,NLTE corrections by \citet{Reggiani2019_K} for three different K abundances, A(K)=1.33, 2.33 and 3.33 (A(K)$_\odot=5.03$). NLTE corrections for individual stars by \citet{Andrievsky2010} are shown as X~symbols, and from \citet{Takeda2009} as circles.}
\label{fig: K_common_lines}
\end{center}
\end{figure*}

\begin{figure*}
\begin{center}
\includegraphics[width=0.9\hsize]{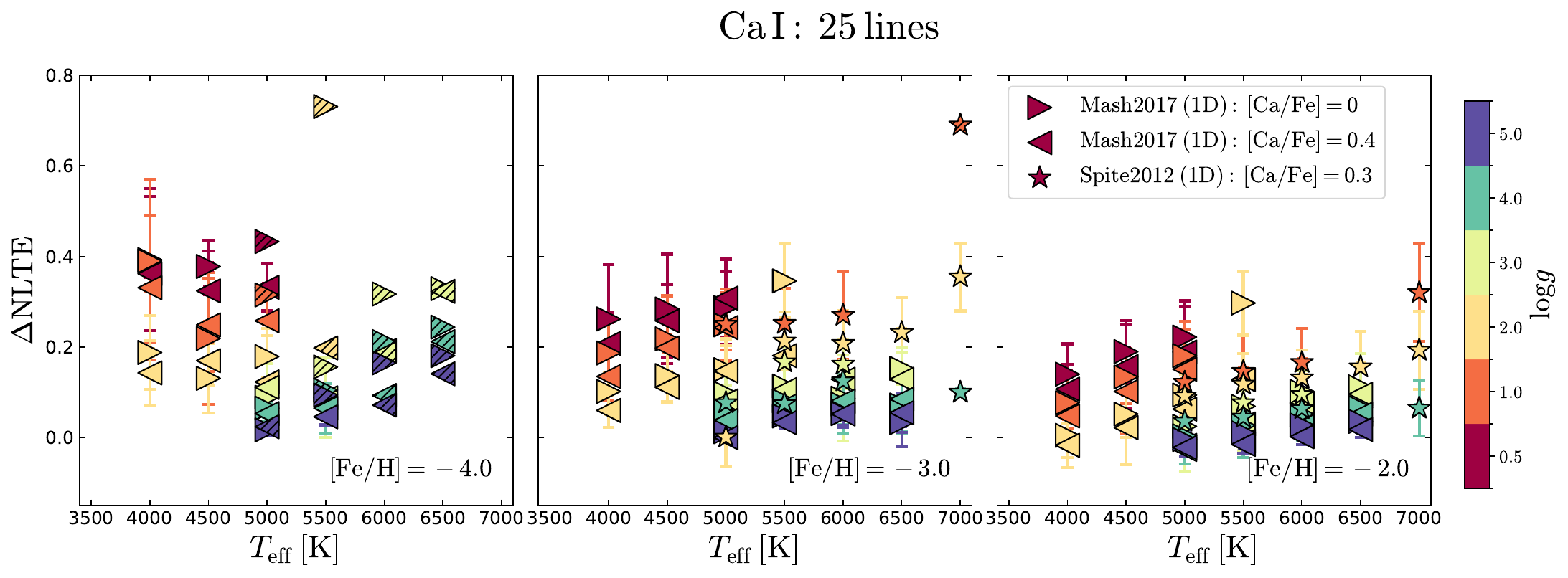} 
\caption{Same as Fig.~\ref{fig: Si_common_lines}, but for \ion{Ca}{I}. Triangles and star symbols show 1D\,NLTE corrections from \citet{Mashonkina2007_Ca1} and  \citet{Spite2012_Ca}, respectively. Symbols are hatched diagonally where only one line is available in the range $5\:{\rm m\AA}<{\rm EW}<200\:{\rm m\AA}$.}
\label{fig: Ca1_common_lines}
\end{center}
\end{figure*}

\begin{figure*}
\begin{center}
\includegraphics[width=0.9\hsize]{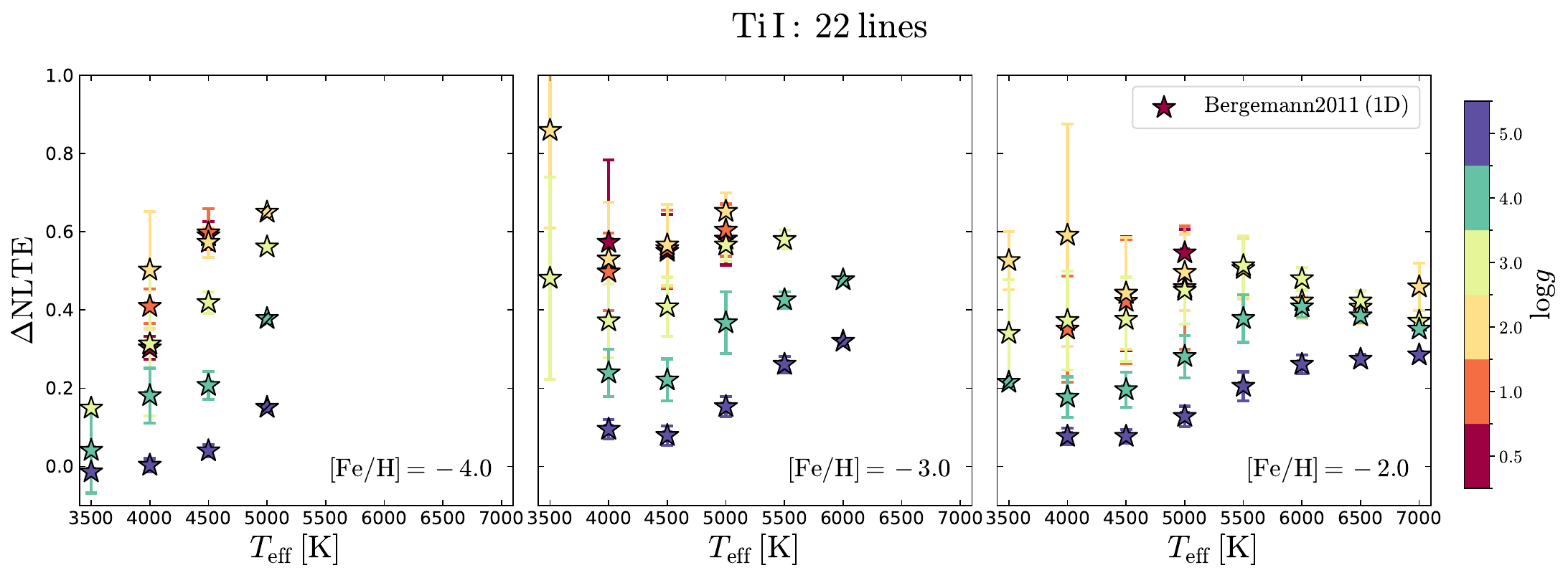} \\
\includegraphics[width=0.9\hsize]{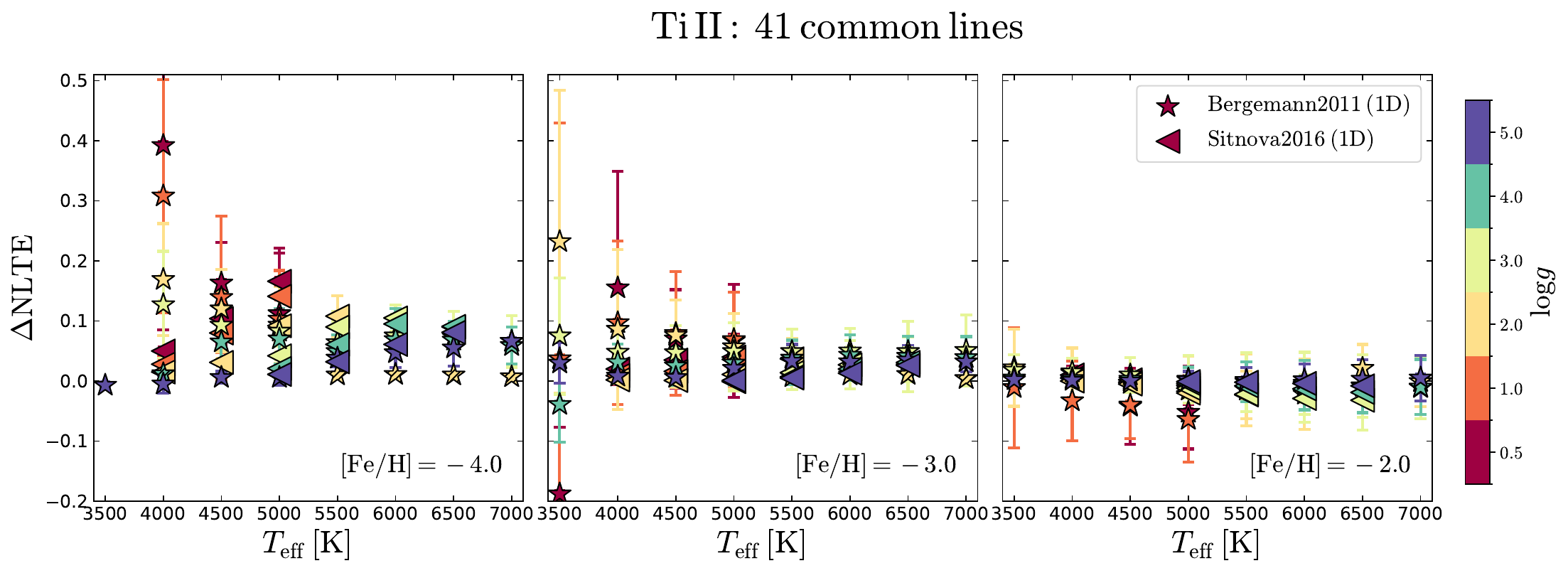} 
\caption{Same as Fig.~\ref{fig: Si_common_lines}, but for \ion{Ti}{I} (top) and \ion{Ti}{II} (bottom). Star symbols and triangles show the 1D\,NLTE corrections by \citet{Bergemann2011_Ti1_Ti2} and \citet{Sitnova2016_Ti2}, respectively.}
\label{fig: Ti1_common_lines}
\end{center}
\end{figure*}

\begin{figure*}
\begin{center}
\includegraphics[width=0.9\hsize]{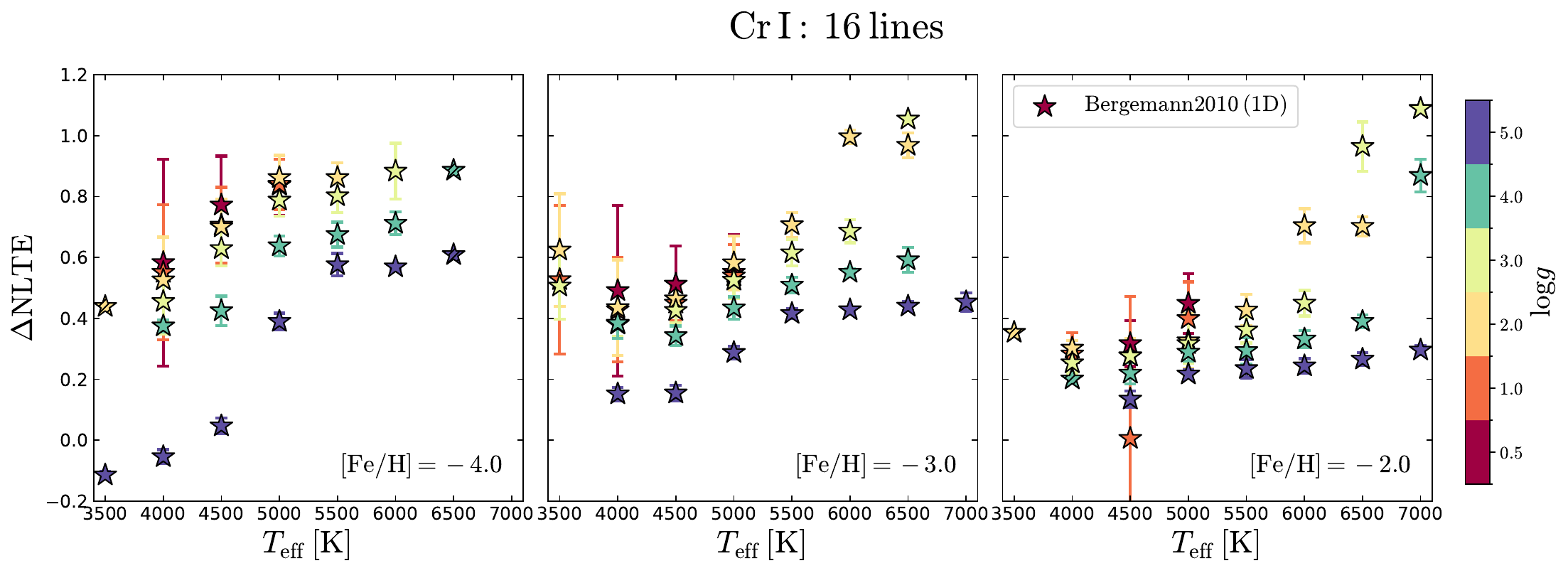} 
\caption{Same as Fig.~\ref{fig: Si_common_lines}, but for \ion{Cr}{I}, using the 1D\,NLTE corrections by \citet{Bergemann2010_Cr}. 
Error bars represent the standard deviation.}
\label{fig: Cr1_common_lines}
\end{center}
\end{figure*}

\begin{figure*}
\begin{center}
\includegraphics[width=0.9\hsize]{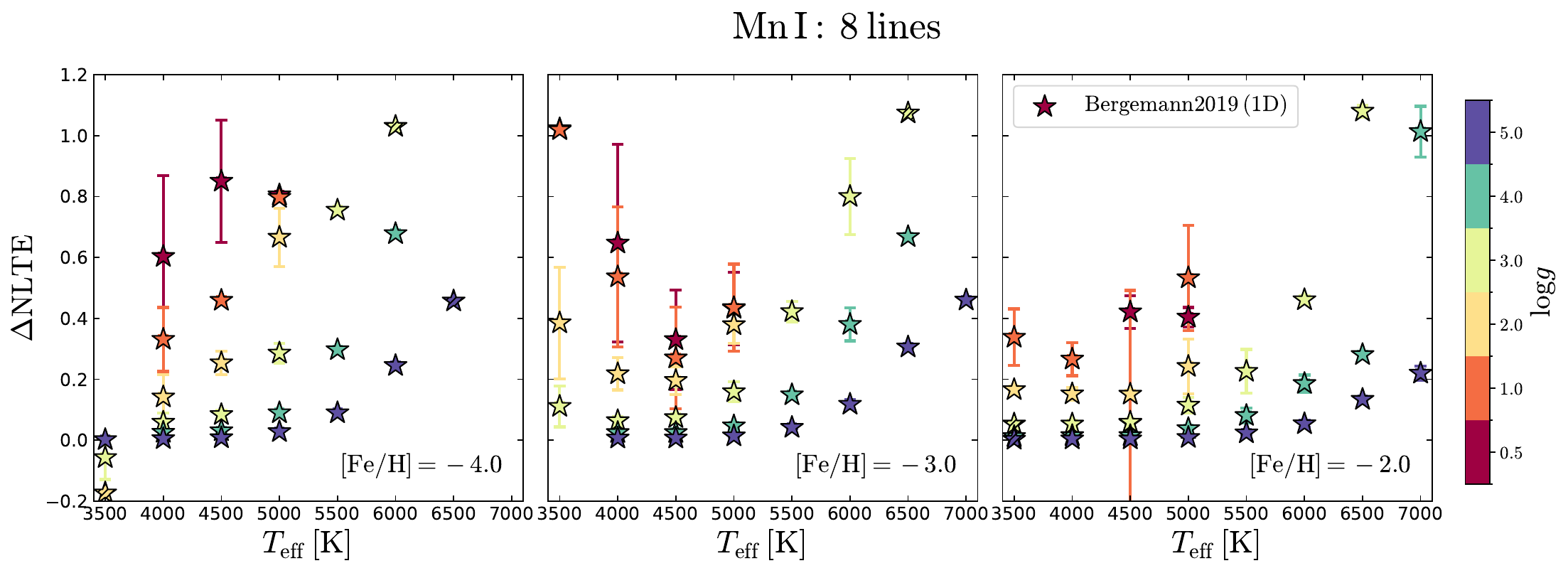} 
\caption{Same as Fig.~\ref{fig: Si_common_lines}, but for \ion{Mn}{I}. Star symbols show 1D\,NLTE corrections by \citet{Bergemann19}.}
\label{fig: Mn1_common_lines}
\end{center}
\end{figure*}

\begin{figure*}
\begin{center}
\includegraphics[width=0.9\hsize]{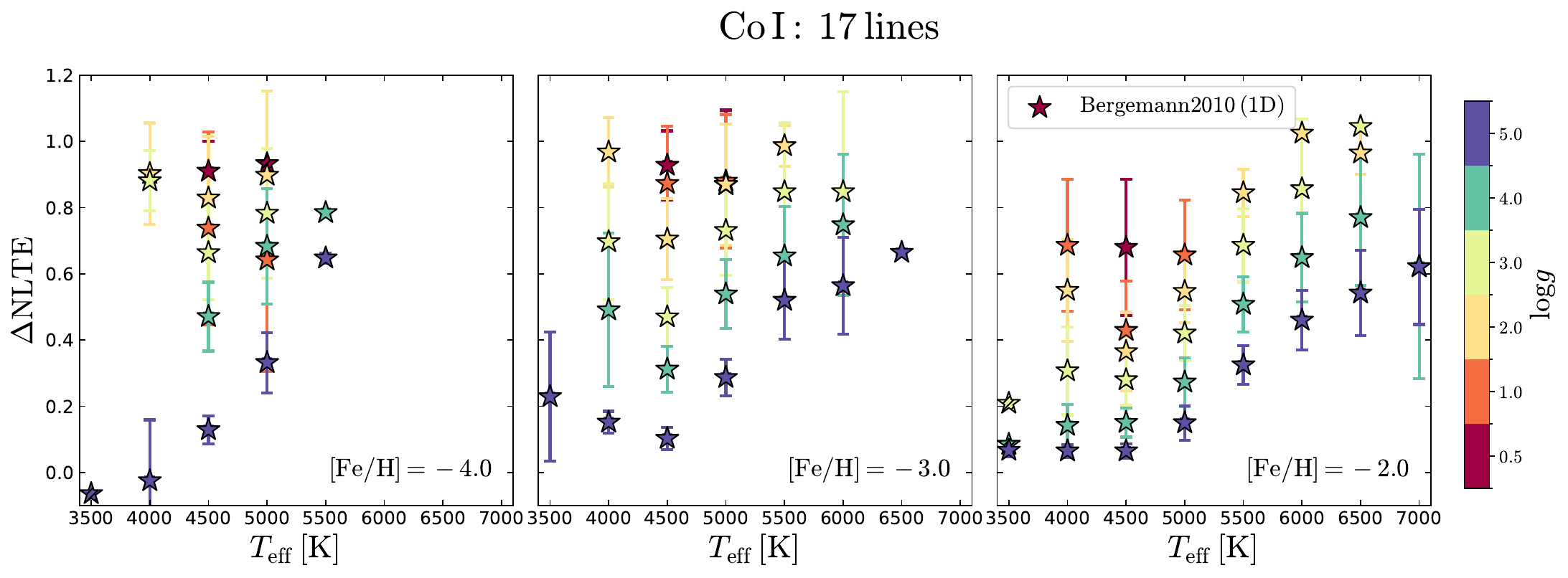} 
\caption{Same as Fig.~\ref{fig: Si_common_lines}, but for \ion{Co}{I}, using the 1D\,NLTE corrections by \citet{Bergemann2010_Co}.}
\label{fig: Co1_lines}
\end{center}
\end{figure*}

\begin{figure*}
\begin{center}
\includegraphics[width=0.9\hsize]{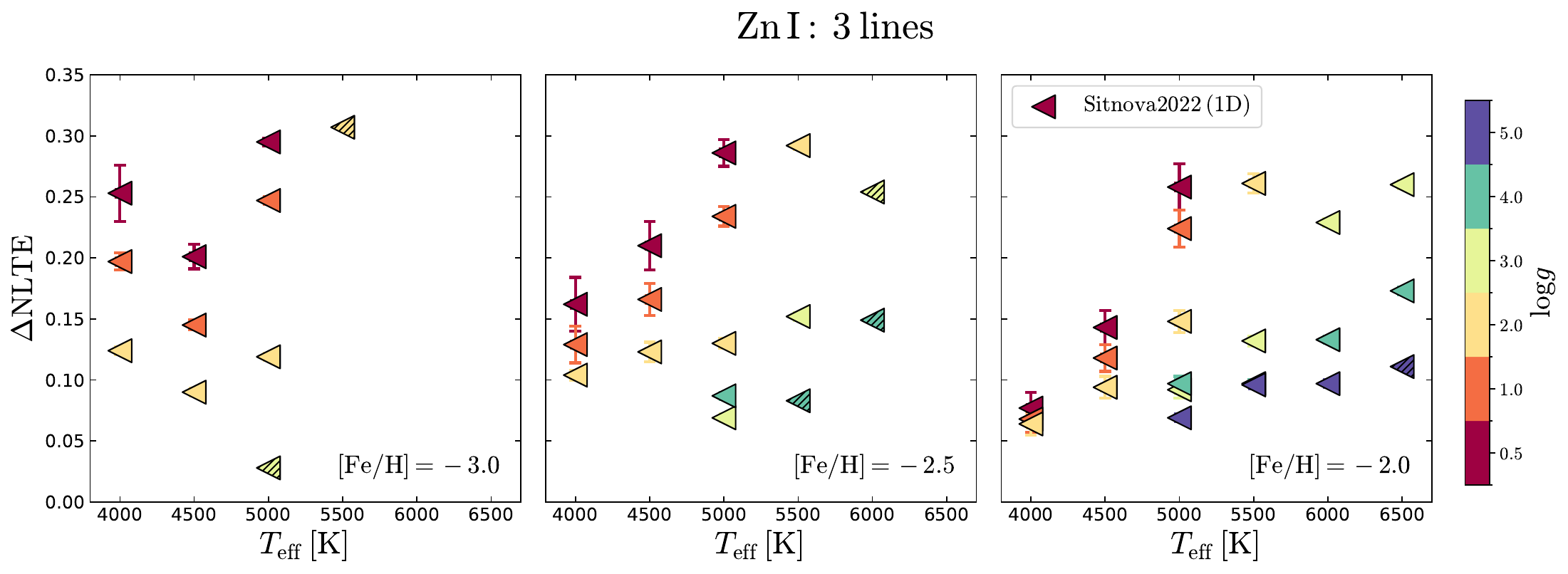} 
\caption{Same as Fig.~\ref{fig: Si_common_lines}, but for \ion{Zn}{I} using the  1D\,NLTE corrections by \citet{Sitnova2022_Zn1_Zn2}.}
\label{fig: Zn_lines}
\end{center}
\end{figure*}

\begin{figure*}
\begin{center}
\includegraphics[width=0.7\hsize]{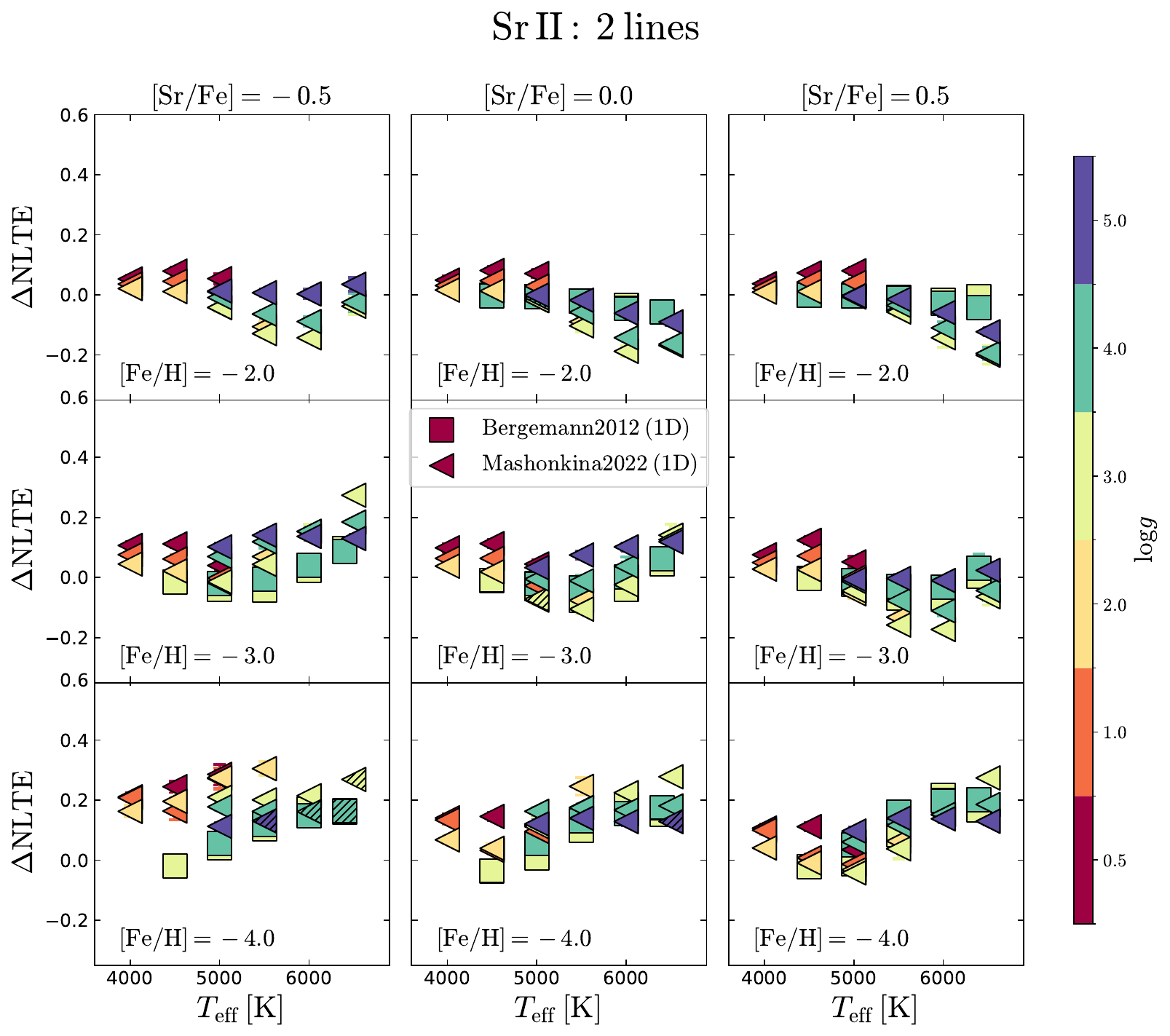} 
\caption{Same as Fig.~\ref{fig: Si_common_lines}, but for \ion{Sr}{II}. Squares and triangles show 1D\,NLTE corrections by \citet{Bergemann2012_Sr} and \citet{Mashonkina2022_Sr}, respectively. Columns show different [\ion{Sr}{II}/Fe] values: ${\rm [\ion{Sr}{II}/Fe]}=-0.5$ (left), 0 (middle) and +0.5 (right).}
\label{fig: Sr2_lines}
\end{center}
\end{figure*}

\begin{figure*}
\begin{center}
\includegraphics[width=0.9\hsize]{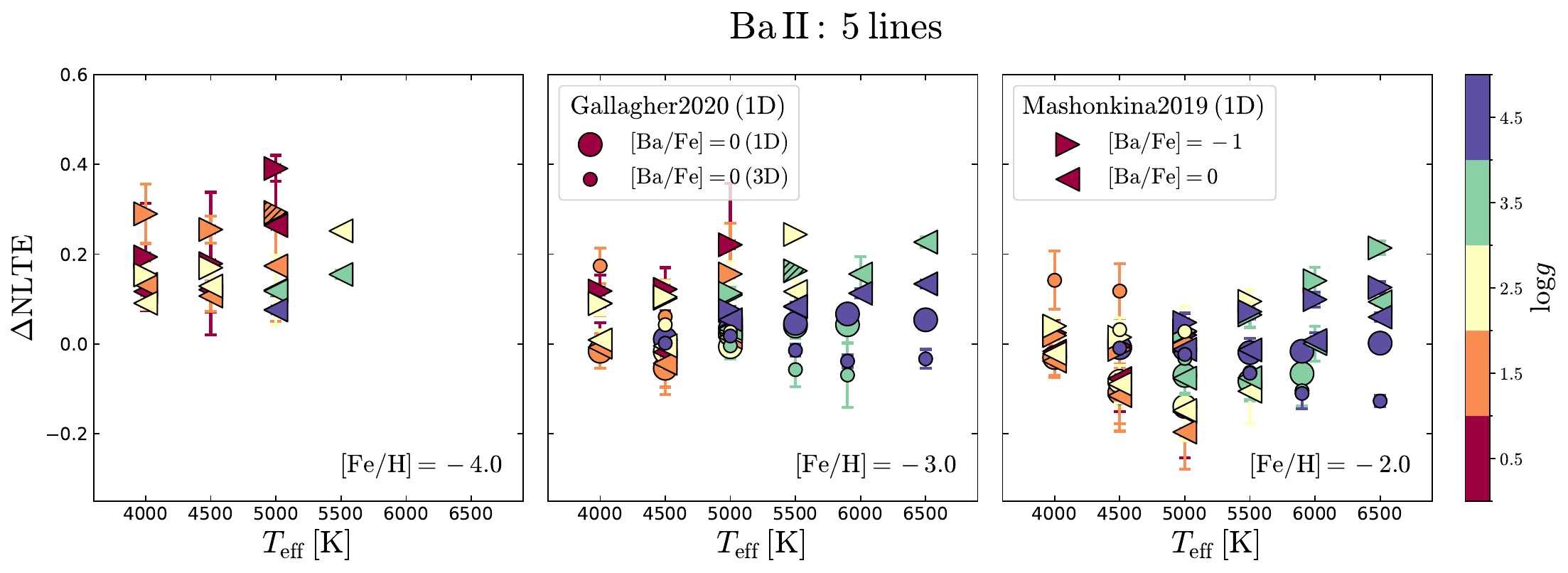} 
\caption{Same as Fig.~\ref{fig: Si_common_lines}, but for \ion{Ba}{II}. Triangles show the 1D\,NLTE corrections by \citet{Mashonkina2019_Ba2} and circles the 1D and 3D\,NLTE corrections by \citet{Gallagher2020}.}
\label{fig: Ba2_lines}
\end{center}
\end{figure*}

\begin{figure*}
\begin{center}
\includegraphics[width=0.9\hsize]{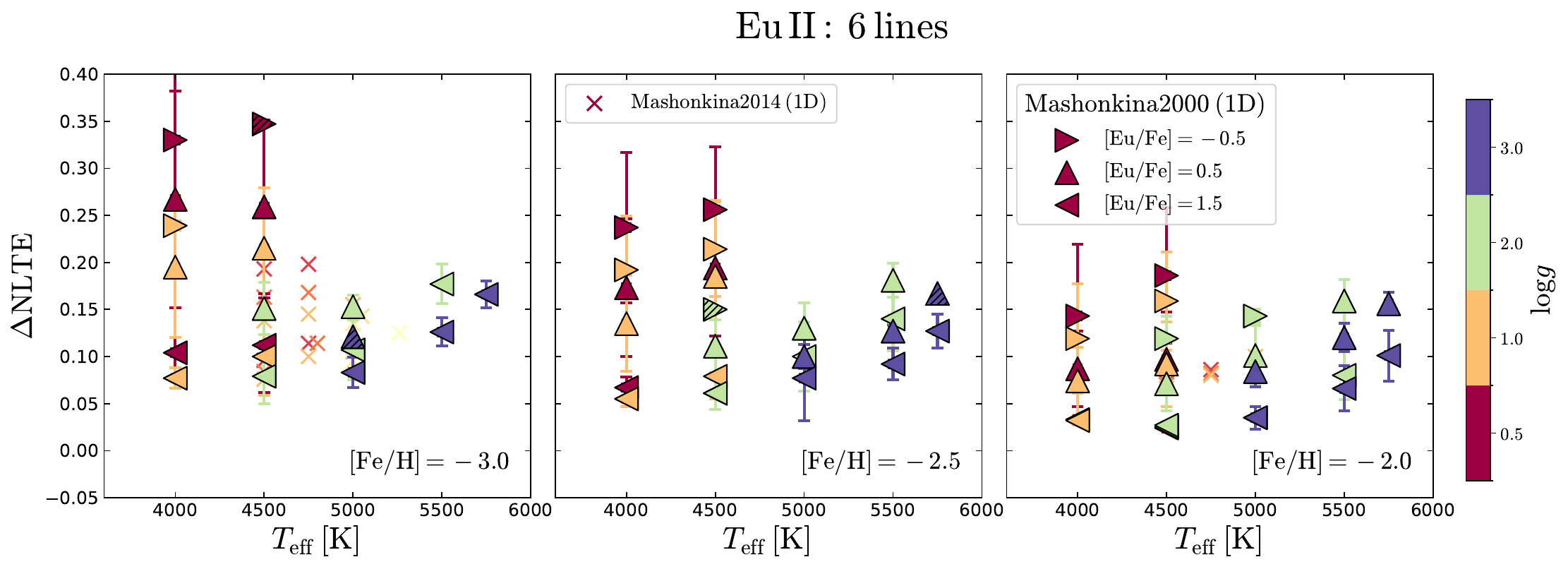} 
\caption{Same as Fig.~\ref{fig: Si_common_lines}, but for \ion{Eu}{II} using the 1D\,NLTE corrections by \citet{Mashonkina2000_Eu2}.}
\label{fig: Eu2_lines}
\end{center}
\end{figure*}

Figures~\ref{fig: C1_lines}-\ref{fig: Eu2_lines} show a comparison of different grids of calculated NLTE corrections for different [Fe/H], \teff, and \logg. The details for each element are discussed in Sec.~\ref{Nlite}.

\section{The applied corrections}
\label{app: applied_corr}

\begin{figure*}
\begin{center}
    \includegraphics[width=0.49\hsize]{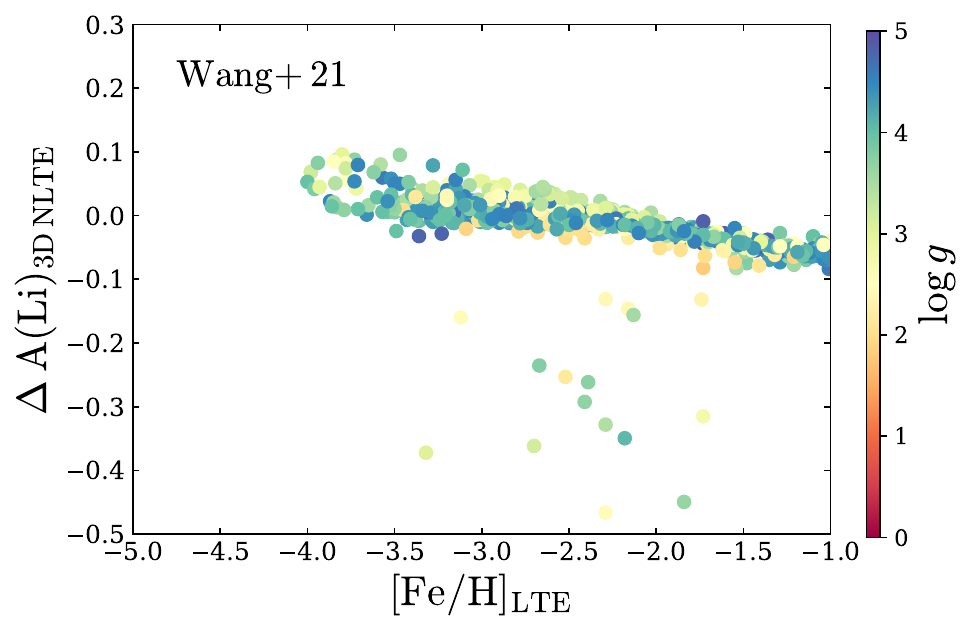} 
    \includegraphics[width=0.49\hsize]{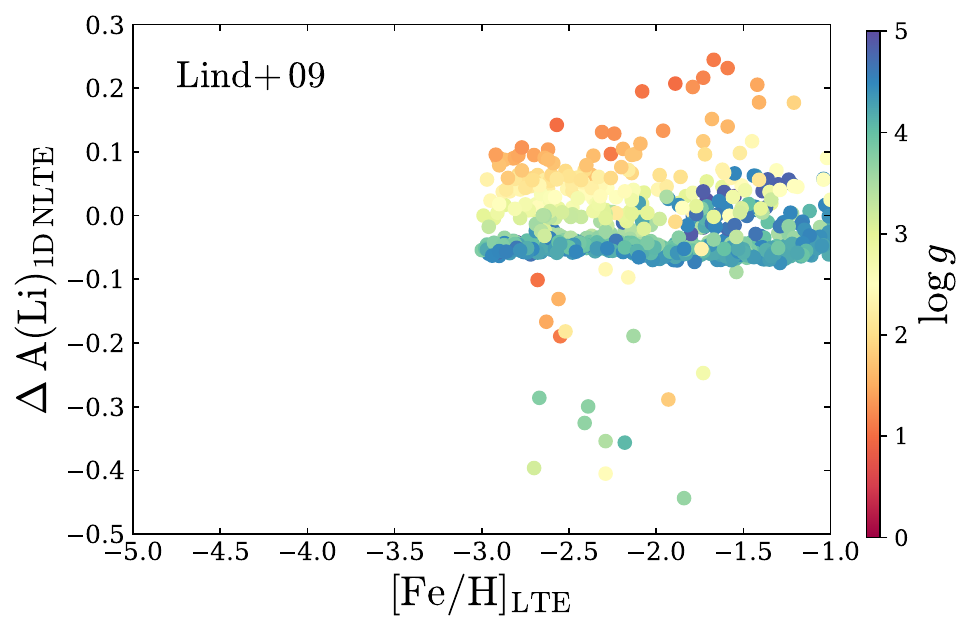} 
\caption{NLTE corrections for A(\ion{Li}{I}), for MP SAGA stars as a function of their $\rm[Fe/H]_{LTE}$, colour-coded according to their $\logg$ values. Li has been corrected using the 3D\,NLTE corrections from our fiducial grid of \citeauthor{Wang2021} (\citeyear{Wang2021}; left), and the 1D\,NLTE corrections from \citeauthor{Lind2009_Li1} (\citeyear{Lind2009_Li1}; right). Only stars with stellar parameters within the respective grid limits are shown.}
\label{fig: LiFe1_logg}
\end{center}
\end{figure*}

\begin{figure*}
\begin{center}
    \includegraphics[width=0.49\hsize]{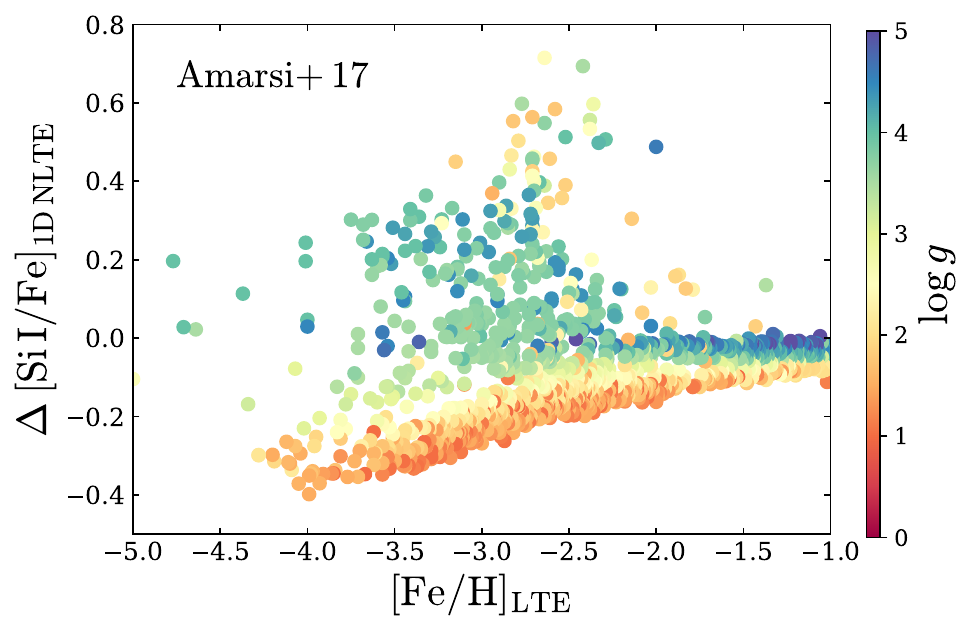} 
    \includegraphics[width=0.49\hsize]{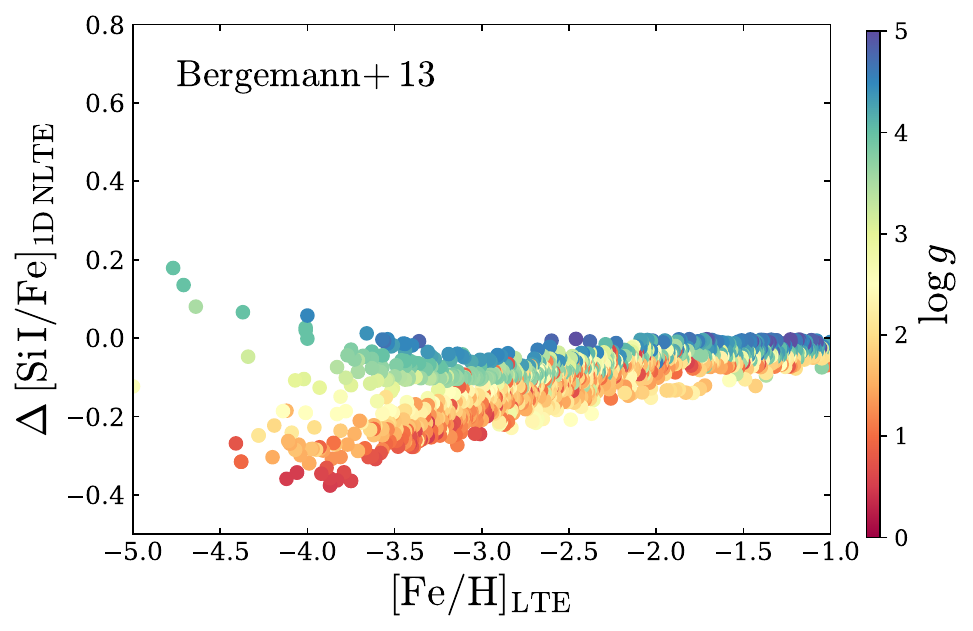} 
\caption{Net NLTE corrections for [\ion{Si}{I}/Fe] in MP SAGA stars, as a function of $\rm[Fe/H]_{LTE}$, colour-coded according to $\logg$. Two sets of Si 1D\,NLTE corrections are shown, from our fiducial grid of \citeauthor{Amarsi2017_Si} (\citeyear{Amarsi2017_Si}; left), and that of \citeauthor{Bergemann2013_Si} (\citeyear{Bergemann2013_Si}; right). The 1D\,NLTE corrections for \ion{Fe}{I} are the mean of the grids of \citet{Mashonkina2011_Fe1} and \citet{Bergemann2012_Fe1_Fe2} in both panels. Only stars with stellar parameters within the respective grid limits are shown.}
\label{fig: SiFe1_logg}
\end{center}
\end{figure*}

\begin{figure*}
\begin{center}
    \includegraphics[width=0.49\hsize]{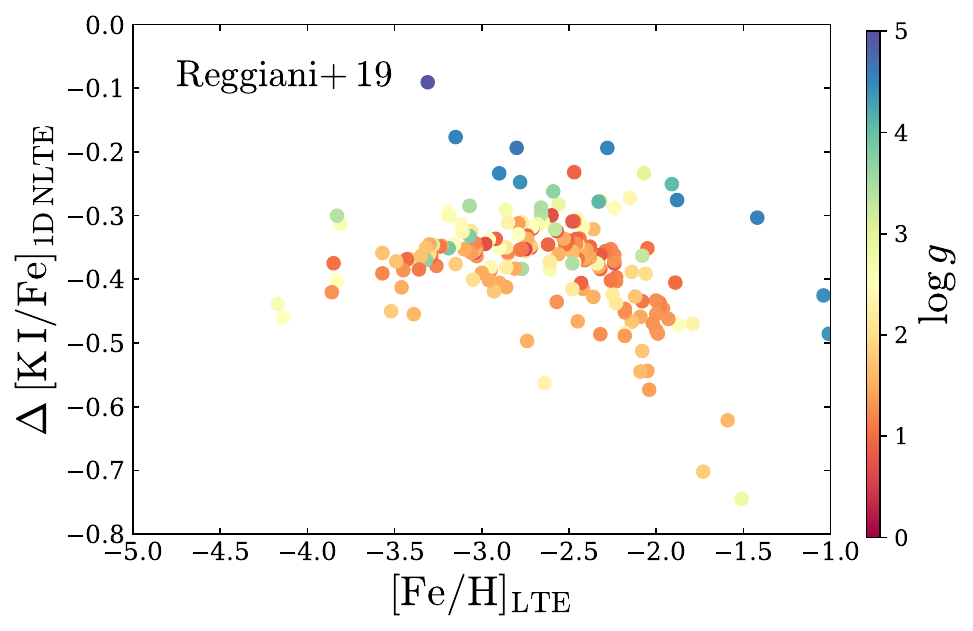} 
    \includegraphics[width=0.49\hsize]{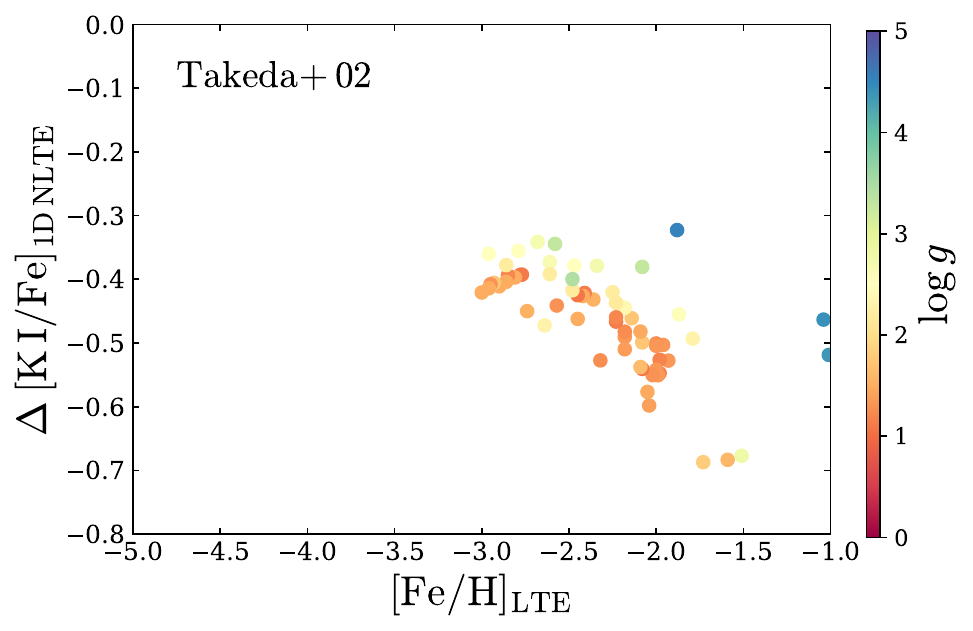} 
\caption{Same as Fig.~\ref{fig: SiFe1_logg}, but for [\ion{K}{I}/Fe], using our fiducial grid of \citeauthor{Reggiani2019_K} (\citeyear{Reggiani2019_K}; left), and that of \citeauthor{Takeda2002} (\citeyear{Takeda2002};right).}
\label{fig: KFe1_logg}
\end{center}
\end{figure*}

\begin{figure*}
\begin{center}
    \includegraphics[width=0.49\hsize]{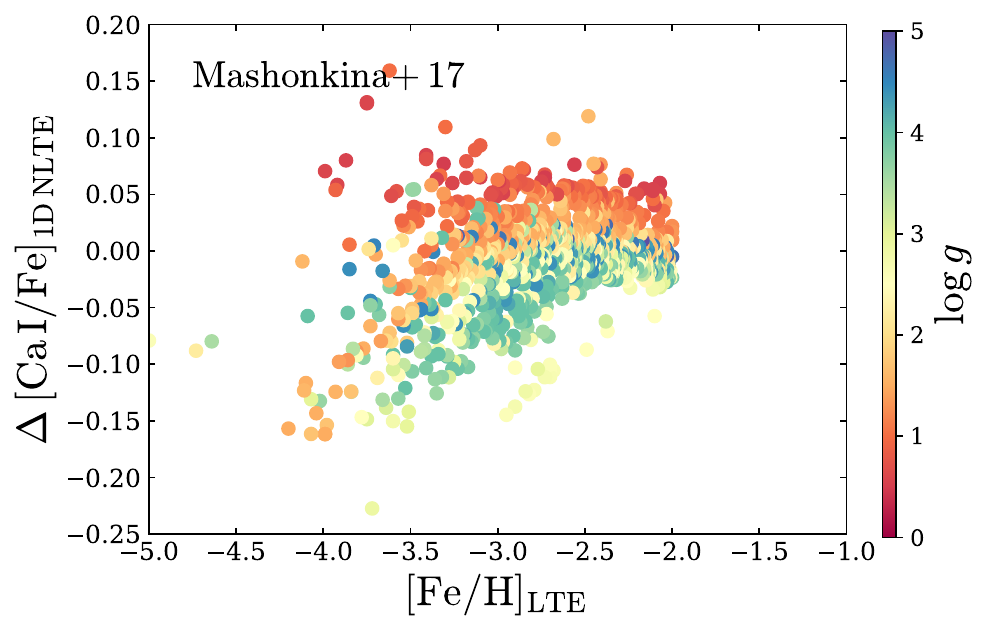} 
    \includegraphics[width=0.49\hsize]{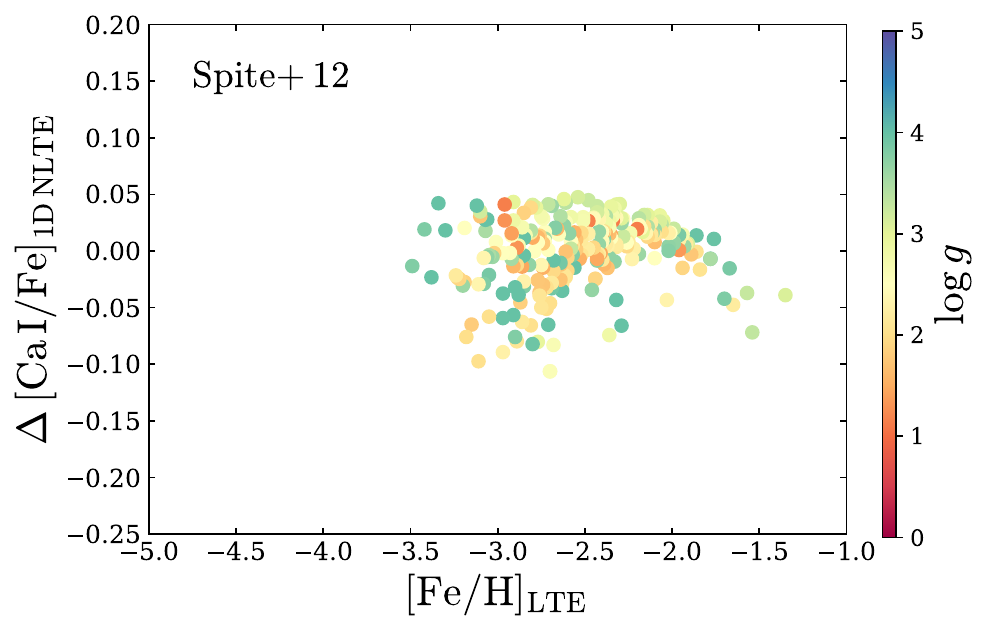} 
\caption{Same as Fig.~\ref{fig: SiFe1_logg}, but for [\ion{Ca}{I}/Fe] using our fiducial grid from \citeauthor{Mashonkina2017_Ca1} (\citeyear{Mashonkina2017_Ca1}; left) and \citeauthor{Spite2012_Ca} (\citeyear{Spite2012_Ca}; right).}
\label{fig: CaFe1_logg}
\end{center}
\end{figure*}

\begin{figure*}
\begin{center}
    \includegraphics[width=0.49\hsize]{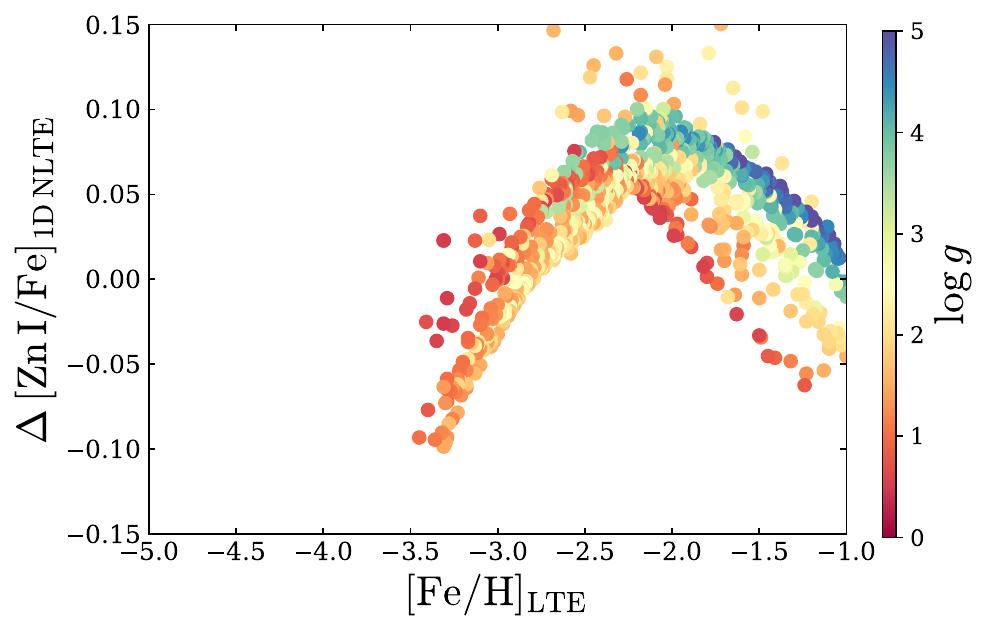} 
    \includegraphics[width=0.49\hsize]{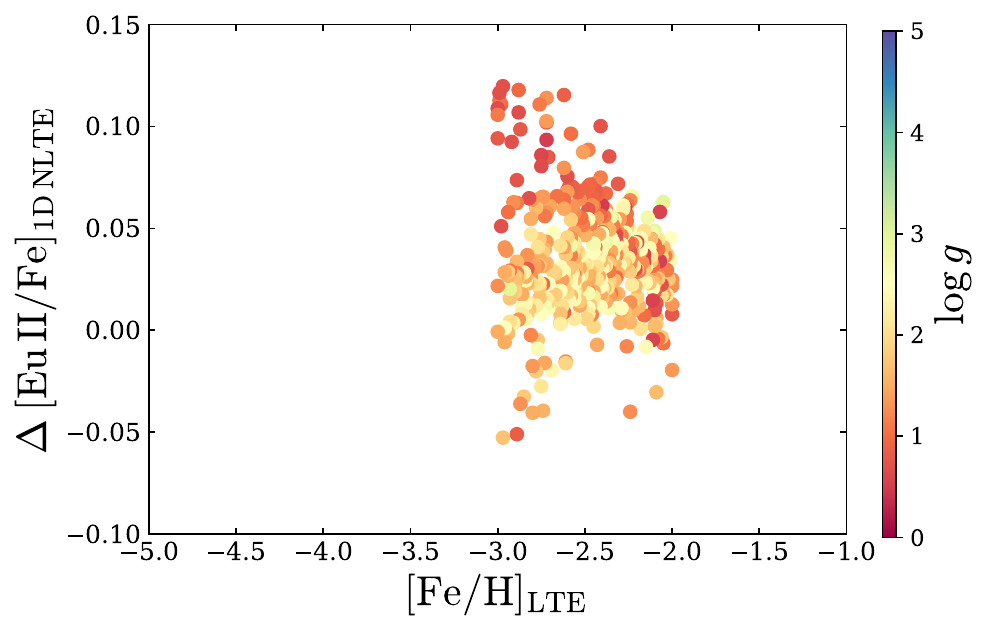} 
\caption{Same as Fig.~\ref{fig: SiFe1_logg}, but for [\ion{Zn}{I}/Fe] (left), and [\ion{Eu}{II}/Fe] (right), using the 1D\,NLTE corrections from \citet{Sitnova2022_Zn1_Zn2}, and \citet{Mashonkina2000_Eu2}, respectively.}
\label{fig: 3Fe1_logg}
\end{center}
\end{figure*}

\begin{figure*}
\begin{center}
    \includegraphics[width=0.49\hsize]{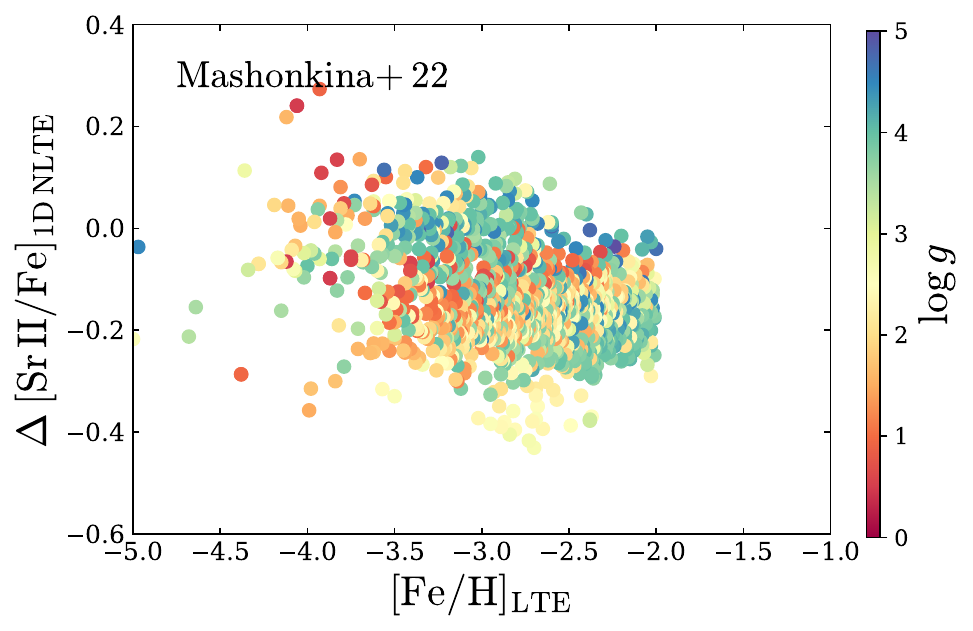} 
    \includegraphics[width=0.49\hsize]{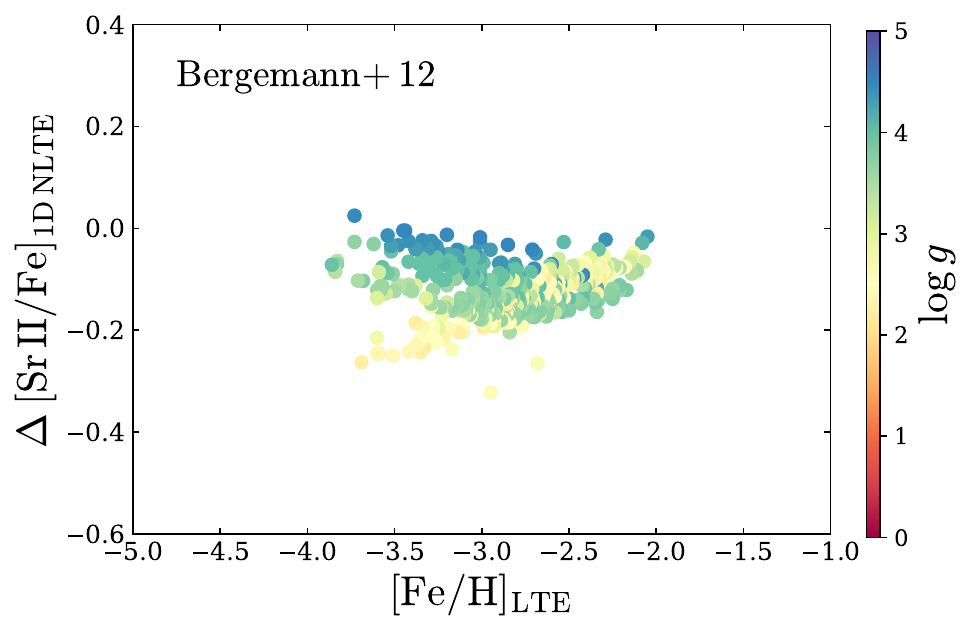} 
\caption{Same as Fig.~\ref{fig: SiFe1_logg}, but for [\ion{Sr}{II}/Fe], using our fiducial grid of \citeauthor{Mashonkina2022_Sr} (\citeyear{Mashonkina2022_Sr}; left), and the grid from \citeauthor{Bergemann2012_Sr} (\citeyear{Bergemann2012_Sr}; right).}
\label{fig: SrFe1_logg}
\end{center}
\end{figure*}

\begin{figure*}
\begin{center}
    \includegraphics[width=0.49\hsize]{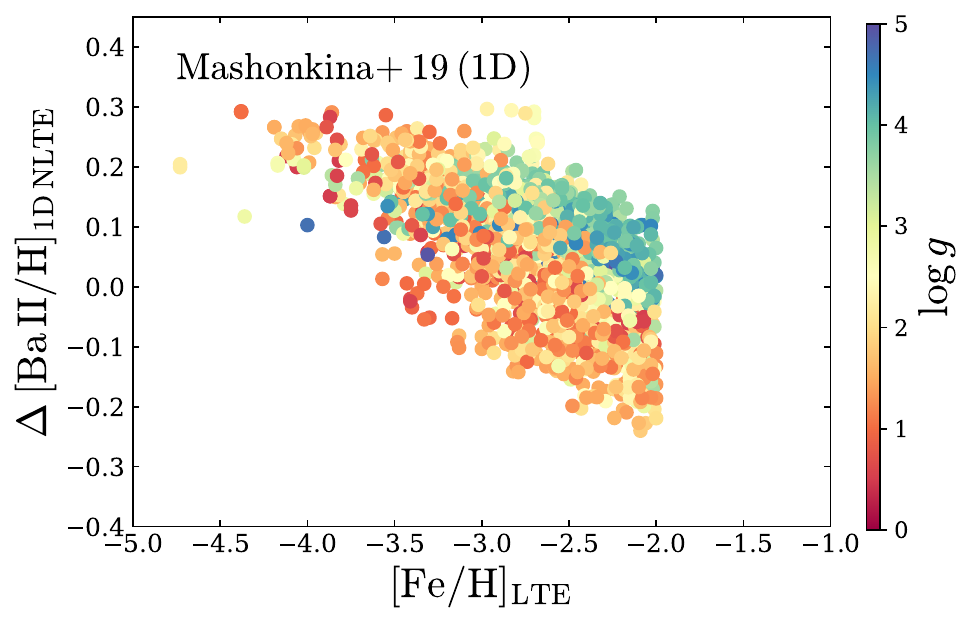} 
    \includegraphics[width=0.49\hsize]{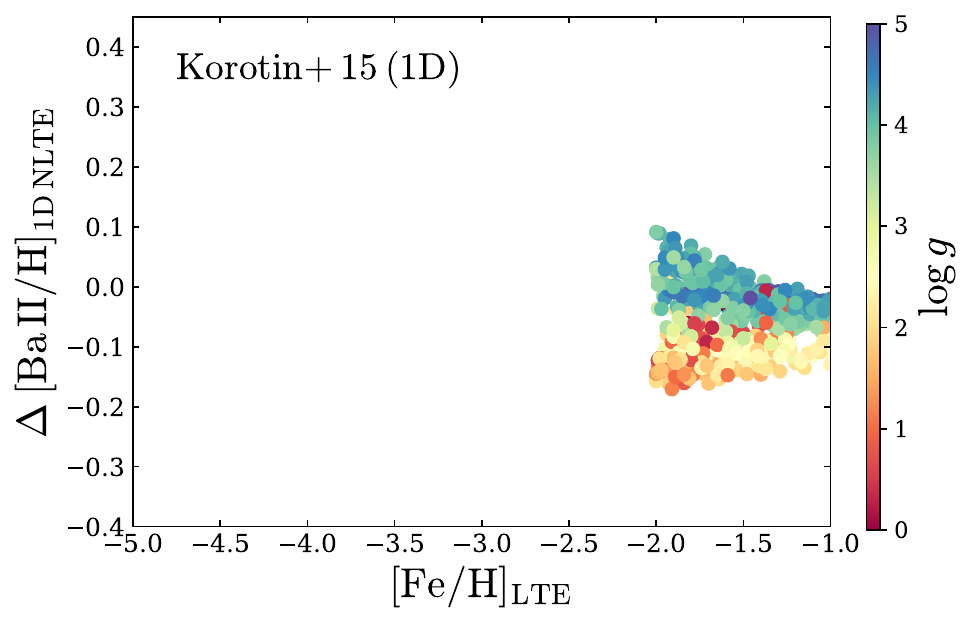} \\
    \includegraphics[width=0.49\hsize]{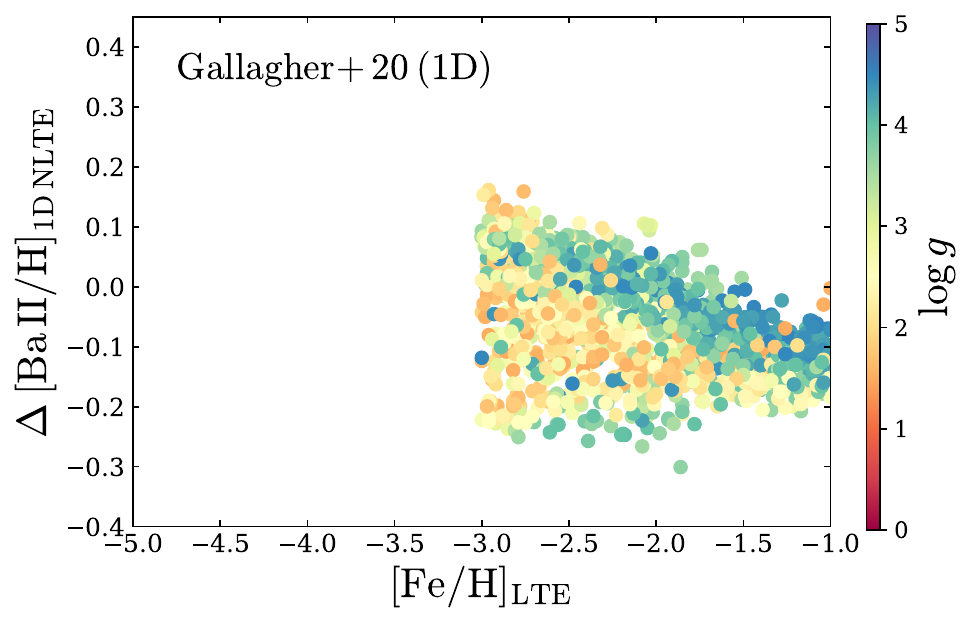} 
    \includegraphics[width=0.49\hsize]{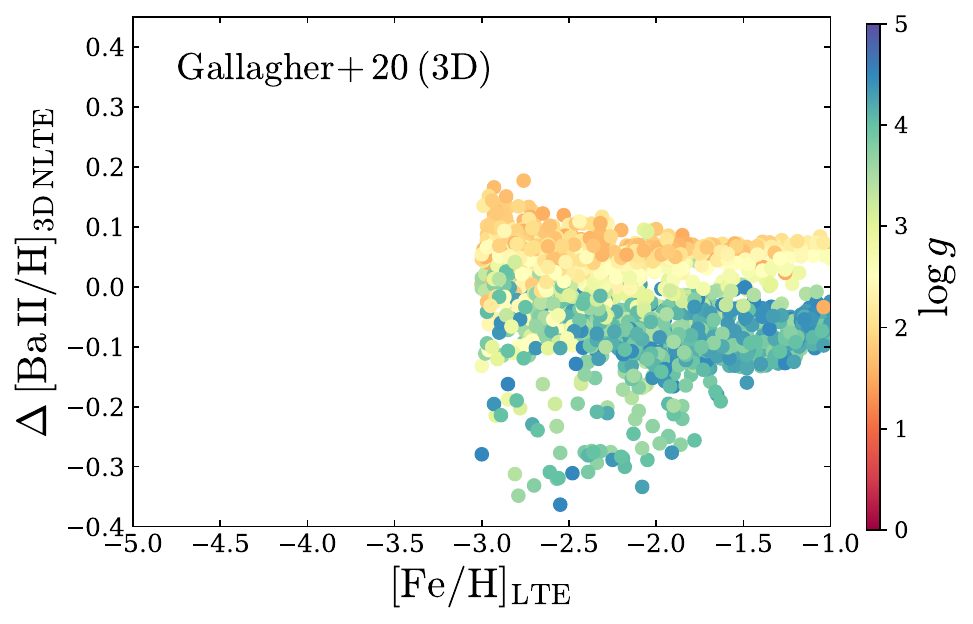} 
\caption{Same as Fig.~\ref{fig: SiFe1_logg}, but for [\ion{Ba}{II}/H], based on the fiducial 1D\,NLTE corrections from \citeauthor{Mashonkina2019_Ba2} (\citeyear{Mashonkina2019_Ba2}; top left), as well as those of \citeauthor{Korotin2015_Ba2} (\citeyear{Korotin2015_Ba2}; top right); and finally the 1D (bottom-left) and 3D (bottom-left) NLTE corrections from \citet{Gallagher2020}.}
\label{fig: BaFe1_logg}
\end{center}
\end{figure*}

For completeness, we present here discussion of the applied corrections for the NLTE-SAGA database for elements which were not included in Sec.~\ref{sec:appcorr}. The results are shown in Figs.~\ref{fig: LiFe1_logg}-\ref{fig: BaFe1_logg}, for Li, Si, K, Ca, Zn, Sr, and Ba. In all cases, the fiducial $\rm[Fe/H]_{1D\,NLTE}$ is adopted, i.e. using the mean corrections of \citet{Mashonkina2011_Fe1} and \citet{Bergemann2012_Fe1_Fe2}, see Table~\ref{table:Grids}. 

The left and right panels of Fig.~\ref{fig: LiFe1_logg} illustrate the NLTE corrections for A(Li) based on the 3D and 1D\,grids from \citet{Wang2021} and \citet{Lind2009_Li1}, respectively. In both cases, the corrections remain above $-0.1$\,dex, except for a handful of stars with high ${\rm A(Li)}>2.28$.
The corrections of \citet{Lind2009_Li1}, reaching close to +0.3\,dex, show a clear dependence on $\logg$ at fixed metallicity. In contrast, the corrections of \citet{Wang2021}, always remaining below +0.1\,dex, have no significant correlation with $\logg$ but instead show an increasing trend with decreasing  ${\rm [Fe/H]_{LTE}}$. For the same stars the largest difference between the two sets reaches 0.24\,dex.
We note that when using the limited grid of \citet{Sbordone2010} instead (see Table~\ref{table:Grids}), we find weak corrections that range from $-0.01$ to $+0.04$\,dex, and from +0.02 to +0.055\,dex for the 1D and 3D cases, respectively.

The net [Si/Fe] corrections (Fig.~\ref{fig: SiFe1_logg}; left panel) are negative for the majority ($\sim70\%$) of stars, showing an increasing trend (in absolute values) with decreasing [Fe/H], especially at low $\logg<3$. However, at $\rm[Fe/H]<-2$ and high $\logg>3$, the [Si/Fe] corrections rise up to +0.7\,dex, as the Si corrections of \citet{Amarsi2017_Si} increase significantly with increasing $T_{\rm eff}$ (see Fig.~\ref{fig: Si_common_lines}). Contrarily, the \citet{Bergemann2013_Si} corrections for Si, being small at all $T_{\rm eff}$ and $\logg$ values, result in negative net [Si/Fe] corrections for more than $99\%$ of the stars (Fig.~\ref{fig: SiFe1_logg}; right panel). The average absolute difference between the two correction sets is 0.075\,dex, with the largest discrepancy reaching 0.87\,dex.

The net corrections for [K/Fe] for all SAGA entries with potassium measurements are negative (Fig.~\ref{fig: KFe1_logg}; left panel). The K corrections increase strongly (in absolute magnitude) with increasing metallicity (see Fig.~\ref{fig: K_common_lines}). However, the net corrections show a downturn at [Fe/H]$\lesssim-2.5$, driven by the $\Delta {\rm [Fe/H]_{1D \: NLTE}}$ dependence on [Fe/H]$_{\rm LTE}$ (See Fig.~\ref{fig: Fe1_logg1}). When using the \citet{Takeda2002} grid instead of our fiducial grid from \citet{Reggiani2019_K}, the corrections remain largely consistent, differing by $<0.1$\,dex.

Both the Ca and Fe corrections exhibit a clear dependence on $\logg$. However, some of this effect cancels out and this trend is less pronounced in the net [Ca/Fe] corrections (Fig.~\ref{fig: CaFe1_logg}). While most Ca corrections are positive (similar to Fe, see Fig.~\ref{fig: Fe1_logg1}), the net [Ca/Fe] corrections vary between $-0.2$ and +0.1\,dex. Using the \citet{Spite2012_Ca} grid, instead of our default from \citet{Mashonkina2007_Ca1}, yields very similar results, with a maximum difference of less than 0.12\,dex between the two grids. 

Fig.~\ref{fig: 3Fe1_logg} displays the net corrections for [\ion{Zn}{I}/Fe] and [\ion{Eu}{II}/Fe]. The corrections for \ion{Zn}{I} are always positive, increasing with decreasing metallicity down to $\rm[Fe/H]\approx-2$, and then leveling off at lower metallicities. Due to the anticorrelation between $\Delta {\rm [Fe/H]_{NLTE}}$ and [Fe/H]$_{\rm LTE}$ (see Fig.~\ref{fig: Fe1_logg1}), the net corrections of [Zn/Fe] peak around $\rm[Fe/H]\approx-2$ and turn negative at $\rm[Fe/H]\lesssim-2.5$. Still, the absolute corrections for [Zn/Fe] remain small, always $\lesssim0.1\,$dex.
The corrections for [\ion{Eu}{II}/Fe], only available for giant stars, are modest ($\rm -0.10\,dex<\Delta NLTE< 0.15\,dex$), and show an increasing trend with decreasing $\logg$.

The net [\ion{Sr}{II}/Fe] corrections from \citet{Mashonkina2022_Sr} lie in the range $-0.45$ to +0.25 dex and show no clear trend with $\logg$ due to their dependence on both [Fe/H]$_{\rm LTE}$ and [\ion{Sr}{II}/Fe]$_{\rm LTE}$ (Fig.~\ref{fig: SrFe1_logg}). When using the \citet{Bergemann2012_Sr} grid we find somewhat weaker corrections, with differences up to $0.2\,$dex (0.06\,dex on average).

Finally, in Fig.~\ref{fig: BaFe1_logg}, we show $\Delta {\rm [\ion{Ba}{II}/H]_{NLTE}}$ based on our fiducial 1D\,NLTE corrections of \citeauthor{Mashonkina2019_Ba2} (\citeyear{Mashonkina2019_Ba2}; top-left), and those of \citeauthor{Korotin2015_Ba2} (\citeyear{Korotin2015_Ba2}; top-right); as well as the 1D and 3D\,NLTE corrections of \citeauthor{Gallagher2020} (\citeyear{Gallagher2020}; bottom panels). We do not plot the net [\ion{Ba}{II}/Fe] corrections here, to facilitate comparison, as 3D-NLTE corrections for Fe are only available for hot dwarf stars (see Table~\ref{table:Grids}). All 1D\,NLTE corrections for Ba exhibit an increasing trend with $\logg$, though the \citet{Mashonkina2019_Ba2} and \citet{Korotin2015_Ba2} corrections are on average 0.04 and 0.06\,dex higher, respectively, than those of \citet{Gallagher2020} in the overlapping regions. In fact, more than 85$\%$ of stars have corrections that differ less than 0.1\,dex among the three sets. The observed trend, however, reverses when we consider the 3D corrections of \citet{Gallagher2020}; here, stars with higher $\logg$ exhibit more negative corrections. In all cases, the corrections span a range of approximately $-0.4$ to $+0.3\,$dex.

\section{Comparison with {\sc NEFERTITI}}
\label{app: nefertiti}

\begin{figure*}
\begin{center}
    \includegraphics[width=0.95\hsize]{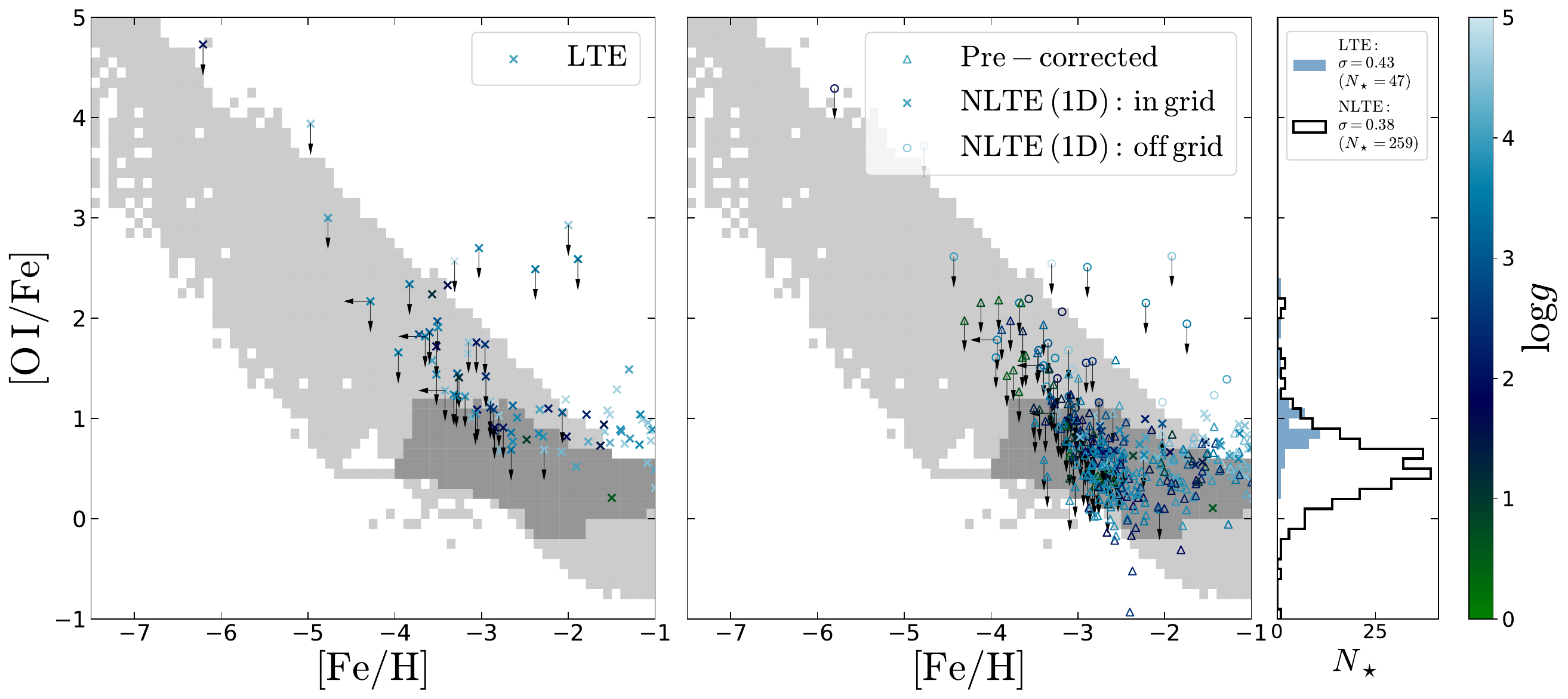} 
\caption{Same as Fig.~\ref{fig: nlte_NaAl}, but for [\ion{O}{I}/Fe]; observed abundances have been corrected using the grid of \citet{Amarsi19} for \ion{O}{I} and the mean corrections of \citet{Mashonkina2011_Fe1} and \citet{Bergemann2012_Fe1_Fe2} for \ion{Fe}{I}.}
\label{fig: nlteO}
\end{center}
\end{figure*}

\begin{figure*}
\begin{center}
    \includegraphics[width=0.95\hsize]{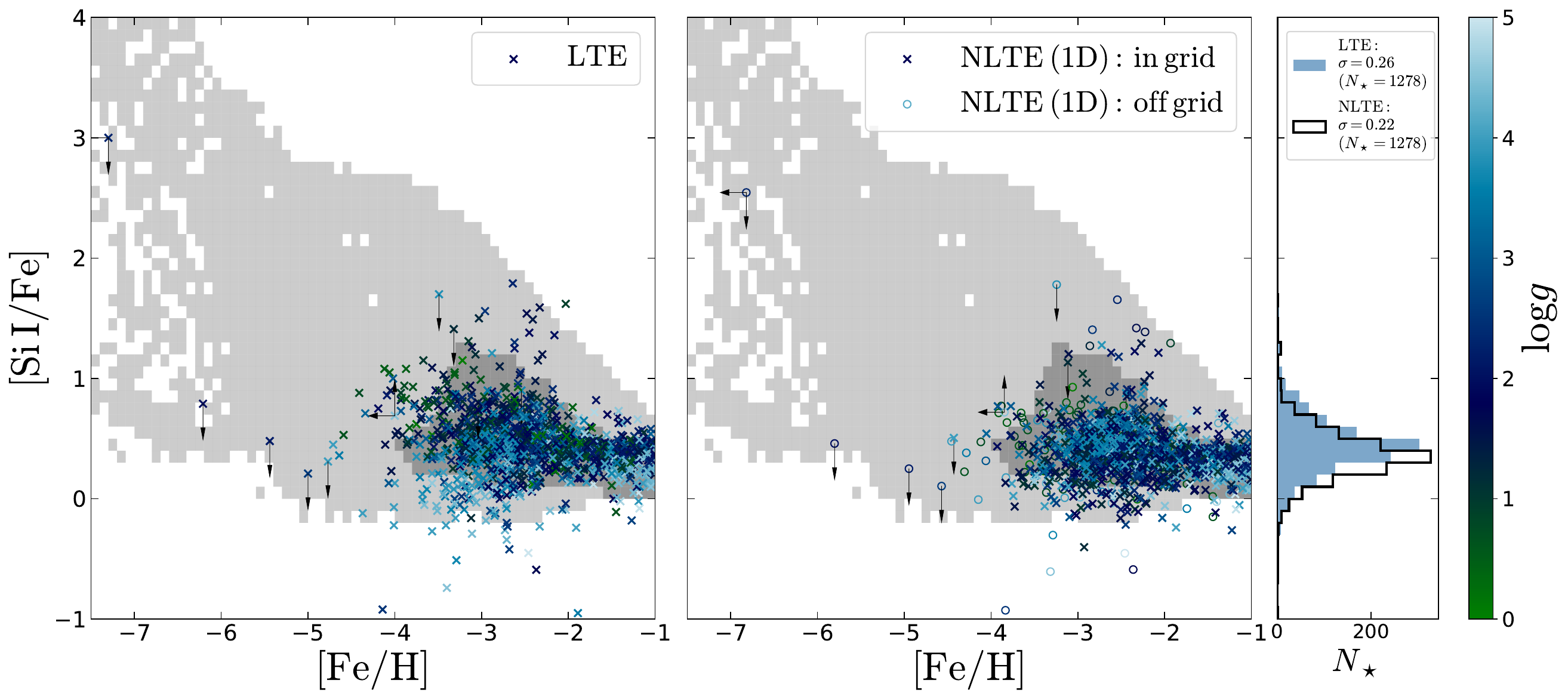} 
\caption{Same as Fig.~\ref{fig: nlte_NaAl}, but for [Si/Fe]; observed abundances have been corrected using the grid from \citet{Amarsi2017_Si} for Si and the mean corrections of \citet{Mashonkina2011_Fe1} and \citet{Bergemann2012_Fe1_Fe2} for \ion{Fe}{I}.}
\label{fig: nlteMgAlSi}
\end{center}
\end{figure*}

\begin{figure*}
\begin{center}
    \includegraphics[width=0.95\hsize]{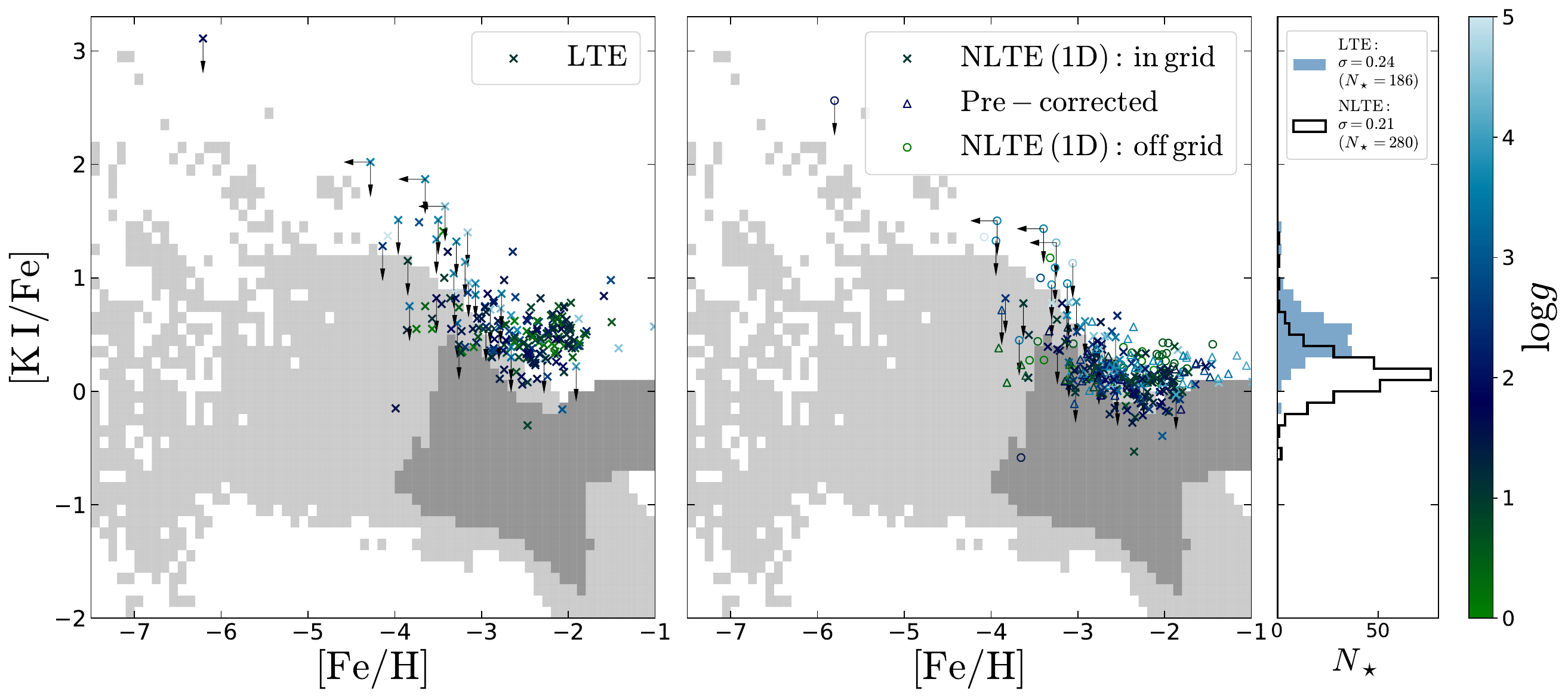} \\
    \includegraphics[width=0.95\hsize]{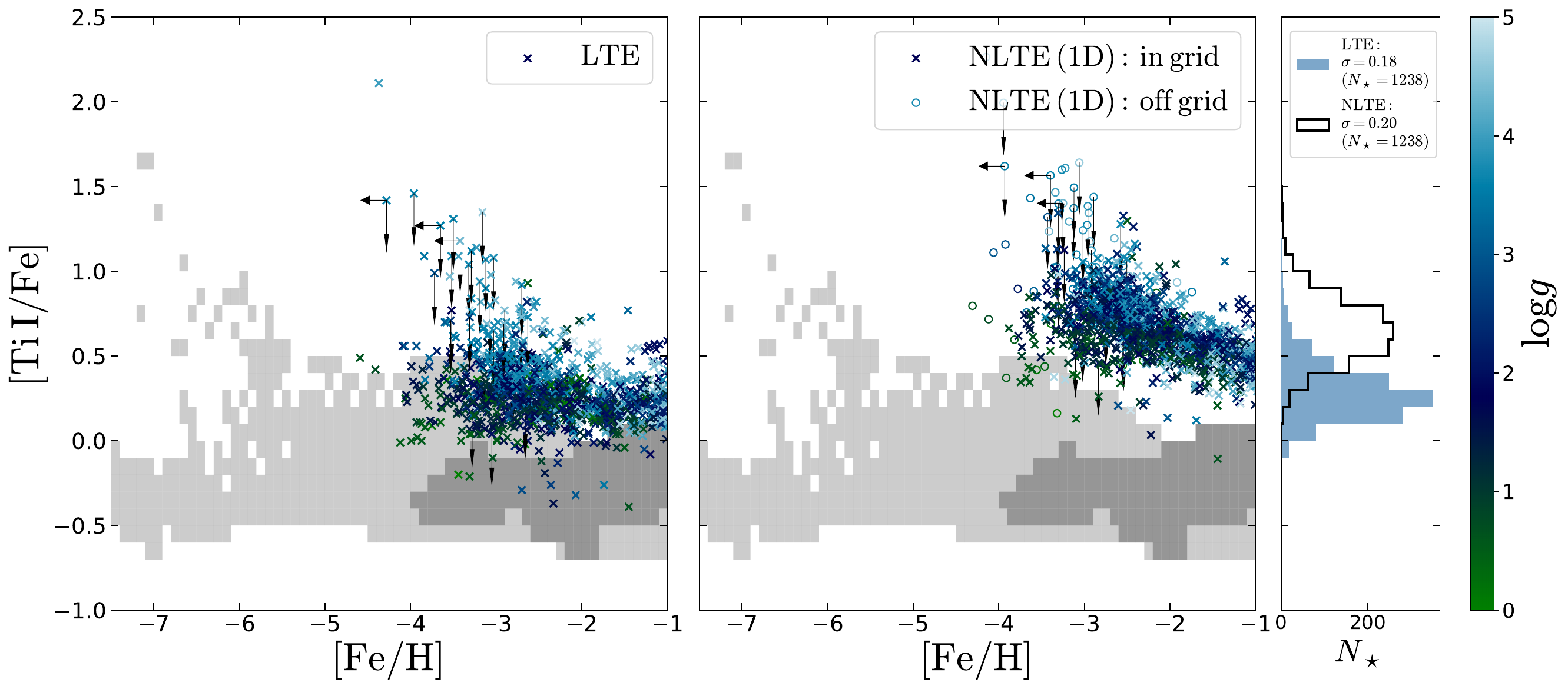} \\
    \includegraphics[width=0.95\hsize]{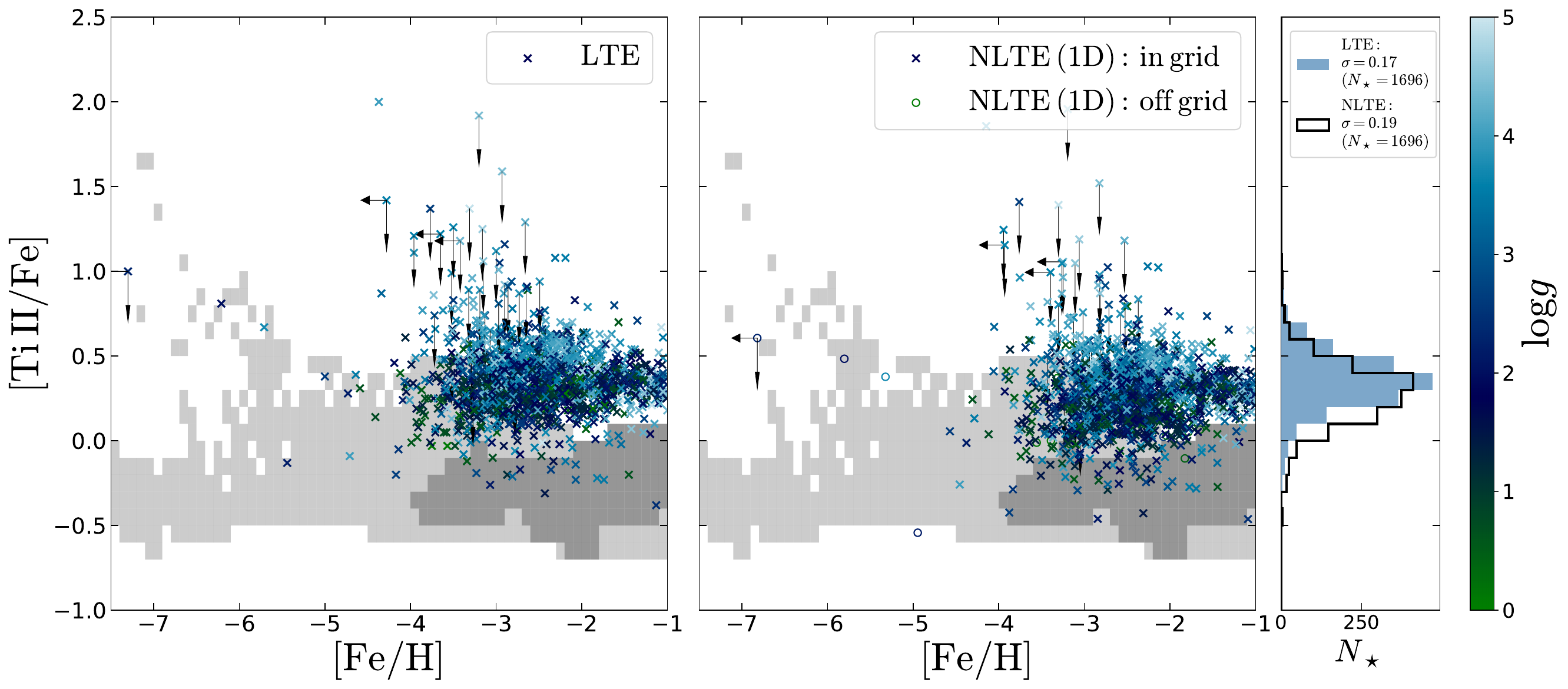} 
\caption{Same as Fig.~\ref{fig: nlte_NaAl}, but for [K/Fe] (top), [\ion{Ti}{I}/Fe] (middle) and [\ion{Ti}{II}/Fe] (bottom). Observed abundances have been corrected using the grids from \citet{Reggiani2019_K} for K, \citet{Bergemann2011_Ti1_Ti2} for \ion{Ti}{I} and \ion{Ti}{II}, and the mean corrections of \citet{Mashonkina2011_Fe1} and \citet{Bergemann2012_Fe1_Fe2} for \ion{Fe}{I}.}
\label{fig: nlteKCaTi}
\end{center}
\end{figure*}

\begin{figure*}
\begin{center}
    \includegraphics[width=0.95\hsize]{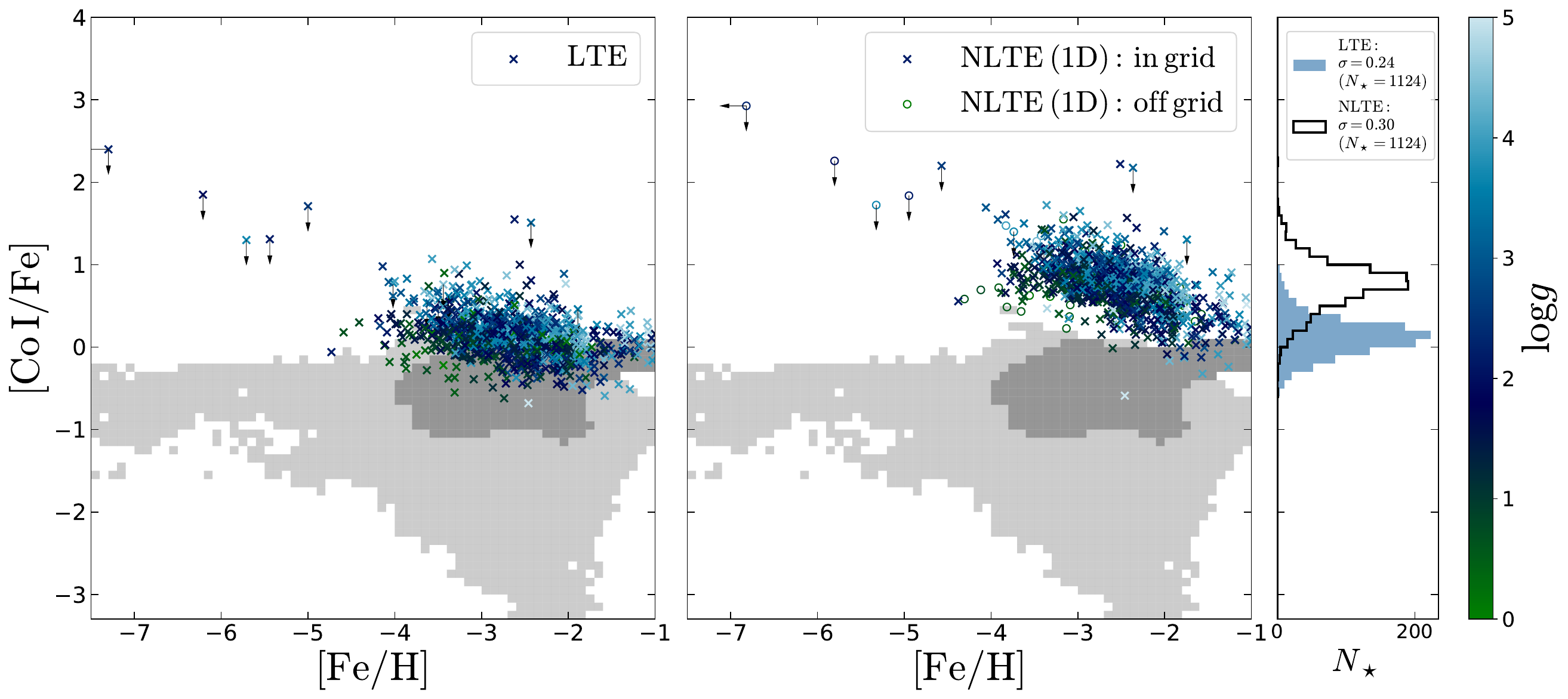} \\
    \includegraphics[width=0.95\hsize]{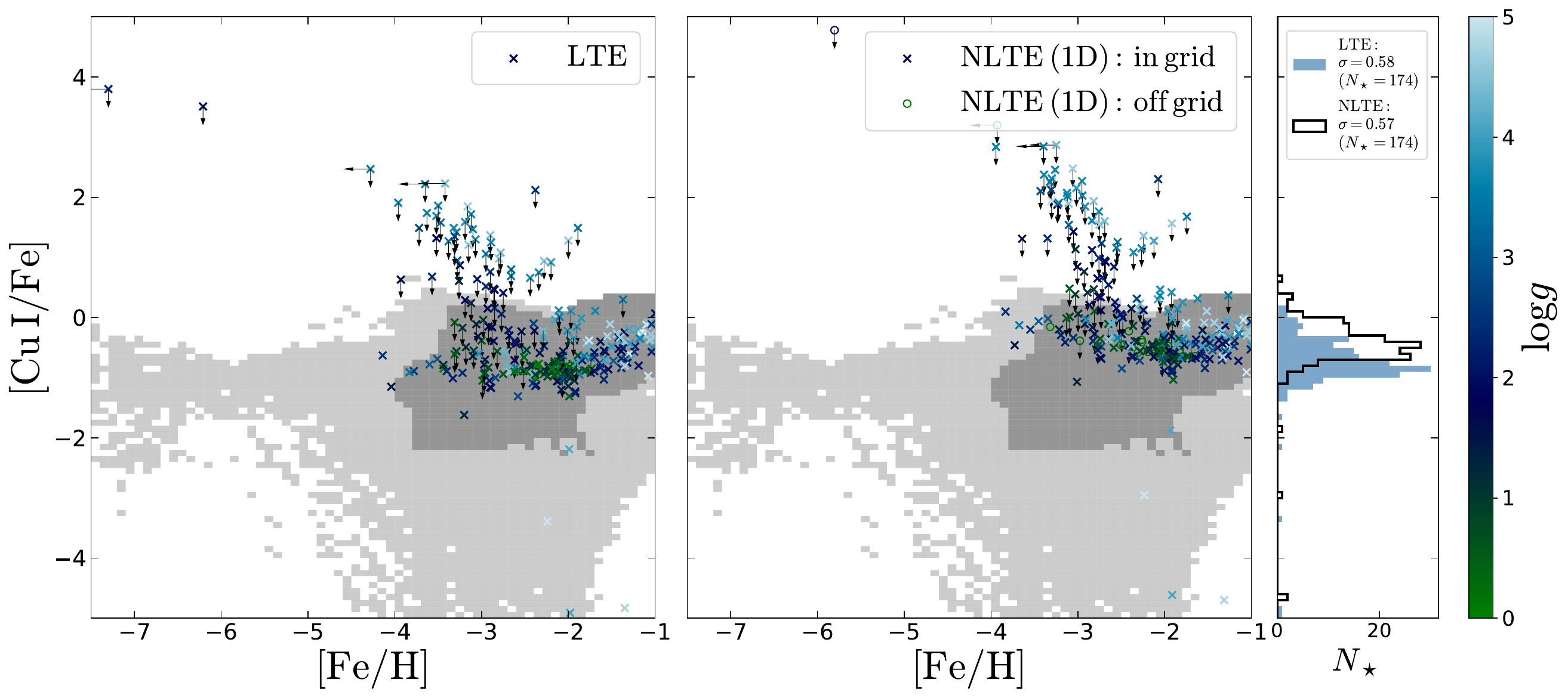} \\
    \includegraphics[width=0.95\hsize]{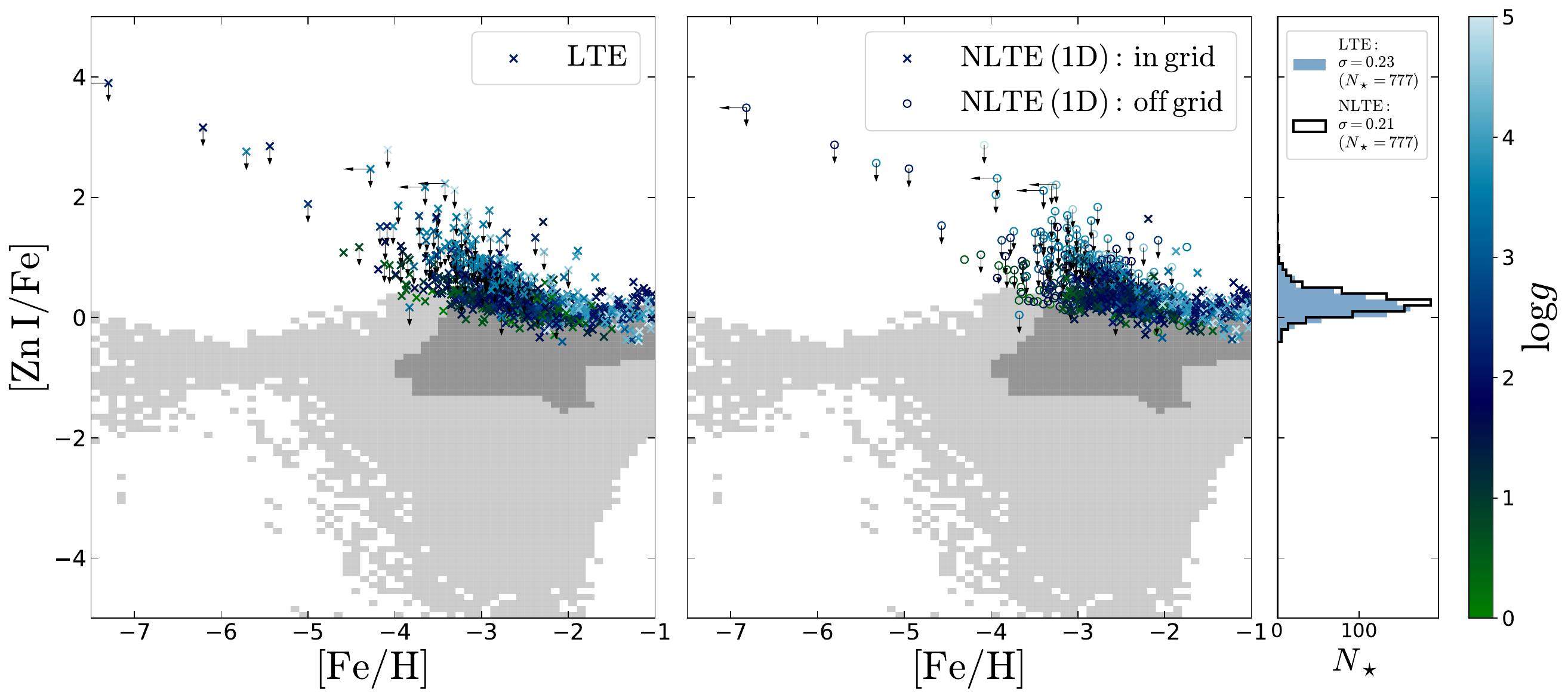} \\
\caption{Same as Fig.~\ref{fig: nlte_NaAl}, but for [Co/Fe] (top), [Cu/Fe] (middle) and [Zn/Fe] (bottom). Observed abundances have been corrected using the grid from \citet{Bergemann2010_Co} for \ion{Co}{I}, the linear least-squares fit to the corrections of \citet{Shi2018_Cu}, \citet{Andrievsky2018_Cu} and \citet{Xu2022_Cu} for Cu (see Sec.~\ref{copper}), the corrections of \citet{Sitnova2022_Zn1_Zn2} for \ion{Zn}{I} and the mean corrections of \citet{Mashonkina2011_Fe1} and \citet{Bergemann2012_Fe1_Fe2} for \ion{Fe}{I}.}
\label{fig: nlteCuZn}
\end{center}
\end{figure*}

In the following, we will discuss the comparison of the NLTE-SAGA database with our NEFERTITI model for elements not included in Sec.~\ref{sec:nefertiti}. The results are given in Figs.~\ref{fig: nlteO}-\ref{fig: nlteCuZn} for O, Si, K, Ti, Co, Cu, and Zn. The heavy elements Sr and Ba are currently outside the scope of the model. 

Fig.~\ref{fig: nlteO} presents the [O/Fe]--[Fe/H] distribution of inner-halo stars in our {\sc NEFERTITI} model, in comparison to all SAGA stars with measured \ion{O}{I} abundances. We find that the observed and predicted distributions are overall in good agreement. Most of the observed stars have been published with \ion{O}{I} abundances already corrected for NLTE effects, e.g. stars from \citet{Roederer14} who used the \ion{O}{I} NLTE corrections of \citet{Fabbian2009_OI}; \citet{Gratton2000} who used the prescription by \citet{Gratton1999}; \citet{Fulbright2003} who used the corrections of \citet{Takeda2000}; and others. The net [\ion{O}{I}/Fe] NLTE corrections for the rest of the stars are strongest at ${\rm [Fe/H]_{LTE}}<-2$, reaching $\sim-0.45\,$dex. In this [Fe/H] range, the \ion{O}{I} abundances of $\sim45\%$ of the stars are reported as upper limits. At $\rm [Fe/H]>-2$ the corrections for [\ion{O}{I}/Fe] are smaller, in the range $-0.2$ to 0\,dex (with the exception of two stars with high ${\rm [O/Fe]>+1}$).  

Fig.~\ref{fig: nlteMgAlSi} shows that the $\rm[Si/Fe]_{LTE}$ ratios from SAGA exhibit an increasing trend with decreasing 
$\logg$ at fixed metallicity. This trend disappears after correcting for NLTE effects. The overall distribution becomes flatter, more in agreement with other $\alpha$-elements such as Mg and Ca (see Fig.~\ref{fig: nlte_MgCa}). Furthermore, after applying the NLTE corrections, the scatter in [Si/Fe] decreases from $\sigma_{\rm LTE}=0.26$ to $\sigma_{\rm NLTE}=0.22$, see Fig.~\ref{fig: nlteMgAlSi}.

Fig.~\ref{fig: nlteKCaTi} displays [K/Fe], [\ion{Ti}{I}/Fe], and [\ion{Ti}{II}/Fe] as a function of [Fe/H]. We find that the observed LTE distributions of K and Ti lie $\sim1 \,$dex and $\sim 0.6\,$dex higher than our model predictions at $-2\leq {\rm [Fe/H]}\leq-1$. These discrepancies are consistent with previous findings that theoretical models systematically underproduce these elements compared to observations at all metallicities (see, e.g. ~\citealt{Nomoto2013, Kobayashi2020}). 
The negative NLTE corrections for [K/Fe] decrease the scatter of the distribution from $\sigma_{\rm LTE}=0.24$ to $\sigma_{\rm NLTE}=0.21$ and lower its peak by 0.4\,dex. However, the observed distribution still lies on the high end of our model predictions. Substituting the \citet{Reggiani2019_K} corrections with those of \citet{Takeda2002} - which differ by less than 0.1\,dex for stars where both grids are defined - lowers the peak by an additional 0.1\,dex, resulting in limited improvement. 
In contrast, the corrections for [\ion{Ti}{I}/Fe] shift the peak upwards by 0.4\,dex, further exacerbating the discrepancy with model predictions. The situation is similar for [\ion{Ti}{II}/Fe], where the LTE peak is 0.1\,dex higher than that of [\ion{Ti}{I}/Fe] and remains unchanged in NLTE.

Finally, Fig.~\ref{fig: nlteCuZn} shows the distributions of [Co/Fe], [Cu/Fe] and [Zn/Fe] as a function of [Fe/H]. 
We find that in LTE, the observed [Co/Fe] distribution lies on the high side of our model predictions. Applying NLTE corrections causes a shift of +0.6\,dex, raising the entire distribution above the predicted one. Furthermore, the scatter is increased from $\sigma_{\rm LTE}=0.24$ to $\sigma_{\rm NLTE}=0.30$, owing to a downward trend of [Co/Fe] with [Fe/H] that appears in NLTE.

The observed [Cu/Fe] distribution aligns well with our model in LTE, while the NLTE corrections shift its peak by +0.4\,dex, moving it towards the high side of our predictions. In addition, the distribution becomes flatter and even shows an upturn in the [Cu/Fe] trend at ${\rm [Fe/H]}\lesssim -2$, in agreement with recent calculations of NLTE Cu abundances in the Milky Way halo \citep{Caliskan2025}.

The observed distribution of Zn lies on the high end of our model predictions. This issue is similarly reported in \citet{Rossi2024}, who also use the \citet{Limongi2018} Pop~II yields, but not in \citet{Kobayashi2020} who use a different set of yields and a large fraction ($=50\%$) of Pop~II hypernovae. Indeed, the \citet{Limongi2018} yields for ccSNe predict a $\rm[Zn/Fe]<0$ at all masses, lower than in other published sets (see Fig. 5 of \citealt{Nomoto2013}). The NLTE corrections here, are minimal and do not improve the agreement; the peak is shifted by +0.1\,dex and the scatter is slightly decreased. 
Note that the predicted distributions for [Cu/Fe] and [Zn/Fe] are significantly extended due to the inclusion of zero-metallicity PISNe, which yield very low abundance ratios for these elements \citep[e.g.][]{Salvadori2019}. This primarily affects the light grey area, representing Pop~III descendants ($\geq50\%$), but also the dark grey area (Pop~II descendants, $>50\%$) in the case of significant Pop~III contribution. 

\end{document}